\documentclass[10pt,a4paper]{article}

\usepackage{amsmath,amsgen,latexsym}
\usepackage{amstext,amssymb,amsfonts,latexsym}
\usepackage{theorem}
\usepackage{pifont}
\usepackage{graphicx}

 \newcommand{\bs}{\bigskip}
 \newcommand{\ms}{\medskip}
 \newcommand{\n}{\noindent}
 \newcommand{\s}{\smallskip}
 \newcommand{\hs}[1]{\hspace*{ #1 mm}}
 \newcommand{\vs}[1]{\vspace*{ #1 mm}}

 \newcommand{\setempty}{\varnothing}
 \newcommand{\real}{\mathbb{R}}
 \newcommand{\nat}{\mathbb{N}}
 
 \newcommand{\integer}{\mathbb{Z}}

 \newcommand{\eg}{\textrm{e.g.},\hspace*{2mm}}
 
 \newcommand{\etalc}{\textrm{et al.}}

 \newcommand{\CC}{{\cal C}}
 
 \newcommand{\DD}{{\cal D}}

 \newcommand{\SSS}{{\cal S}}
 \newcommand{\TT}{{\cal T}}
 \newcommand{\PP}{{\cal P}}

 \newcommand{\dl}{\mathrm{L}}
 \newcommand{\nl}{\mathrm{NL}}
 \newcommand{\p}{\mathrm{P}}
 \newcommand{\np}{\mathrm{NP}}

 \newcommand{\fl}{\mathrm{FL}}

 \newcommand{\cfl}{\mathrm{CFL}}
 
 \newcommand{\dcfl}{\mathrm{DCFL}}

\theoremstyle{plain}
\theoremheaderfont{\bfseries}
 \newtheorem{theorem}{Theorem}[section]
 \newtheorem{lemma}[theorem]{Lemma}
\newtheorem{proposition}[theorem]{{\bf Proposition}}
 \newtheorem{corollary}[theorem]{Corollary}

 {\theorembodyfont{\rmfamily}
  \newtheorem{definition}[theorem]{Definition}}
 {\theorembodyfont{\rmfamily} \newtheorem{example}[theorem]{Example}}
 {\theorembodyfont{\rmfamily} }

 \newenvironment{proofof}[1]{\vspace*{5mm} \par \noindent
         {\bf Proof of #1.\hs{2}}}{\hfill$\Box$ \vspace*{3mm}}

 \newenvironment{proof}{\par \noindent
            {\bf Proof. \hs{2}}}{\hfill$\Box$ \vspace*{3mm}}

 \newcommand{\ceilings}[1]{\lceil #1 \rceil}
 
 \newcommand{\pair}[1]{\langle #1 \rangle}

\newcommand{\ignore}[1]{}

\newcommand{\lo}{\mathrm{LO}}
\newcommand{\nlo}{\mathrm{NLO}}

\newcommand{\lsas}{\mathrm{LSAS}}

\newcommand{\auxl}{\mathrm{auxL}}
\newcommand{\auxfl}{\mathrm{auxFL}}

\newcommand{\psublin}{\mathrm{PsubLIN}}

\newcommand{\search}{\mathrm{Search\mbox{-}}}
\newcommand{\dstcon}{\mathrm{DSTCON}}

\newcommand{\Lreduces}{\leq^{\mathrm{L}}_{m}}
\newcommand{\LTreduces}{\leq^{\mathrm{L}}_{T}}
\newcommand{\sLreduces}{\leq^{\mathrm{sL}}_{m}}
\newcommand{\sLTreduces}{\leq^{\mathrm{sL}}_{T}}
\newcommand{\SLRFreduces}{\leq^{\mathrm{SLRF}}_{T}}
\newcommand{\sSLRFreduces}{\leq^{\mathrm{sSLRF}}_{T}}

\newcommand{\sLequiv}{\equiv^{\mathrm{sL}}_{m}}
\newcommand{\sLTequiv}{\equiv^{\mathrm{sL}}_{T}}

\newcommand{\sSLRFequiv}{\equiv^{\mathrm{sSLRF}}_{T}}

\newcommand{\pdtimespace}{\mathrm{PDTIME},\!\mathrm{SPACE}}

\newcommand{\dtimespace}{\mathrm{DTIME},\!\mathrm{SPACE}}

\newcommand{\para}{\mathrm{para}\mbox{-}}
\newcommand{\boldvec}[1]{\mbox{\boldmath $ #1 $}}
\newcommand{\logdcfl}{\mathrm{LOGDCFL}}
\newcommand{\logcfl}{\mathrm{LOGCFL}}

\begin{document}
%%%%%%%%%%%%%%%%%%
%%%%%%%%%%%%%%%%%%
%%%
%%%
\pagestyle{plain}
\pagenumbering{arabic}
\setcounter{page}{1}
\setcounter{footnote}{1}

\begin{center}
{\Large {\bf The 2CNF Boolean Formula Satisfiability Problem and \s\\
the Linear Space Hypothesis}}\footnote{This work was done in part while serving as a visiting professor at the University of Toronto between August 1, 2016 and March 30, 2017 and was supported by the Natural Sciences and Engineering Research Council of Canada.}\footnote{This current article corrects and extends its preliminary report that has appeared in the Proceedings of the
42nd International Symposium on
Mathematical Foundations of Computer Science (MFCS 2017),
August 21--25, 2017, Aalborg, Denmark, Leibniz International Proceedings in Informatics (LIPIcs), Schloss Dagstuhl - Leibniz-Zentrum f\"{u}r Informatik 2017, vol. 83,  pp. 62:1--62:14, 2017.}
\bs\s\\

{\sc Tomoyuki Yamakami}\footnote{The author is grateful to Stephen A. Cook for his brief comments on this work while visiting the University of Toronto.}\footnote{Present Affiliation: Faculty of Engineering, University of Fukui, 3-9-1 Bunkyo, Fukui 910-8507, Japan}
\ms\\
\end{center}

%%%%%%%%%%%%%%%%%%%%%%%%%%%%%%%%%%
%%%%%%%%%%%%%%%%%%
\sloppy

\begin{abstract}
We aim at investigating the solvability/insolvability of nondeterministic logarithmic-space (NL) decision, search, and optimization problems parameterized by natural size parameters using simultaneously polynomial time and sub-linear space. We are particularly focused on $\mathrm{2SAT}_3$---a restricted variant of the 2CNF Boolean (propositional) formula satisfiability problem in which each variable of a given 2CNF formula appears at most 3 times in the form of literals---parameterized by the total number $m_{vbl}(\phi)$ of variables of each given Boolean formula $\phi$.
We propose a new, practical working hypothesis, called the linear space hypothesis (LSH), which asserts that $(\mathrm{2SAT}_3,m_{vbl})$ cannot be solved in polynomial time using only ``sub-linear'' space (i.e., $m_{vbl}(x)^{\varepsilon}\, polylog(|x|)$ space for a constant $\varepsilon\in[0,1)$) on all instances $x$. Immediate consequences of LSH include $\dl\neq\nl$, $\logdcfl\neq\logcfl$, and $\mathrm{SC}\neq \mathrm{NSC}$.
For our investigation, we fully utilize a key notion of ``short reductions'', under which the class $\mathrm{PsubLIN}$ of all parameterized polynomial-time sub-linear-space solvable problems is indeed closed.

\ms
\n{\bf keywords}: 
2CNF Boolean formula satisfiability, 
parameterized decision problem, 
sub-linear-space computability, 
linear space hypothesis, 
short reduction, 
syntactic NL
\end{abstract}

%%%%%%%%%%%%%%%%%
%%%%%%%%%%%%%%%%%
\sloppy
\section{Motivational Discussion and Quick Overview}

%%%%
\subsection{Space Complexity of Parameterized 2SAT}\label{sec:introduction}

Since Cook \cite{Coo71} demonstrated its $\np$-completeness in 1971, the \emph{Boolean (propositional) formula satisfiability problem} (SAT) of determining whether or not a given Boolean formula is satisfied by a suitably-chosen variable assignment has been studied extensively for about 50 years. As its restricted variant, the \emph{$k$CNF Boolean (propositional) formula satisfiability problem} ($k$SAT) for an integer index $k\geq3$, whose input formulas are of  $k$-conjunctive normal form ($k$CNF),  has also been a centerpiece of computational complexity theory. Since $k$SAT is complete for $\np$ (nondeterministic polynomial time) \cite{Coo71}, its solvability is linked to the computational complexity of all other $\np$ problems in such a way that, if $k$SAT is solvable in polynomial time, then so are all other NP problems. It is also known that $k$SAT is polynomial-time reducible to $3$SAT and vice versa.
A recent study has been focused on the solvability of $k$SAT with $n$ Boolean variables and $m$ clauses within  ``sub-exponential'' (which means $2^{\varepsilon n} poly(n+m)$  for an absolute constant $\varepsilon\in(0,1)$ and a suitable polynomial $poly(\cdot)$) runtime.
In this line of study,  Impagliazzo, Paturi, and Zane \cite{IPZ01} took a new approach toward $k$SAT and its search version, $\search{k}\mathrm{SAT}$, \emph{parameterized by} the number $m_{vbl}(x)$ of all Boolean variables and the number $m_{cls}(x)$ of all clauses in a given $k$CNF formula $x$ as natural ``size parameters'' (which were called ``complexity parameters'' in \cite{IPZ01}).
To discuss such sub-exponential-time solvability for a wide range of $\np$-complete problems, Impagliazzo et al. further  devised  a crucial notion of  \emph{sub-exponential-time reduction family (or SERF-reduction)}, which preserves the sub-exponential-time complexity, and they cleverly demonstrated that the aforementioned size parameters $m_{vbl}(x)$ and $m_{cls}(x)$  make $\search{k}\mathrm{SAT}$ \emph{SERF-equivalent} (namely, the two problems are SERF-reducible to each other). Motivated by a fine role of SERF-reducibility,  Impagliazzo and Paturi \cite{IP01} formally proposed, as a working hypothesis, the \emph{exponential time hypothesis} (ETH), and later Impagliazzo, Paturi, and Zane \cite{IPZ01} extended it into the \emph{strong exponential time hypothesis} (SETH), which  asserts the insolvability of $k$SAT parameterized by $m_{vbl}(x)$ (succinctly denoted in this paper by $(k\mathrm{SAT},m_{vbl})$) within sub-exponential time for all indices $k\geq3$.  Their hypothesis turned out to be a stronger assertion than   $\p\neq\np$ and it has  led  to intriguing consequences, including finer lower bounds on the solvability of various parameterized NP problems (see, e.g., a survey \cite{LMS11}).

Whereas SETH concerns with $k$SAT for $k\geq3$, we are focused on the remaining case of $k=2$. The decision problem $2$SAT is known to be complete\footnote{This is because Jones, Lien, and Laaser \cite{JLL76} demonstrated the $\nl$-completeness of the complement of $\mathrm{SAT}$ (called $\mathrm{UNSAT}_2$ in \cite{JLL76}) and Immerman \cite{Imm88} and Szelepcs\'{e}nyi \cite{Sze88} proved the closure of $\nl$ under complementation.} for $\nl$ (nondeterministic logarithmic space) under log-space reductions.
Since  $\mathrm{2SAT}$ already enjoys a polynomial-time algorithm (because of  $\nl\subseteq\p$), we are more concerned with how much memory space such an algorithm  demands to run. An elaborate algorithm solves $\mathrm{2SAT}$ with $n$ variables and $m$ clauses using simultaneously polynomial time and  $(n/2^{c\sqrt{\log{n}}})\, polylog(m+n)$ space (Theorem \ref{2SAT-solvable}), where $c>0$ is a fixed constant and $polylog(\cdot)$ is a suitable polylogarithmic function. This space bound is slightly below $n$; however, it is not yet known that  2SAT parameterized by $m_{vbl}(x)$ or $m_{cls}(x)$ can be solved in polynomial time using strictly ``sub-linear'' space, where the informal term ``sub-linear'' for a size parameter $m(x)$ refers to  a function of the form $m(x)^{\varepsilon} \ell(|x|)$ on input instances $x$  for a certain absolute constant $\varepsilon\in[0,1)$ and an appropriately-chosen polylogarithmic function $\ell(n)$. Of course, this multiplicative factor $\ell(|x|)$ becomes redundant if $m(x)$ is relatively large (for example, $m(x)\geq \log^{k}|x|$ for a large constant $k>0$) and thus ``sub-linear'' turns out to be simply $m(x)^{\varepsilon}$.

In parallel to a restriction of SAT onto $k$SAT, we further limit $\mathrm{2SAT}$ to $\mathrm{2SAT}_k$ for polynomial-time sub-linear-space solvability, where $\mathrm{2SAT}_k$ (\emph{$k$-bounded $\mathrm{2SAT}$}) consists of all satisfiable formulas in which  each variable appears as literals in at most $k$ clauses. Notice that $\mathrm{2SAT}_k$ for each $k\geq3$ is also $\nl$-complete (Proposition \ref{2SAT_k-complete}) as 2SAT is; by contrast, $k\mathrm{SAT}_2$  already falls into $\dl$ for any index $k\geq2$.

%%%%%
\subsection{Sub-Linear Space and Short Reductions}\label{sec:sub-linear-short}

A decision problem $L$ parameterized by a size parameter $m$ forms a \emph{parameterized decision problem} expressed as $(L,m)$.
Naturally, we can expand  $\dl$ (logarithmic-space complexity class) and $\p$ (polynomial-time complexity class) by parameterizing their associated  decision problems and, for clarity reason, we respectively denote these  parameterized complexity classes by $\para\dl$ and $\para\p$.
Similarly, all parameterized decision problems solvable in polynomial time using sub-linear space then form a \emph{parameterized complexity class} $\psublin$  (whose prefix ``P'' refers to ``polynomial time'' and suffix ``subLIN'' stands for ``sub-linear space''), which is located between $\para\dl$ and $\para\p$.
This new complexity class $\psublin$ naturally includes, for example, DCFL (deterministic context-free) because Cook \cite{Coo79} earlier showed that every language in DCFL  is recognized in polynomial time using $O(\log^2n)$-space,\footnote{All decision problems recognized in polynomial time using polylogarithmic space form Steve's class $\mathrm{SC}$ \cite{Coo79}.}
where $n$ indicates input size.
Unfortunately, there is no known separation among $\para\dl$, $\para\nl$, $\psublin$, and $\para\p$.

It turns out that $\psublin$ does not seem to be closed under standard log-space reductions; thus, those reductions  are no longer useful tools to discuss the solvability of NL-complete problems within polynomial time and  sub-linear space. Therefore, we need to introduce a much weaker form of reductions, called \emph{short reductions}, which preserve the polynomial-time sub-linear-space complexity.
Intuitively speaking, a short reduction is a reduction between two parameterized  decision problems computed by  a \emph{reduction machine} (or a \emph{reduction function}) that can generate strings of certain sizes proportional to or less than sizes of its input strings.
In particular, we will define three types of such short reductions in Section \ref{sec:complete}: \emph{short L-m-reducibility} ($\sLreduces$), \emph{short L-T-reducibility} ($\sLTreduces$), and \emph{short SLRF-T-reducibility} ($\sSLRFreduces$), where SLRF stands for \emph{sub-linear space reduction family}.
Note that any $\sSLRFreduces$-reduction is an $\SLRFreduces$-reduction but the converse is not true because there is a pair of problems reducible by $\SLRFreduces$-reductions but not by $\sSLRFreduces$-reductions (Lemmas  \ref{L-vs-SLRF-reduction}(2) and \ref{reduction-property}(3)).

As noted earlier, Impagliazzo et al. demonstrated in \cite[Corollary 2]{IPZ01} that, for any number $k\geq3$,  $(k\mathrm{SAT},m_{vbl})$ is SERF-equivalent to $(k\mathrm{SAT},m_{cls})$.  Similarly, we can construct a short reduction from $\mathrm{2SAT}_3$ with $m_{vbl}$ to $\mathrm{2SAT}_3$ with $m_{cls}$, and vice verse; in other words, $(\mathrm{2SAT}_3,m_{vbl})$ and $(\mathrm{2SAT}_3,m_{cls})$ are equivalent under short L-T-reductions (Lemma \ref{basic-property-SAT}(2)). On the contrary, such equivalence is not known for $\mathrm{2SAT}$ and this circumstance signifies the importance of $\mathrm{2SAT}_3$ in the study of polynomial-time sub-linear-space computability.

Another importance of $\mathrm{2SAT}_3$ can be demonstrated by showing that $\mathrm{2SAT}_3$ is hard for a natural subclass of $\nl$, which we call \emph{Syntactic NL} or simply \emph{SNL} under short $\mathrm{SLRF}$-T-reductions.
An \emph{SNL sentence} $\Phi$ is of the form $\exists^f P_1\cdots \exists^f P_l \forall i_1\cdots \forall i_r \forall y_1\cdots \forall y_s \: \psi$, starting with second-order existential ``functional'' quantifiers, followed by first-order universal quantifiers, ranging over specified domains. Their precise definitions will be given in Section \ref{sec:2SAT-definition}. Such an SNL sentence can capture essential, syntactic structures of some of $\nl$-complete problems.
Let $\para\mathrm{SNL}$ stand for the collection of all parameterized decision problems $(P,m)$ in which $P$ is expressed ``syntactically'' by an appropriate SNL-formula $\Phi$ and $m(x)$ equals the size of one of the domains associated with $\Phi$.
For its slightly restricted subclass $\para\mathrm{SNL}_{\omega}$,
we further claim in Proposition \ref{hardness-SNL} that $(\mathrm{2SAT}_3,m_{vbl})$ is hard for $\para\mathrm{SNL}_{\omega}$ under short SLRF-T-reductions.

%%%
%%%
\subsection{A New, Practical Working Hypothesis for 2SAT$_3$}\label{sec:new-hypothesis}

Since its introduction in 2001, ETH and SETH for $k$SAT ($k\geq3$) have served as a driving force to obtain finer lower bounds on the sub-exponential-time computability of various parameterized NP problems, since those bounds do not seem to be obtained directly from the popular assumption of $\p\neq\np$.
In a similar vein, we wish to propose a new, practical working hypothesis, called the {\em linear space hypothesis} (LSH) for $\mathrm{2SAT}_3$, in which no deterministic algorithm solves $(\mathrm{2SAT}_3,m_{vbl})$ simultaneously in polynomial time and sub-linear space. More precisely:

\begin{definition}\label{LSH-definition}
[{\sc The Linear Space Hypothesis (LSH) for $\mathrm{2SAT}_3$ with $m_{vbl}$}]
For any choice of $\varepsilon\in [0,1)$ and any polylogarithmic function $\ell$, no deterministic Turing machine solves $\mathrm{2SAT}_3$ parameterized by $m_{vbl}$ in time polynomial in $m(x)$ using $m_{vbl}(x)^{\varepsilon} \ell(|x|)$ space, where $x$ refers to an input instance to $\mathrm{2SAT}_3$.
\end{definition}

It turns out that $m_{vbl}$ in the above definition can be replaced by $m_{cls}$ (Proposition \ref{m_vbl-equivalent-m_cls}(2)), and we therefore tend to omit the reference to $m_{vbl}$ and refer to the hypothesis as ``LSH for $2\mathrm{SAT}_3$.''
From this working hypothesis LSH for $\mathrm{2SAT}_3$, we can draw various  conclusions, including $\dl\neq\nl$, $\logdcfl\neq\logcfl$, and $\mathrm{SC}\neq\mathrm{NSC}$ (Theorem \ref{LSH-implies-L-NL}), where $\mathrm{NSC}$ is the nondeterministic counterpart of Steve's class $\mathrm{SC}$.
This consequence allows us to expect that LSH for $\mathrm{2SAT}_3$ will lead to finer, better consequences than what the popular assumption like  $\dl\neq\nl$ can lead to.

We present in Section \ref{sec:alternative} two useful characterizations of the hypothesis LSH for $\mathrm{2SAT}_3$. The first characterization concerns with the \emph{$(2,k)$-entry $\{0,1\}$-linear  programming problem} ($\mathrm{LP}_{2,k}$), in which we check whether there exists a $\{0,1\}$-vector $x$ satisfying $Ax\geq b$ for
a rational $m\times n$ matrix $A$ and a rational vector $b$ of dimension $n$, provided that each row of $A$ has at most two nonzero entries and  each column of $A$ has at most $k$ non-zero entries.
For this problem, we use two size parameters $m_{col}(x)$ and $m_{row}(x)$, which respectively indicate the numbers of columns and of rows of $A$ for any instance $x=(A,b)$ given to $\mathrm{LP}_{2,k}$.

The second characterization uses the \emph{degree-$k$ directed $s$-$t$ connectivity problem} ($k\dstcon$)
of asking whether a path from a given vertex $s$ to another vertex $t$ exists in a given directed graph $G$ whose vertices have degree  (i.e., indegree plus outdegree) at most $k$.
For an instance $x=(G,s,t)$ to $k\dstcon$, we concentrate on two size parameters $m_{ver}(x)$ and $m_{edg}(x)$, which respectively denote the number of vertices and that of edges in $G$.

As a characterization theorem (Theorem \ref{LSH-equiv}), we prove that the following two statements are logically equivalent to LSH for $\mathrm{2SAT}_3$: (i) LSH for $\mathrm{LP}_{2,3}$ with either $m_{row}$ or $m_{col}$ and (ii) LSH for ${3}\dstcon$ with either $m_{ver}$ or $m_{edg}$. These characterizations are shown by the use of the aforementioned short  reducibilities, which are specifically designed to preserve the polynomial-time sub-linear-space computability.

%%%%

Naturally, we can ask what consequences we can draw from our working hypothesis LSH for $2\mathrm{SAT}_3$.
In Section  \ref{sec:application}, we will present four examples of such applications.

(1) As the first example, we look into \emph{NL search problems}. A special  NL search problem,  called  $\search\mathrm{1NFA}$, is to find an input of length $n$ accepted by a given 1nfa $M$ that disallows any $\lambda$-move (or $\lambda$-transition). We use the size parameter $m_{nfa}(x)=|Q||\Sigma|n$ defined for each instance $x = (M,1^n)$ given to $\search\mathrm{1NFA}$. We show that, for any constant $\varepsilon\in(0,1/2)$,  no $|x|^{O(1)}$-time $O(m_{nfa}(x)^{1/2-\varepsilon})$-space algorithm solves $\search\mathrm{1NFA}$ on all instances $x$ (Theorem \ref{Search-1NFA}).

(2) In another NL search problem, $\search\mathrm{UOCK}$, we are asked for  finding, for a given string $w$ and for a sequence $(w_1,w_2,\ldots,w_n)$ of $n$ strings, an index sequence $(i_1,\ldots,i_k)$ in increasing order that makes the concatenation $w_{i_1}\cdots w_{i_k}$  equal to $w$. As a natural size parameter, we set $m_{elm}(x)=n$ for each instance $x$ of the form $(w,(w_1,w_2,\ldots,w_n))$. We then show that,  for any constant $\varepsilon>0$, there is no $|x|^{O(1)}$-time $O(m_{elm}(x)^{1/2-\varepsilon})$-space algorithm working for $\search\mathrm{UOCK}$ on all instances $x$ (Theorem \ref{search-UOCK}).

(3) Let us consider an \emph{NL optimization (or NLO) problem}, called $\mathrm{Max\mbox{-}HPP}$, studied in \cite{Tan07,Yam13}. By setting $m_{col}(x)=n$ for each instance $x$, we show that  no $|x|^{O(1)}$-time $O(k^{1/3}\log{m_{col}(x)})$-space algorithm that finds $(1+\frac{1}{k})$-approximate solutions for  $\mathrm{Max\mbox{-}HPP}$ on all instances $x$ (Theorem  \ref{Max-CPath-Weight-solver}).

(4) As the fourth example, we are focused on unary 1nfa's, which work on single-letter input alphabets. For any given $n$-state unary 1nfa, there is always a unary 1dfa of $O(n\log{n})$ states that produces the same outputs as the 1nfa does (see, \eg  \cite{HU79}). We show that, for any constant $\varepsilon\in(0,1)$,  there is no $|x|^{O(1)}$-time $O(n^{\varepsilon})$-space algorithm that converts each $n$-state unary 1nfa into an equivalent 1dfa of $O(n\log{n})$ state (Theorem \ref{1nfa-1dfa}).

%%%%
\subsection{A Recent Progress after Our Preliminary Report}

Since the first publication of this paper as a preliminary report \cite{Yam17a}, there have been five papers \cite{Yam17b,Yam18a,Yam19,Yam22a,Yam22b} addressing fascinating  results relevant to the linear space hypothesis.

In \cite{Yam18a}, the parameterized decision problems are proven to be closely connected to state-complexity classes defined by finite automata. In particular, the author presented a characterization of the hypothesis LSH for $2\mathrm{SAT}_3$ using the state complexity of transformation of certain types of nondeterministic finite automata to deterministic ones. More importantly, the ``nonuniform'' version of LSH for $2\mathrm{SAT}_3$ (called the \emph{nonuniform linear space hypothesis}) was further introduced and characterized in terms of certain types of nonuniform state complexity.
In \cite{Yam17b}, the author explored the properties of the short reducibilities, further expanded the scope of these reducibilities to search problems, and drew a picture of a large landscape of the relative  complexity of various combinatorial problems via short reductions.
In \cite{Yam22a}, three additional parameterized decision problems were introduced and proven to be ``equivalent'' to $(\mathrm{2SAT}_3,m_{vbl})$ by  short reductions.
In \cite{Yam22b}, the author further studied \emph{fine-grained space complexities} of various parameterized decision problems in direct connection to LSH for $\mathrm{2SAT}_3$.
In \cite{Yam19}, the author provided circumstantial evidences that the validity of LSH for $\mathrm{2SAT}_3$ might be quite difficult to verify by demonstrating two contradictory relativized worlds in which LSH holds and LSH fails.

We hope that the working hypothesis LSH for $\mathrm{2SAT}_3$ (and also its nonuniform version) will stimulate the study on the space complexity of ``tractable'' problems in general and eventually lead us to a rich research area that we have not yet seen in the past literature.

%%%%%%%%%%%%%%%%%
%%%%%%%%%%%%%%%%%
%%%%%%%%%%%%%%%%%
\section{Basic Concepts and Notation}\label{sec:preliminaries}

This section will provide fundamental concepts and notation necessary to read through the rest of the paper.
Let $\nat$ denote the set of all {\em natural numbers} (i.e., nonnegative integers) and set $\nat^{+}=\nat-\{0\}$. Two notations $\real$ and $\real^{\geq0}$ denote respectively the set of all {\em real numbers} and that of all {\em nonnegative real numbers}. For any two integers $m$ and $n$ with $m\leq n$, the \emph{integer interval} $[m,n]_{\integer}$ refers to
the set $\{m,m+1,m+2,\ldots,n\}$. We particularly abbreviate $[1,n]_{\integer}$ as $[n]$ when $n\geq1$. For technical convenience, when $n<1$, we additionally set $[n]=\setempty$.
Given a set $S$, $\PP(S)$ expresses the \emph{power set} of $S$, namely, the set of all subsets of $S$. For two sets $A$ and $B$, their \emph{symmetric difference} $A\triangle B$ is the set $(A-B)\cup (B-A)$. For a basic knowledge on graphs, see textbooks,  e.g., \cite{GYZ14}. Additionally, for a vertex $v$ of a graph, we use the notation $deg(v)$ to denote the degree of $v$, where the generic term ``{degree}'' refers to the sum of the indegree and outdegree.

In this paper, all {\em polynomials} are assumed to have nonnegative integer coefficients and all {\em logarithms} are taken to the base $2$, provided that ``$\log{0}$'' is conventionally set to be $0$.
A {\em polylogarithmic (or polylog) function} $\ell$ is a function mapping $\nat$ to $\real^{\geq0}$ such that there exists a polynomial $p$ for which $\ell(n)=p(\log{n})$ holds for all $n\in\nat$.
For the sake of readability, we abuse the informal notation ``$polylog(n)$'' to indicate an appropriately-chosen polylog function $\ell$ in $n$. Let $\mathrm{Polylog}$ denote the set of all polylog functions.

An \emph{alphabet} is a finite nonempty set $\Sigma$ of ``symbols'' or ``letters,'' and a \emph{string} over $\Sigma$ is a finite sequence of symbols in $\Sigma$. The \emph{length} of a string $x$ is the total number of symbols used in $x$ and we use the notation $|x|$ to denote it. The \emph{empty string} has  length $0$ and it is always denoted by $\lambda$. Let $\Sigma^*$ denote the set of all strings over $\Sigma$. A \emph{language} over $\Sigma$ is a subset of $\Sigma^*$. Hereafter, we freely identify a decision problem $P$ with its associated language. Given a number $n\in\nat$, the notation $\Sigma^n$ (resp., $\Sigma^{\leq n}$, $\Sigma^{\geq n}$) denotes the set of all strings of length exactly $n$ (resp., length at most $n$, length at least $n$) over $\Sigma$. Moreover, we set $\Sigma^{<n} = \Sigma^{\leq n} - \Sigma^n$ and $\Sigma^{>n} = \Sigma^{\geq n} - \Sigma^n$.
The notation $\cfl$ refers to the family of all \emph{context-free languages} and $\dcfl$ is the deterministic counterpart of $\cfl$. Refer to textbooks, e.g., \cite{HU79} for more detail.

\emph{Turing machines} are generally used to \emph{solve} a mathematical problem. Throughout this paper, we use the following form of Turing machines. Our Turing machine consists of a read-only input tape, (possibly) a write-once output tape, and a constant number of read/write work tapes. A tape head of a \emph{write-once} tape moves to the right if it writes a non-blank symbol, and it stays still on a blank tape cell otherwise. All other tape heads always move in both directions (to the right and to the left). We sometimes abbreviate a nondeterministic Turing machine as an NTM and a deterministic Turing machine as a DTM. As customary, we define $\p$ as the complexity class of all decision problems solved by DTMs in polynomial time. As its subclasses, $\mathrm{SC}$ (resp., $\dl$) is defined in terms of polynomial-time, polylogarithmic-space (resp., logarithmic-space) DTMs.  Moreover, $\mathrm{NSC}$ and $\nl$ denote respectively the nondeterministic counterparts of $\mathrm{SC}$ and $\dl$.

In a course of our study on polynomial-time sub-linear-space computability, it is convenient to expand the standard framework of decision problems to problems that are \emph{parameterized} by properly chosen ``size parameters'' (called ``complexity parameters'' in \cite{IPZ01}), which serve as a basis unit of the time/space complexity of an underlying algorithm.
In this respect, we naturally follow a framework of Impagliazzo, Paturi, and Zane \cite{IPZ01} to work with a flexible choice of size parameters.

Given an alphabet $\Sigma$ of a target decision problem, a \emph{size parameter} is formally a function from $\Sigma^*$ to $\nat$.
A \emph{standard size parameter} from $\Sigma^*$ to $\nat$ measures the total number $|x|$ of symbols in an instance $x$ over $\Sigma$ and it is often denoted by $||$.
For any size parameter $m:\Sigma^*\to\nat$, we say that $m$ is \emph{polynomially bounded} if there exists a polynomial $p$ satisfying $m(x)\leq p(|x|)$ for all strings $x\in\Sigma^*$ and that $m$ is \emph{polynomially honest} if there is a polynomial $p$ such that $|x|\leq p(m(x))$ for all $x\in\Sigma^*$. In addition, $m$ is said to be \emph{almost non-zero} if there is a finite set $S\subseteq\Sigma^*$ satisfying $m(x)>0$ for all instances $x$ in $\Sigma^*-S$.

In the rest of this paper, we focus our attention on the following ``log-space'' size parameters.

\begin{definition}
A {\em log-space size parameter} $m(x)$ for a mathematical problem $P$ is a function mapping $\Sigma^*$  (where $\Sigma$ is an input alphabet used for $P$) to $\nat$ such that
(1) $m$ is polynomially bounded for all instances of $P$ and
(2) $m$ must be computed by a certain DTM that takes input $x$ and outputs $m(x)$ \emph{in unary} (i.e., produces a string $1^{m(x)}$) on a write-once output tape using at most $c\log{|x|}+d$ space for certain fixed constants $c,d>0$.
\end{definition}

As key examples of log-space size parameters, for any graph-related problem (such as $3\dstcon$ in Section \ref{sec:alternative}), we define  $m_{edg}(x)$ and $m_{ver}(x)$ respectively to be the total number of edges and that of vertices in each graph instance $x$ given to the problem. It is not difficult to show that $m_{ver}$ and $m_{edg}$ are indeed  log-space size parameters.
To emphasize the use of size parameter $m$, we explicitly write $(P,m)$ instead of $P$.
For clarity reason, we call this pair $(P,m)$ a \emph{parameterized decision problem} and any collection of such parameterized decision problems is conveniently called a \emph{parameterized complexity class}.
Given a size parameter $m$, we say that a multi-tape Turing machine $M$ uses {\em logarithmic space} (or {\em log space}, for short) \emph{with respect to} $m$ if there exist two absolute constants $c,d\geq0$ for which each of the work tapes (not including input and output tapes) used by $M$ on $x$ are upper-bounded by $c\log{m(x)}+d$
on every input $x$. Similarly, $M$ runs in \emph{polynomial time with respect to} $m$ if there is a polynomial $p$ such that $M$ halts within $p(m(x))$ steps for all instances $x$.
As long as $m$ is polynomially bounded and almost non-zero, those two concepts with respect to $m$ coincide with the ones with respect to the standard size parameter $||$.

Two specific notations $\dl$ and $\nl$ respectively stand for the classes of all decision problems solvable on multi-tape deterministic and nondeterministic Turing machines using log space with respect to $||$. It is known that the additional requirement of ``polynomial runtime'' does not change those classes. To distinguish from a standard complexity class $\CC$, we use the prefix ``para-'' as in $\para\CC$ to denote the corresponding parameterized complexity class.
By appropriately parameterizing all standard decision problems, we obtain a parameterization of standard complexity classes. For example, the \emph{parameterized NL}, denoted by $\para\nl$, is composed of all parameterized decision problems $(K,m)$ with decision problems (or equivalently, languages) $K$ and log-space size parameters $m$ for which there are polynomials $p$, constants $c,d>0$, and  NTMs $M$ solving (or recognizing) $K$ in time at most $p(m(x))$
using space at most $c\log{m(x)}+d$ on all instances $x$ given to $K$.  Another example is the \emph{parameterized L}, denoted by $\para\dl$, which consists of all parameterized decision problems $(K,m)$ with log-space size parameters $m$ solvable deterministically in time $m(x)^{O(1)}$ using space $O(\log{m(x)})$ on all instances $x$ to $K$. For time-bounded computing,
$\para\p$ refers to the collection of all parameterized decision problems $(L,m)$ with log-space size parameters $m$ for which a certain DTM recognizes $L$ in time polynomial in $m(x)$.

More generally, we formulate a parameterized complexity class $\para\pdtimespace(\cdot)$ as follows.

\begin{definition}\label{PDTIME-SPACE}
Given a function $s:\nat\times \nat\to \nat$, the notation $\para\pdtimespace(s(|x|,m(x)))$, where $x$ is a ``symbolic'' instance,   expresses the collection of all parameterized decision problems $(P,m)$ with  languages $L$ and log-space size parameters $m$ that are solvable deterministically in time $m(x)^{O(1)}$
(i.e., polynomial time) using space $O(s(|x|,m(x)))$ on any instance $x$ given to $P$.
\end{definition}

With the use of the above notation, for example, $\para\dl$ can be expressed concisely as $\para\pdtimespace(\log{m(x)})$.
From a different perspective, many standard complexity classes, such as $\dl$, $\nl$, and $\p$, can be treated as special cases of $\para\dl$, $\para\nl$, and $\para\p$ when restricted to the standard size parameter $||$.
Hereafter, we freely identify a language $K$ with its parameterization $(K,||)$.

To define \emph{NL search problems} and \emph{NL optimization problems} in Section \ref{sec:application}, it is convenient for us to use a practical model of ``auxiliary Turing machine'' (see, e.g., \cite{Yam13} for details). An \emph{auxiliary Turing machine} is a multi-tape deterministic Turing machine equipped with an extra ``read-once''  \emph{auxiliary input tape}, in which a tape head scans each auxiliary input symbol only once by moving from left to right. Following \cite{Yam13}, let us define two auxiliary complexity classes, $\auxl$ and $\auxfl$. Given two alphabets $\Sigma$ and $\Gamma$, a decision problem  $P$  with $P\subseteq \Sigma^*\times\Gamma^*$ is in $\auxl$ if there exist a polynomial $p$ and an auxiliary Turing machine $M$ that takes a standard input $x$ and an auxiliary input $y$ of length $p(|x|)$ and decides whether $M$ accepts $(x,y)$ or rejects it in time polynomial in $|x|$ using space logarithmic in $|x|$. Its functional version is denoted by $\auxfl$, provided that each underlying auxiliary Turing machine is equipped with an extra \emph{write-once} output tape and that the machine produces output strings of at most polynomial length by moving from left to right.
We call a function $f:\Sigma_1^*\to\Sigma_2^*$ (for two alphabets $\Sigma_1$ and $\Sigma_2$) \emph{polynomially bounded} if there exists a polynomial $p$ satisfying $|f(x)|\leq p(|x|)$ for all instances $x\in\Sigma_1^*$.
With this terminology, the last requirement of $\auxfl$ actually states that all functions are polynomially bounded. Remember that $\fl$ is defined by Turing machines with no auxiliary tapes or auxiliary inputs.

%%%%%%%%%%%%%%%%%
%%%%%%%%%%%%%%%%%
\section{Sub-Linear Space and Short Reductions}\label{sec:complete}

Let us recall from \cite{IP01,IPZ01} that the term ``sub-exponential'' means $2^{\varepsilon m(x)}\, poly(|x|)$ on all instances $x$ for a certain fixed constant $\varepsilon\in(0,1)$  and a certain polynomially bounded function  $poly(n)$.
As a good analogy, we wish to use the term ``sub-linear'' to address functions of the form $m(x)^{\varepsilon}\, polylog(|x|)$ on instances $x$ for a certain constant  $\varepsilon\in(0,1)$ and a certain polylog function $polylog(n)$. As noted in Section \ref{sec:sub-linear-short}, the multiplicative factor $polylog(|x|)$ can be eliminated whenever  $m(x)$ is relatively large. Our main subject of this paper is sub-linear-space computability restricted to polynomially-bounded runtime.

Firstly, we wish to provide basic concepts associated with parameterized decision problems. A parameterized decision problem $(P,m)$ is said to be {\em solvable in polynomial time using sub-linear space} if, for a certain choice of constant $\varepsilon\in(0,1)$, there exist a DTM $M_{\varepsilon}$, a polynomial $p_{\varepsilon}$, and a polylog function $\ell_{\varepsilon}$ for which $M$ solves $P$ simultaneously taking at most $p_{\varepsilon}(m(x))$ steps and using at most $m(x)^{\varepsilon} \ell_{\varepsilon}(|x|)$ tape cells for all instances $x$ given to $P$. Notice that, when $m$ is polynomially bounded (i.e., there exists a polynomial $r$ satisfying $m(x)\leq r(|x|)$ for all $x$), we can replace $p_{\varepsilon}(|x|m(x))$ by $p_{\varepsilon}(|x|)$.

The notation $\psublin$ expresses the collection of all parameterized decision problems $(P,m)$ with log-space size parameters $m$ that are solvable  in polynomial time using sub-linear space. Using the notation $\para\pdtimespace(\cdot)$ given in Definition \ref{PDTIME-SPACE}, $\psublin$ is formally defined as
\[
\psublin = \bigcup_{\ell\in \mathrm{Polylog}}\; \bigcup_{\varepsilon\in [0,1)} \para\pdtimespace(m(x)^{\varepsilon} \ell(|x|)),
\]
where $x$ is a ``symbolic'' instance and $m$ is a ``symbolic'' size parameter.
It thus follows that $\para\dl\subseteq \psublin \subseteq \para\p$
but none of these inclusions is known to be proper.

The notion of ``reducibility'' among decision problems is quite useful in measuring the relative complexity of the problems. For the parameterized complexity class $\psublin$,  in particular, we need a restricted form of reducibility, which we call ``short'' reducibility, satisfying a special property that any outcome of the reduction is linearly upper-bounded in size by an  input of the reduction.
Let us formally introduce such restricted reductions for parameterized decision problems of our interest.

We begin with a description of {\em L-m-reducibility}  for  parameterized decision problems. This is a straightforward extension of the $\dl$-m-reducibility for decision problems (or languages). Given two parameterized decision problems $(P_1,m_1)$ and $(P_2,m_2)$, we say that $(P_1,m_1)$ is {\em L-m-reducible to} $(P_2,m_2)$, denoted by $(P_1,m_1)\Lreduces(P_2,m_2)$, if there exists a function $f\in\fl$ (which is called a \emph{reduction function}) and two constants $k_1,k_2>0$ such that, for any input string $x$,  (i) $x\in P_1$ iff $f(x)\in P_2$ and (iii) $m_2(f(x))\leq m_1(x)^{k_1}+k_1$. Notice that, since all functions in $\fl$ are, by their definition, polynomially bounded, the inequality $|f(x)|\leq |x|^{c_1}+c_2$ holds for two fixed constants $c_1,c_2>0$.

Concerning  polynomial-time sub-linear-space solvability, a restricted variant of $\dl$-m-reducibility, which we call the {\em short L-m-reducibility} (or sL-m-reducibility, in short), is defined by replacing the equality $m_2(f(x))\leq m_1(x)^{k_1}+k_1$  in the above definition of $\Lreduces$ with $m_2(f(x))\leq k_1 m_1(x)+k_1$. To express this new reducibility, we use a new  notation of  $\sLreduces$.
We further say that $(P_1,m_1)$ is \emph{equivalent} to $(P_2,m_2)$ via short L-m-reductions if $(P_1,m_1)$ is reducible to $(P_2,m_2)$ and vice visa via appropriate short L-m-reductions.

When many-one reducibility is too restrictive to use, we need a stronger notion of Turing reduction. To fit into a framework of polynomial-time  sub-linear-space computability, we use a \emph{polynomial-time sub-linear-space reduction family} (SLRF, for short), performed by oracle Turing machines, which is called a \emph{reduction machine}.
Our oracle Turing machine $M$ is equipped with an extra write-once  \emph{query tape}.
Once $M$ finishes writing a query word, say, $z$ on the query tape from left to right, $M$ enters a \emph{query state} $q_{query}$. The query word is then sent to an oracle, which is an external information source, and the query tape is automatically erased and its tape head returns to the initial cell. The oracle informs the machine of its answer by changing the machine's inner state from $q_{query}$ to either $q_{yes}$ (an affirmative answer) or $q_{no}$ (a negative answer).
In the case where the oracle returns strings instead of yes/no answers, we further equip $M$ with a read-once \emph{answer tape}, on which the oracle writes its answer and $M$ reads this answer only once by moving its tape head from left to right with (possibly) stopping at any time.
The above process is generally referred to as an \emph{oracle query} and this process is assumed to take only a single step.

\begin{definition}\label{def:sSLRF-reduction}
A parameterized decision problem $(P_1,m_1)$ is {\em SLRF-T-reducible to} another problem $(P_2,m_2)$, denoted by $(P_1,m_1)\SLRFreduces(P_2,m_2)$, if, for any fixed value  $\varepsilon>0$, there exist an  oracle Turing machine $M_{\varepsilon}$, a polynomial $p_{\varepsilon}$, a polylog function $\ell_{\varepsilon}$, and two constants $k_1,k_2\geq1$ such that,  for every  instance $x$ to $P_1$, (1) $M_{\varepsilon}^{P_2}$ runs in at most $p_{\varepsilon}(|x|m_1(x))$ time using at most $m_1(x)^{\varepsilon}  \ell_{\varepsilon}(|x|)$ space, provided that its query tape is not subject to this space bound, and (2) if $M_{\varepsilon}^{P_2}(x)$ makes a query to $P_2$ with a query word $z$, then $z$ must satisfy both $m_2(z)\leq  m_1(x)^{k_1}+k_1$ and $|z|\leq |x|^{k_2}+k_2$.
The {\em short SLRF-T-reducibility} (or sSLRF-T-reducibility, for short) is obtained from the SLRF-reducibility by substituting $m_2(z)\leq  k_1 m_1(x)+k_1$ for the above inequality $m_2(z)\leq  m_1(x)^{k_1}+k_1$. We use the notation $\sSLRFreduces$ to address this restricted reducibility.
\end{definition}

In the following quick example, we see how a short SLRF-T-reduction works between two parameterized decision problems.

\begin{example}
We show a simple example of short SLRF-T-reduction from $(\mathrm{CLIQUE},||)$ to $(\mathrm{ExactCLIQUE},||)$. The decision problem $\mathrm{CLIQUE}$ asks whether a given undirected graph $G$ has a clique (i.e., a complete subgraph) of size $k$. The problem $\mathrm{ExactCLIQUE}$ is similar but checks if the graph has the maximal clique of size $k$. The desired reduction machine $M$ works as follows. Given an instance $\pair{G,k}$ of $\mathrm{CLIQUE}$ with a graph $G=(V,E)$ and a positive integer $k$, $M$ inductively picks a number $i$ in $[k,|v|]_{\integer}$, generates $\pair{G,i}$ on a query tape, and makes a query to the oracle $\mathrm{ExactCLIQUE}$, assuming a suitable encoding $\pair{G,i}$ of $G$ and $i$.
As soon as an oracle answer becomes affirmative during this inductive process, $M$ accepts the input. If the oracle never returns any positive answer, then $M$ rejects the input. To produce each query word, $M$ needs only $O(1)$ work  space. Moreover, the binary size of each query word $\pair{G,i}$ is at most the binary size of $\pair{G,k}$ plus $O(\log{|V|})$ bits. Since $V$ is a part of the input $\pair{G,k}$, we obtain $|\pair{G,i}|\leq 2|\pair{G,k}|$. Thus, this reduction is indeed short SLRF-T-reduction.
\end{example}

In the case where the space usage of $M_{\varepsilon}$ in Definition \ref{def:sSLRF-reduction} is limited to $O(\log|x|)$, since the value $\varepsilon$ becomes irrelevant, we use a different notation of $\sLTreduces$ for $\sSLRFreduces$.

For any reduction $\leq_{r}$, a decision problem $P$ is said to be \emph{$\leq_{r}$-hard} for a given class $\CC$ of problems if every problem $Q$ in $\CC$ is $\leq_{r}$-reducible to $P$. When $P$ further belongs to  $\CC$, it is said to be \emph{$\leq_r$-complete} for $\CC$. We use the notation $\leq_{r}\!\!(\CC)$ to express the collection of all problems that are $\leq_{r}$-reducible to certain problems in $\CC$. When $\CC$ is a singleton, say, $\CC=\{A\}$, we write $\leq_{r}\!\!(A)$ instead of $\leq_{r}\!\!(\{A\})$.

We begin with proving simple relationships between short reductions and non-short reductions.

\begin{lemma}\label{L-vs-SLRF-reduction}
Let $P_1,P_2$ be decision problems and let $m_1,m_2$ be size parameters.
\renewcommand{\labelitemi}{$\circ$}
\begin{enumerate}\vs{-1}
  \setlength{\topsep}{-2mm}%
  \setlength{\itemsep}{1mm}%
  \setlength{\parskip}{0cm}%

\item $(P_1,m_1)\Lreduces(P_2,m_2)$ implies $(P_1,m_1)\LTreduces(P_2,m_2)$, which further implies $(P_1,m_1)\SLRFreduces(P_2,m_2)$. The same statement  holds also for $\sLreduces$, $\sLTreduces$,  and $\sSLRFreduces$.

\item $(P_1,m_1)\sLreduces (P_2,m_2)$ implies $(P_1,m_1)\Lreduces (P_2,m_2)$. The same holds for $\sSLRFreduces$ and $\SLRFreduces$.
\end{enumerate}
\end{lemma}

\begin{proof}
(1) The first implication is obtained by substituting a reduction machine $M$ for the reduction function $f$ in the definition of $\LTreduces$.
For the second implication, we further need to replace $M$ by $M_{\varepsilon}$. This is possible because, for each value $\varepsilon>0$ and for any fixed constants $c,d>0$, the inequality $c\log{n}+d\leq n^{\varepsilon}$ holds for all but finitely many numbers  $n$ in $\nat$.

(2) This comes from the definitions of $\Lreduces$ and $\sLreduces$, as well as $\SLRFreduces$ and $\sSLRFreduces$.
\end{proof}

Next, we show three statements, which concern with basic properties of SLRF-T- and sSLRF-T-reductions.

\begin{lemma}\label{reduction-property}
\begin{enumerate}%\vs{-2}
  \setlength{\topsep}{-2mm}%
  \setlength{\itemsep}{1mm}%
  \setlength{\parskip}{0cm}%

\item The reducibilities $\SLRFreduces$ and $\sSLRFreduces$ are reflexive and transitive. A similar statement holds for $\Lreduces$ and $\sLreduces$, and $\LTreduces$ and $\sLTreduces$.

\item The parameterized complexity class $\psublin$ is closed under $\sSLRFreduces$-reductions.

\item There exist two parameterized problems $(X,m)$ and $(Y,m)$ with a log-space size parameter $m$ and recursive decision problems $X$ and $Y$ such that $(X,m)\SLRFreduces (Y,m)$ and $(X,m)\not\sSLRFreduces (Y,m)$. A similar statement holds also for $\Lreduces$ and $\sLreduces$.
\end{enumerate}
\end{lemma}

\begin{proof}
(1)  We intend to show the target statement only for the reducibility $\sSLRFreduces$ over parameterized decision problems because the other reducibility $\SLRFreduces$ can be
treated with only a slight modification.

The \emph{reflexivity} of $\sSLRFreduces$ is trivial by considering the ``identity'' reduction. As for the \emph{transitivity} of $\SLRFreduces$, assume that $(P_1,m_1)\sSLRFreduces(P_2,m_2)$ and $(P_2,m_2)\sSLRFreduces(P_3,m_3)$ for three  parameterized decision problems $(P_1,m_1)$, $(P_2,m_2)$, and $(P_3,m_3)$. Given any constant $\varepsilon>0$, we take two oracle Turing machines $M_{1,\varepsilon}$ and $M_{2,\varepsilon}$ that solve $(P_1,m_1)$ and $(P_2,m_2)$ with the help of oracles $(P_2,m_2)$ and $(P_3,m_3)$, respectively. Moreover, we take positive constants $k_1$ and $k_2$, and a polylog function $\ell_1$ for $M_{1,\varepsilon}$, and $k'_1$, $k'_2$, and $\ell_2$ for $M_{2,\varepsilon}$, as specified in the definition of $\sSLRFreduces$.
In what follows, we wish to construct a new reduction machine from $(P_1,m_1)$ to $(P_3,m_3)$.

The desired reduction machine $N$ works as follows. On input $x$, $N$ simulates $M_{1,\varepsilon}$ on $x$. When $M_{1,\varepsilon}$ makes a query of word $z_i$ to $P_2$, instead of making an actual query, $N$ starts simulating $M_{2,\varepsilon}$ on $z_i$ with an appropriate access to $P_3$ used as an oracle.
In this simulation, whenever $M_{2,\varepsilon}$ queries a word $u_j$, we force $P_3$ to return its oracle answer directly to $N$. Hence, the actual query words of $N$ match those of $M_{2,\varepsilon}$. If $M_{2,\varepsilon}$ finally enters either an accepting state or a rejecting state, then $N$ regards it as an oracle answer of $P_2$ to the query of $z_i$. Notice that $m_2(z_i)\leq k_1 m_1(x)+k_1 \leq 2k_1m_1(x)$.
Here, it is important to note that the space usage for query words of $M_{2,\varepsilon}$ is not included here, because these words are directly written down on the query tape of $N$.

For any sufficiently long string $x$, the space usage $\ell(x)$ of $N$ on the instance  $x$ is upper-bounded by the sum of $m_1(x)^{\varepsilon}\ell_{1}(|x|)$ (work space used by $M_{1,\varepsilon}$ on $x$), $\max_{i}\{\log{|z_i|}\}$ (work space needed to produce query words $z_i$ of $M_{1,\varepsilon}$), and $\max_{i}\{m_2(z_i)^{\varepsilon} \ell_2(|z_i|)\}$ (work space used by $M_{2,\varepsilon}$ on $z_i$).
Since $|z_i|\leq |x|^{k_2}+k_2 \leq |x|^{k_3}$ holds for an appropriate constant $k_3>0$, it then follows that
\begin{eqnarray*}
\ell(x) &\leq& m_1^{\varepsilon}\ell_{1}(|x|) + \log|x|^{k_3} + (2k_1m_1(x))^{\varepsilon} \ell_{2}(\log|x|^{k_3}) \\
&\leq& m_1(x)^{\varepsilon} [ \ell_1(|x|)+k_3\log{|x|}+2k_1\ell_2(|x|^{k_3})] \;\;\leq\;\;
m_1(x)^{\varepsilon}\ell'(|x|)
\end{eqnarray*}
for a suitably chosen polylog function  $\ell'$. Moreover, the size of query words $u_j$ generated by $N$ is upper-bounded by $m_3(u_j)$, which is at most  $\max_{i}\{k'_2m_2(z_i)+k'_2\}\leq k_1k'_2m_1(x)+k_1k'_2+k'_2 \leq km_1(x)+k$ for an appropriate constant $k>0$.

Since $N$ correctly solves (or recognizes) $P_1$ using $P_3$ as an oracle, $N$ is therefore a short SLRF-reduction machine, as requested.

(2)  To prove the statement, we assume that $(P_1,m_1)\sSLRFreduces(P_2,m_2)$ and that $(P_2,m_2)\in\psublin$. Our goal is to show that $(P_1,m_1)$ also belongs to $\psublin$.

Since $(P_2,m_2)\in\psublin$, we take a constant $\varepsilon\in[0,1)$ and a DTM $M_{2}$ solving $(P_2,m_2)$ in polynomial time using $m_2(x)^{\varepsilon} \ell_2(|x|)$ space, where $\ell_2(\cdot)$ is a certain polylog function. For this constant $\varepsilon$, we take a reduction machine $M_{1}$ that reduces   $(P_1,m_1)$ to $(P_2,m_2)$ in polynomial time using space at most $m_1(x)^{\varepsilon}\ell_1(|x|)$ for a certain polylog function $\ell_1$. Moreover, there are two constants $k_1,k_2>0$ satisfying both $m_2(z)\leq k_1m_1(x)+k_1$ and $|z|\leq |x|^{k_2}+k_2$ for any query word $z$ of $M_1$ on every instance $x$.

Let us define a new DTM $N$ that solves $(P_1,m_1)$ by simulating $M_1$.
On input $x$, during the simulation of $M_{1}$ on $x$, whenever $M_1$ makes the $i$th query of the form $y_i$, $N$ stops the simulation and starts to  simulate $M_2$ on $y_i$. After $M_2$ enters either an accepting state or a rejecting state, $N$ treats it as an oracle answer $a_{y_i}$ to the query and then resumes the simulation of $M_1$ using this value $a_{y_i}$.

Clearly, $N$ runs in polynomial time. An appropriate implementation of $N$ requires work space of at most $c m_1(x)^{\varepsilon}\ell_1(|x|) + c \max_{i}\{m_2(y_i)^{\varepsilon}\ell_2(|y_i|)\}$ for a fixed constant $c>0$.   Since $m_2(y_i)\leq k_1 m_1(x)+k_1$ and $|y_i|\leq |x|^{k_2}+k_2\leq k_2|x|^{k_2}$, the space usage of $N$ is upper-bounded by
\begin{eqnarray*}
  \lefteqn{c m_1(x)^{\varepsilon}\ell_1(|x|) + c (2k_1m_1(x))^{\varepsilon}\ell_2(k_2|x|^{k_2})}\hs{10} \\
  &\leq& c m_1(x)^{\varepsilon} [ \ell_1(|x|)+2k_1\ell_2(k_2|x|^{k_2}) ] \;\;\leq\;\; m_1(x)^{\varepsilon}\ell(|x|),
\end{eqnarray*}
where $\ell$ is an appropriately chosen polylog function.
Therefore, $(P_1,m_1)$ is solvable in polynomial time using sub-linear space; in other words, $(P_1,m_1)$ belongs to $\psublin$.

(3) We will show the statement only for the case of $\leq^{\dl}_{m}$ and $\leq^{\mathrm{sL}}_{m}$ since the other cases are similarly treated in essence. Therefore, our goal here is to construct inductively two decision problems $X$ and $Y$ satisfying that $(X,m)\Lreduces (Y,m)$ and $(X,m)\not\sLreduces (Y,m)$ for a certain log-space size parameter $m$. Fix our alphabet $\Sigma=\{1\}$ and set $m(x)=|x|$ for all $x\in\Sigma^*$. We take two effective enumerations of all $\sLreduces$-reduction functions  over the alphabet $\Sigma$ as $g_1,g_2,\ldots$ (with possible repetitions) and of all natural numbers as $k_1,k_2,\ldots$ (with possible repetitions) so that the set $\{(g_i,k_i)\mid i\geq1\}$ of pairs satisfies the condition of $|g_i(x)|\leq k_i|x|+k_i$ for all indices $i\in\nat^{+}$ and for all strings $x\in\Sigma^*$. For the construction of $X$ and $Y$, we use the special  function $f:\Sigma^*\to\Sigma^*$ defined by setting $f(1^n)=1^{n^2}$ for any index $n\in\nat$ and $f(y)=\lambda$
for any binary string $y$ not in $\Sigma^*$.

The construction of the desired decision problems $X$ and $Y$ proceeds by stages. At Stage $0$, we set $n_0=0$, $x_0=\lambda$, and $X_0=Y_0=\setempty$. At Stage $i\geq1$, by induction hypothesis, we assume that $n_{i-1}$, $x_{i-1}$, $X_{i-1}$, and $Y_{i-1}$ have been already defined. We further assume that $X_{i-1}\subseteq \Sigma^{\leq n_{i-1}}$ and $Y_{i-1}\subseteq \Sigma^{n_{i-1}^2}$. We then define $n_i$ to be $\min\{e\in\nat^{+}\mid e>n_{2i-1}, k_ie+k_i<e^2\}$ and then set $x_i=1^{n_i}$ and $y_i=g_i(x_i)$. Clearly, $x_i\notin X_{i-1}$ and $|g_i(x_i)|\leq k_in_i+k_i <n_i^2 = |f(x_i)|$. There are two cases to consider separately.
(i) If $y_i\in Y_{i-1}$, then we set $X_i=X_{i-1}$ and $Y_{i}=Y_{i-1}$.
(ii) If $y_i\notin Y_{i-1}$, then we set $X_i=X_{i-1}\cup\{x_i\}$ and $Y_i=Y_{i-1}\cup\{f(x_i)\}$. In the end, $X$ and $Y$ are defined as $X=\bigcup_{i\in\nat}X_i$ and $Y=\bigcup_{i\in\nat}Y_i$.

Next, we prove that $(X,m)\Lreduces (Y,m)$ via $f$; that is, for any $x$, $x\in X$ iff $f(x)\in Y$. Since $X$ is a subset of $\{x_i\mid i\in\nat\}$, if $x\in X$, then there is an index $i$ such that $x=x_i\in X_i-X_{i-1}$. In this case, $f(x)$ is included in $Y_i$, and thus $f(x)\in Y$ follows. Conversely, if $f(y)\in Y$, then $f(x)\in Y_i-Y_{i-1}$ for a certain index $i$. Hence, $x = x_i\in X_i$ follows. This implies $x\in X$.

Finally, we verify that $(X,m)\not\sLreduces (Y,m)$. Assume that $(X,m)\sLreduces (Y,m)$ via a certain reduction $g$. Take the minimum index $i$ satisfying $g=g_i$ and consider Stage $i$. Let $x=x_i$. Either $x_i\in X_i$ or $x_i\notin X_i$ holds. If $x_i\in X_i$, then case (ii) happens, and thus $y_i\notin Y_{i-1}$. If $x_i\notin X_i$, then case (i) occurs, leading to  $y_i\in Y_{i-1}$. Thus, it follows that either (a) $x_i\notin X$ and $y_i\in Y$  or (b) $x_i\in X$ and $y_i\notin Y$. Since $y_i=f(x_i)$, a contradiction follows. We therefore conclude that $(X,m)\not\sLreduces (Y,m)$.
\end{proof}

%%%%%%%%%%%%%%%%%
%%%%%%%%%%%%%%%%%
\section{The 2CNF Boolean Formula Satisfiability Problem and Syntactic NL}\label{sec:2SAT-definition}

We will make a brief discussion on the space complexity of $\mathrm{2SAT}$ and $\mathrm{2SAT}_3$ and the complexity class $\mathrm{SNL}$. As noted in Section \ref{sec:introduction}, $\mathrm{2SAT}$ is NL-complete under L-m-reductions.
Here, a \emph{Boolean (propositional) formula} is made up by \emph{variables} and logical connectives $\{\vee, \wedge, \neg\}$ and a \emph{variable assignment} assigns either True (1) or False (0) to each variable. A {\em literal} is either a  variable $x$ or its negation $\overline{x}$.
Given such a literal $v$, the succinct notation $\overline{v}$ indicates $\overline{x}$ if $v$ is a variable $x$, and $\overline{v}$ indicates $x$ if $v$ is the negation of $x$. This notation is useful because we do not need to worry about whether or not $v$ is a variable.

For practical reason, we always assume the existence of an efficient encoding scheme of Boolean formulas into unique binary strings for which each variable (as well as logical connectives) is encoded into an $O(\log{mn})$-bit string, where $n$ is the total number of variables and $m$ is the total number of clauses.

%%%
\subsection{Properties of 2SAT$_3$}\label{sec:properties-2SAT3}

We further restrict $\mathrm{2SAT}$ by limiting the number of occurrences of variables in the form of literals in each Boolean formula. Fix $k\in\nat^{+}$ arbitrarily. We denote by $\mathrm{2SAT}_k$ the collection of all formulas $\phi$ in $\mathrm{2SAT}$ such that, for each variable $v$ in $\phi$, the total number of occurrences of $v$ and $\overline{v}$ is at most $k$.
Since $\mathrm{2SAT}_1$ and $\mathrm{2SAT}_2$ are solvable using only log space, we hereafter force our attention on the case of $k\geq3$.

In what follows, we are focused on two specific size parameters:  $m_{vbl}(x)$ and $m_{cls}(x)$, which  respectively denote the total number of variables and that of clauses appearing in formula-related instance $x$ (not necessarily limited to instances of $\mathrm{2SAT}$).
Based on the fact that 2SAT is NL-complete \cite{JLL76}, we derive the following conclusion: $(\mathrm{2SAT},||)\Lreduces(\mathrm{2SAT}_3,||)$.

\begin{proposition}\label{2SAT_k-complete}
For each index $k\geq3$, $\mathrm{2SAT}_k$ is $\nl$-complete.
\end{proposition}

\begin{proof}
It is obvious that $(\mathrm{2SAT}_3,m_{vbl})\sLreduces (\mathrm{2SAT}_k,m_{vbl})$ for any $k\geq3$. Since $\mathrm{2SAT}$ is known to be $\nl$-complete under L-m-reduction, it suffices to show that $\mathrm{2SAT}\Lreduces \mathrm{2SAT}_3$,  or equivalently $(2\mathrm{SAT},||)\Lreduces (2\mathrm{SAT}_3,||)$.

Let $\phi$ denote any 2CNF formula given to $\mathrm{2SAT}$. If $\phi$ contains a unit clause of $x$ and another unit clause of $\overline{x}$ for a certain variable $x$, then we define $f(\phi)=x\wedge \overline{x}$. Hereafter, we assume otherwise. We recursively modify $\phi$ according to the following rules. If $x\vee\overline{x}$ is a clause of $\phi$, then we delete it. If $\phi$ contains a clause of either $x$ or $x\vee x$ (resp., $\overline{x}$ or $\overline{x}\vee \overline{x}$) for a certain variable $x$, then we first delete the unit clause of $x$ and all clauses of the form $x\vee y$ (resp., $\overline{x}\vee y$) for any literal $y\notin\{x,\overline{x}\}$ and then include $x\vee x$ (resp., $\overline{x}\vee \overline{x}$).

After the above modification of $\phi$, let $V_3$ denote the set of all variables, each of which appears as literals in more than $3$ distinct clauses. For each variable $x\in V_3$, we set $C_x$ to be the set of all clauses containing literals of $x$.
For readability, for any clause $x\vee y$ in $C_x$, we also include $y\vee x$ into $C_x$ if $x$ and $y$ are distinct literals.
Let us assume that $C_x$ has the form
$\{x\vee y_i, y_i\vee x, \overline{x}\vee z_j, z_j\vee \overline{x} \mid i\in[l],j\in[m]\}$
for certain numbers $l,m\in\nat$. Note that $|C_x|>3$.
We prepare new variables $\{x_i^{(1)},\overline{x}_i^{(1)},x_j^{(2)},\overline{x}_j^{(2)} \mid i\in[l],j\in[m]\}$ and replace $C_x$ by the following set of new clauses: $x_i^{(2)}\vee y_i$, $\overline{x}_j^{(1)}\vee z_j$, $\overline{x}_i^{(1)}\vee x_{i+1}^{(1)}$,  $\overline{x}_j^{(2)}\vee x_{j+1}^{(2)}$, $\overline{x}_l^{(2)}\vee x_1^{(1)}$, and  $\overline{x}_l^{(1)}\vee x_1^{(2)}$  for all indices $i\in[l]$ and $j\in[m]$. Let $\phi'$ be the conjunction of all clauses obtained by this modification.

By the above modification procedure, $\phi'$ is a 2CNF formula whose clauses have ``exactly'' two distinct literals each. Such a formula is called an \emph{exact 2CNF formula}. It also follows that, for each variable, its literals appear at most $3$ times. Thus, $\phi'$ is indeed an instance to $\mathrm{2SAT}_3$. It is also clear that $\phi'$ is satisfiable iff $\phi$ is satisfiable. Moreover, we can produce  $\phi'$ from $\phi$ using only log space.
Therefore, $2\mathrm{SAT}$ is L-m-reducible to $2\mathrm{SAT}_3$,
as requested.
\end{proof}

Concerning $2\mathrm{SAT}_k$, the following important properties hold.
For any reduction $\leq_r$ defined in Section \ref{sec:complete}, we write $(P_1,m_1)\equiv_{r}(P_2,m_2)$ if $(P_1,m_1)$ is equivalent to $(P_2,m_2)$ via $\leq_r$-reductions; in other words, both $(P_1,m_1)\leq_{r}(P_2,m_2)$ and $(P_2,m_2)\leq_{r}(P_1,m_1)$ hold.

\begin{lemma}\label{basic-property-SAT}
Let $m\in\{m_{vbl},m_{cls}\}$ and $k\geq3$.
\begin{enumerate}\vs{-1}
  \setlength{\topsep}{-2mm}%
  \setlength{\itemsep}{1mm}%
  \setlength{\parskip}{0cm}%

\item $(\mathrm{2SAT}_k,m)\sLequiv (\mathrm{2SAT}_3,m)$.

\item $(\mathrm{2SAT}_3,m_{vbl})\sLequiv (\mathrm{2SAT}_3,m_{cls})$.
\end{enumerate}
\end{lemma}

\begin{proof}
Let $m\in\{m_{vbl},m_{cls}\}$ and $k\geq3$.

(1) Firstly, the short reducibility $(\mathrm{2SAT}_{3},m)\sLreduces(\mathrm{2SAT}_k,m)$ trivially holds by considering the ``identity'' reduction.
Next, we want to show that $(\mathrm{2SAT}_{k},m)\sLreduces(\mathrm{2SAT}_3,m)$. Let us recall the proof of Proposition \ref{2SAT_k-complete}, in which any 2CNF Boolean formula $\phi$ given to $2\mathrm{SAT}_k$ is transformed into another formula $\phi'$, which is an instance to $2\mathrm{SAT}_3$. This modification implies that $m_{vbl}(\phi') \leq (m_{vbl}(\phi)-|V_3|) + \sum_{x\in V_3}4|C_x|$ and $m_{cls}(\phi') \leq m_{cls}(\phi)+\sum_{x\in V_3}6|C_x|$. Since $|C_x|\leq 2k$ and $|V_3|\leq m_{vbl}(\phi)\leq 2m_{cls}(\phi)$, we conclude that  $m_{vbl}(\phi')\leq m_{vbl}(\phi)+8k|V_3|\leq (8k+1)m_{vbl}(\phi)$ and that $m_{cls}(\phi')\leq m_{cls}(\phi)+12k|V_3| \leq (24k+1)m_{cls}(\phi)$. From these inequalities, we conclude that $(2\mathrm{SAT}_k,m)$ is short L-m-reducible to $(2\mathrm{SAT}_3,m)$.

(2) Here, we simply take the identity function as a reduction function. This function is a short L-m-reduction function because of the fact that, for every 2CNF Boolean formula $\phi$ given to $\mathrm{2SAT}_3$, two inequalities $m_{cls}(\phi)\leq 6 m_{vbl}(\phi)$ and  $m_{vbl}(\phi)\leq 2 m_{cls}(\phi)$ follow.
\end{proof}

Contrary to Lemma \ref{basic-property-SAT}(2), it is still unknown whether $(\mathrm{2SAT},m_{vbl})\sLequiv (\mathrm{2SAT},m_{cls})$ or even $(\mathrm{2SAT},m_{vbl})\sLTequiv (\mathrm{2SAT},m_{cls})$. This situation signifies the difference between $\mathrm{2SAT}$ and $\mathrm{2SAT}_3$.

%%%

To solve 2SAT in polynomial time, we need the space usage slightly larger than sub-linear space.
The time-space complexity of $\mathrm{2SAT}$ can be  derived from a deterministic algorithm  designed to solve the \emph{directed $s$-$t$ connectivity problem}\footnote{This is also known as the graph accessibility problem and the graph reachability problem in the literature.} $\dstcon$ of determining whether there exists a path from a vertex $s$ to another vertex $t$ (succinctly called an \emph{$s$-$t$ path}) in a given directed graph $G=(V,E)$.
For practicality, every vertex
is assumed to be expressed as an $O(\log|V||E|)$-bit string.
In 1998, Barnes \etalc~\cite{BBRS98} presented such a polynomial-time algorithm for $\dstcon$, which uses slightly more than sub-linear space, or more precisely, $n/2^{O(\sqrt{\log{n}})}$ ($\leq n^{1-c/\sqrt{\log{n}}}$ for an appropriate  constant $c>0$). Using the assertion that $(\mathrm{2SAT},m_{vbl})$ is short L-T-reducible to $(\dstcon,m_{ver})$, shown in Proposition \ref{2SAT-DSTCON-equiv}(1), we can solve  $(\mathrm{2SAT},m_{vbl})$ in polynomial time using $n/2^{O(\sqrt{\log{n}})}$ space.

\begin{theorem}\label{2SAT-solvable}
For a certain constant $c>0$ and a certain polylog function $\ell$, $\mathrm{2SAT}$ with $n$ variables and $m$ clauses is solvable in polynomial time using $n^{1-c/\sqrt{\log{n}}} \ell(m+n)$ space.
\end{theorem}

As we have noted earlier, our proof of this theorem needs Proposition \ref{2SAT-DSTCON-equiv}(1), which will be proven later in Section \ref{sec:hypothesis}. Given a Boolean formula $\phi$, the notation $|\phi|$ denotes the size of $\phi$ expressed \emph{in binary} by an appropriate encoding.

\begin{proofof}{Theorem \ref{2SAT-solvable}}
Proposition \ref{2SAT-DSTCON-equiv}(1) states that $(\dstcon,m_{ver}) \sLTequiv (\mathrm{2SAT},m_{vbl})$ and  $(\dstcon,m_{edg}) \sLTequiv (\mathrm{2SAT},m_{cls})$. Therefore, the time/space complexity of $\mathrm{2SAT}$ can be  derived directly from a deterministic algorithm that solves $\dstcon$.
Earlier, Barnes \etalc~\cite{BBRS98} presented a polynomial-time algorithm, say, $M$ for $\dstcon$ with $n$ vertices, which operates using at most $n^{1-c/\sqrt{\log{n}}}$ space for a certain fixed constant $c>0$.

Let $\phi$ be any 2CNF Boolean formula, given as an instance to 2SAT, with $n$ variables and $m$ clauses. Here, we consider only the case where $\phi$ contains more than one clause and one variable since, otherwise, the case is trivial. It then follows that $|\phi|\leq (m+n)^k$ for a certain constant $k>0$. In the case (b) of the proof of Proposition \ref{2SAT-DSTCON-equiv}(1), a directed graph $G_{\phi}=(V_{\phi},E_{\phi})$ is defined from $\phi$ with $|V_{\phi}|=2 m_{vbl}(\phi)$ and $|E_{\phi}|\leq 2m_{cls}(\phi)$ in such a way that making queries of the form $(G_{\phi},v,w)$ to  $(\dstcon,m_{ver})$ solves $(\mathrm{2SAT},m_{vbl})$. The space usage for each oracle query is at most $c\log{|\phi|}$ for a certain constant $c>0$; thus, this value is further upper-bounded by $ck\log(m+n)$.
In order to remove the oracle, we further replace ``making an oracle query'' by ``running $M$ on the same query word as an input''. The space required for running $M$ is at most $|V_{\phi}|^{1-c/\sqrt{|V_{\phi}|}}$, which is further bounded from above by $2m_{vbl}(\phi)^{1-c/\sqrt{\log{2m_{vbl}(\phi)}}} = 2n^{1-c/\sqrt{\log{n}+1}}$, as requested.

Therefore, $(\mathrm{2SAT},m_{vbl})$ can be solved in polynomial time using $c' n^{1-c'/\sqrt{\log{n}}}\log^{k'}(m+n)$ space for appropriate constants $c',k'>0$.
\end{proofof}

%%%%
%%%%
\subsection{Syntactic NL or SNL}\label{sec:SNL}

Among all natural subclasses of $\np$, the complexity class $\mathrm{SNP}$ has played a special role in capturing many graph-theoretical features of $\np$ problems in terms of second-order predicate sentences \cite{FM93,PY91}. By taking a different approach, Immerman \cite{Imm87} provided a first-order logical framework to capture $\dl$ and $\nl$. In a similar spirit, we turn our attention to a syntactically restricted subclass of $\para\nl$ defined by \emph{syntactic NL sentences} or \emph{SNL sentences}.
Each SNL sentence gives rise to a logical framework to a parameterized decision problem.

We begin with an introduction of the syntax of our logical system. A \emph{vocabulary} (an \emph{input signature} or an \emph{input relation})  $\TT$ is a finite set $\{(S_i,k_i), c_j, 0, n, suc, pred \mid i\in[d],j\in[d']\}$ of
\emph{predicate symbols} $S_i$ of arity $k_i\geq0$ (or $k_i$-arity predicate symbol)
and \emph{constant symbols} $0$, $n$, and other specific symbols $c_j$  expressing ``input objects'' (such as particular vertices or edges of a graph and particular sizes of columns and rows of a matrix) of the target mathematical problem for certain constants $d,d'\in\nat^{+}$.
We also allow the use of two designated function symbols: the \emph{successor function} $suc(\cdot)$ and the \emph{predecessor function} $pred(\cdot)$ whose meanings are $suc(i)=i+1$ and $pred(i)=\max\{0,i-1\}$ for any $i\in\nat$. We abbreviate $suc(suc(i))$ as $suc^2(i)$ and $suc(suc^2(i))$ as $suc^3(i)$, etc.

Here, we use two types of variables. \emph{First-order variables}, denoted by $i,j,\ldots, u,v,\ldots$, range over natural numbers and input objects  (such as vertices or edges of a graph and entries of a matrix) of the  mathematical problem.
\emph{Second-order variables}, denoted by $P,Q,\ldots$, range over  \emph{binary relations} whose first argument takes a natural number and the second argument takes an input object.

\emph{Atomic formulas} have the following forms:  $S_j(u_1,\ldots,u_{k_i})$, $P(i,v)$, $u=v$, and $i\leq j$, where $i,j,u,v,u_1,\ldots,u_{k_i}$ are \emph{terms} composed of first-order variables, constant symbols, and function symbols, and $P$ is a second-order variable.
In particular, $i,j$ are number-related terms and $u,v,u_1,\ldots,u_{k_i}$ are terms composed of all other input objects.
\emph{Formulas} are built inductively from atomic formulas by connecting them with logical connectives ($\to$, $\neg$, $\vee$, $\wedge$) and first/second-order quantifiers ($\forall$, $\exists$). Notice that $\to$ and $pred$ are included here for our convenience although they are redundant.

In what follows, we concentrate on the specific case where second-order variables represent ``functions''. Hence, it is convenient to introduce a variant of the second-order quantifier.
We use the special notation $\exists^fP[\psi(P)]$ with a formula $\psi$ containing no second-order quantifiers as a shorthand for $\exists P[\psi(P) \wedge Func(P)]$, where
$Func(P)$ is a unique sentence over a second-order variable $P$ that expresses that $P(\cdot,\cdot)$ works as a ``function''; namely, $Func(P) \equiv Func_1(P)\wedge Func_2(P)$, where $Func_1(P) \equiv \forall i \exists w[P(i,w)]$ and  $Func_2(P) \equiv \forall i\forall u\forall v\: [P(i,u)\wedge P(i,v)\rightarrow u=v]$. We particularly call ``$\exists^f$'' the \emph{functional existential quantifier}.

\begin{definition}\label{SNL-sentence}
Let $\TT=\{(S_i,k_i),c_j, 0, n, suc, pred \mid i\in[d],j\in[d']\}$ denote a  vocabulary. An \emph{syntactic NL sentence} (or an \emph{SNL sentence}) over $\TT$ is a second-order sentence $\Phi$ of the form:
\begin{eqnarray*}
\lefteqn{\Phi \equiv \exists^f P_1\cdots \exists^f P_l\: \forall i_1\cdots \forall i_r \: \forall y_1\cdots \forall y_s} \hs{10}\\
&& [ \bigwedge_{j=1}^{t} \psi_j(P_1,\ldots,P_l,i_1, \ldots,i_r,y_1, \ldots,y_s,S_1,\ldots, S_d,c_1,\ldots,c_{d'})],
\end{eqnarray*}
where $l,r,s,t\in\nat$, each $\psi_j$, $j\in[t]$, is a quantifier-free second-order formula such that no two $\psi_j$'s share any common first-order variables.
Here, $P_1,\ldots,P_l$ are second-order binary variables representing functions, $i_1,\ldots,i_r$ are first-order variables representing natural numbers, and $y_1,\ldots,y_s$ are also first-order variables representing all other input objects. Each $\psi_j$ should satisfies the following two \emph{second-order variable requirements}: (i) each $\psi_j$ contains only second-order variables of the form $P_j(i,v_1),P_j(suc(i),v_2),P_j(suc^2(i),v_3), \ldots, P_j(suc^a(i),v_{a+1})$ for a fixed constant $a\in\nat^{+}$ and (ii) $\psi_j$ can be rewritten in the logically equivalent form of finite disjunctions (not necessarily in the normal form) satisfying the following condition: there are only at most two disjuncts containing second-order variables and each of them must have the form $(\bigwedge_{k,i,v} P_k(i,v))\wedge (\bigwedge_{k',i',v'} \neg P_{k'}(i',v')) \wedge R$, where $R$ is an appropriate subformula including no second-order variable.
\end{definition}

Since second-order variables represent ``functions'', the second-order variable requirement limits the number of simultaneous accesses to those functions by allowing only inputs of the functions to be taken from a small set $\{i,suc(i),suc^2(i),\ldots,suc^a(i)\}$.

Next, we explain semantics of SNL sentences $\Phi$.

\begin{definition}
Given a vocabulary $\TT=\{(S_i,k_i),c_j, 0, n, suc, pred \mid i\in[d],j\in[d']\}$, a \emph{relational structure} $\SSS$ over $\TT$ is a set of tuples $(U_i,D_i,k_i)$ and $(\bar{c}_j,V_j)$ with finite universes $U_i$ and $V_j$ of ``input objects'' (including natural numbers) and domains $D_i$ associated with predicate symbols $S_i$ in $\TT$ satisfying $D_i\subseteq U_i^{k_i}$, and constants $\bar{c}_j$ in $V_j$.
The constant symbols $c_j$ are interpreted as $\bar{c}_j$ and the predicate symbols $S_i$ are interpreted as $D_i$ in the following way: when objects $\bar{s}_1,\bar{s}_2,\ldots,\bar{s}_k$ in $U_i$ are assigned respectively to variables $x_1,x_2,\ldots,x_k$, $S_i(x_1,x_2,\ldots,x_k)$ is evaluated to be true (resp., false) iff $(\bar{s}_1,\bar{s}_2,\ldots,\bar{s}_k)\in D_i$ (resp., $\notin D_i$).
If a value in $\{|U_i|, |U_i^{k_i}|, |D_i|, |V_j| :i\in[d],j\in[d']\}$ serves as a size parameter of the underlying mathematical problem, then it is briefly referred to as a  \emph{structure size} of $\SSS$.
\end{definition}

A relational structure $\SSS$ over $\TT$ is said to \emph{describe} (or \emph{represent}) an instance $x$ of the target mathematical problem if every input object appearing in $x$ has its associated predicate symbol in $\TT$ in its universe and domain in $\SSS$ or its associated constant symbol in $\TT$ in its universe. For example, an instance $x=(G,\bar{s},\bar{t})$ given to $\dstcon$ with $G=(V_G,E_G)$ can be represented by $\TT=\{(V,1),(E,2),s,t\}$ and $\SSS=\{(V_G,V_G,1), (V_G,E_G,2), (s,V_G), (\hat{t},V_G)\}$, where $V_G$ and $E_G$ are expressed by  predicate symbols $V$ and $E$, respectively, and $\bar{s}$ and $\bar{t}$ are expressed by constant symbols $s$ and $t$, respectively.

\begin{definition}\label{SNL-model}
Let $\Phi$ denote an SNL sentence described in Definition \ref{SNL-sentence} with variables $P_1,\ldots,P_l,i_1,\ldots,i_r,y_1,\ldots,y_s$.
A \emph{domain structure} $\DD$ for $\Phi$ is the union of three sets $\{(P_j,[e_1]\times U'_j,2)\}_{j\in[l]}$, $\{(i_j,[e'_j])\}_{j\in[r]}$, and $\{(y_j,U''_j)\}_{j\in[s]}$, which provide the scopes of ``variables'' of $\Phi$ in the following manner, where $e_j$ and $e'_j$ are fixed constants in $\nat^{+}$. Each second-order variable $P_j$ ($j\in[l]$) ranges over $[e_j]\times U'_j$. Each first-order variable $i_j$ ($j\in[r]$) ranges over $[e'_j]$ and each variable $y_j$ ($j\in[s]$) ranges over  $U''_{j}$.
\end{definition}

Since a domain structure $\DD$ provides sets of input objects, which semantically limit the scopes of quantifiers in a given SNL sentence $\Phi$, we explicitly write such sets inside  $\Phi$ to clarify the scope of quantifiers. For example, using  sets $I$, $U$, and $U'$, we write $\exists P\subseteq I\times U\: [B(P)]$ and $\forall z\in U'\: [B(z)]$ instead of $\exists P [B(P)]$ and $\forall z [B(z)]$, respectively.

When a relational structure $\SSS$ and a domain structure $\DD$ are given, it is possible to evaluate the \emph{validity} of an SNL sentence $\Phi$ by interpreting all predicate symbols and all constant symbols of $\Phi$ as domains and constants in $\SSS$ by assigning input objects in $\SSS$  and $\DD$ to variables appropriately.
This evaluation makes $\Phi$ either ``true'' or ``false''.
Given a mathematical problem $A$ and an SNL sentence $\Phi$, we say that $\Phi$ \emph{syntactically expresses} $A$ if, for any instance $x$ to $A$, there are a relational structure $\SSS_x$ describing $x$ and a domain structure $\DD_x$ for $\Phi$ satisfying the following: $x\in A$ iff $\Phi$ is true on $\SSS_x$ and $\DD_x$.
We denote by $\para\mathrm{SNL}$ the collection of all parameterized decision problems $(A,m)$ with log-space size parameters $m$ such that (1) there exist a vocabulary $\TT$ and an SNL sentence $\Phi$ for which $\Phi$ syntactically expresses $A$ and (2) for each instance $x$ given to $A$, $m(x)$ equals the structure size of the corresponding relational structure $\SSS_x$.

%%%%

Let us present a simple example of $\para\mathrm{SNL}$ problem. Here, we consider the parameterized decision problem $(\dstcon,m_{ver})$.

\begin{example}\label{example-SNL}
Take an instance $x=(G,\bar{s},\bar{t})$ given to $\dstcon$, where $G=(V_G,E_G)$, $\bar{s},\bar{t}\in V_G$, $\bar{s}$ is a source, and $\bar{t}$ is a sink. Let $n=m_{ver}(x)=|V_G|$.
For convenience, we expand an $s$-$t$ path to an \emph{extended $s$-$t$ path} by padding extra $t$'s to the tail of the original $s$-$t$ path to make the obtained sequence have length exactly $n$.
Recall the aforementioned vocabulary $\TT=\{(V,1),(E,2),s,t\}$ for $\dstcon$.
Let us consider $\Phi$ defined as:
\begin{eqnarray*}
\Phi \equiv \exists^f P\subseteq [n]\times V_G \forall i,j\in[n] \forall u,w \in V_G \: [ F_0(P) \wedge F_1(P,j)  \wedge F_2(P,i,u,w) ],
\end{eqnarray*}
where three formulas $F_0$, $F_1$, and $F_2$ are defined as follows.
\begin{enumerate}\vs{-1}
  \setlength{\topsep}{-2mm}%
  \setlength{\itemsep}{1mm}%
  \setlength{\parskip}{0cm}%

\item $F_0(P) \equiv P(1,s) \wedge P(n,t)$.

\item $F_1(P,j) \equiv 1\leq j <n \wedge P(j,t) \rightarrow P(suc(j),t)$.

\item $F_2(P,i,u,w) \equiv 1\leq i <n \wedge  u\neq t \wedge P(i,u) \wedge P(suc(i),w) \rightarrow E(u,w)$.
\end{enumerate}
The predicate $P(i,v)$ asserts that $v$ is the $i$th vertex of an extended $s$-$t$ path of length $n$. Notice that $F_0$, $F_1$, and $F_2$ share no common first-order variables. It is not difficult to see that $(\dstcon,m_{ver})$ is syntactically expressed by $\Phi$.
Since each quantifier-free formula $F_i$ contains only two occurrences of $P$, the second-order variable requirements are satisfied; thus, $\Phi$ is indeed an SNL sentence.
Therefore, $(\dstcon,m_{ver})$ belongs to $\para\mathrm{SNL}$.
\end{example}

To make a connection to $(\mathrm{2SAT}_3,m_{vbl})$, we further introduce an extra requirement imposed on SNL sentences with relational structures and domain structures.

Let us consider the subformula $F_2(P,i,u,w)$ used in Example \ref{example-SNL} associated with an instance $x=(G,s,t)$, where $G=(V_G,E_G)$ and $n=|V_G|$. It follows that, for each pair $(\bar{\imath},\bar{u})\in [n]\times V_G$ of input objects, (i) if $(\bar{u},\bar{w})\in E_G$, then $F_2(P,\bar{\imath},\bar{u},\bar{w})$ is clearly true but (ii)  if $(\bar{u},\bar{w})\notin E_G$, then $F_2(P,\bar{\imath},\bar{u},\bar{w})$ is  ``undefined'' (i.e., evaluated to be neither true or false). Moreover, the cardinality of the set $\{ \bar{w}\in V_G \mid \text{ $F_2(P,\bar{\imath},\bar{u},\bar{w})$ is undefined }\}$ equals the outdegree of $\bar{u}$. Thus, this set gets larger as $n$ grows. We wish to limit the cardinality of such a set.

\begin{definition}\label{def:SNL-omega}
The parameterized complexity class $\para\mathrm{SNL}_{\omega}$ is composed of all parameterized decision problems $(A,m)$ in $\para\mathrm{SNL}$ that enjoys the following extra requirement. Let $\Phi$ denote any SNL-sentence of the form given in Definition \ref{SNL-sentence} with $t$ quantifier-free subformulas $\psi_j(P_1,\ldots,P_l,\boldvec{i}, \boldvec{y}, S_1,\ldots,S_d,c_1, \ldots,c_{d'})$ together with (hidden) sentence $Func(P_i)$ for all $i\in[l]$, where $\boldvec{i}= (i_1, \ldots,i_r)$ and $\boldvec{y}=(y_1, \ldots,y_s)$.
Assume that $\Phi$ syntactically expresses $A$ by a certain relational structure $\SSS_x$ and a certain domain structure $\DD_x$ associated with each instance $x$ given to $A$ and that $m(x)$
equals a certain structure size of $\SSS_x$.
We demand that the sentence $(\bigwedge_{h=1}^{l} Func(P_h))$ must be  ``expressed'' inside $\Phi$ with no use of existential quantifiers ``$\exists$'' in the following sense: $\exists P_1\cdots \exists P_l \forall \boldvec{i}\forall \boldvec{y} [\bigwedge_{j=1}^{t}\psi_j] \wedge (\bigwedge_{h=1}^{l} Func(P_h))$ is true iff $\exists P_1\cdots \exists P_l \forall \boldvec{i}
\forall \boldvec{y} [\bigwedge_{j=1}^{l}\psi_j]$ is true. Notice that each $\psi_j$ must satisfy the second-order variable requirements.
\end{definition}

As an example, we demonstrate that a simple variant of $(3\dstcon,m_{ver})$ falls in $\para\mathrm{SNL}_{\omega}$. This variant is  $\mathrm{exact3}\dstcon$, which is obtained by limiting $3\dstcon$ to only directed graphs in which $s$ is a source, $t$ is a sink, every non-source, non-sink vertex has \emph{exactly} three edges, and any source (i.e., of indegree $0$) has \emph{exactly} two outgoing edges. It is easy to show that $(\mathrm{exact3}\dstcon,m_{ver})$ is $\leq^{\mathrm{sL}}_{m}$-equivalent to $(3\dstcon,m_{ver})$.

\begin{lemma}\label{exact-vs-3DSTCON}
$(\mathrm{exact3}\dstcon,m_{ver}) \equiv^{\mathrm{sL}}_{m} (3\dstcon,m_{ver})$.
\end{lemma}

\begin{proof}
It is trivial that $(\mathrm{exact3}\dstcon,m_{ver}) \leq^{\mathrm{sL}}_{m} (3\dstcon,m_{ver})$. To show the converse reducibility, let $x=(G,\bar{s},\bar{t})$ be any instance given to $3\dstcon$. If $s$ is not a source, then we remove all incoming edges to $s$. If $t$ is not a sink, then we also remove all outgoing edges from $t$.
Hereafter, we assume that $s$ is a source and $t$ is a sink. In the case where $s$ is adjacent to 3 vertices $v_1$, $v_2$, and $v_3$, we remove all edges from $s$ and then add a new vertex $v_s$ and new edges $(s,v_S),(s,v_1),(v_s,v_2),(v_s,v_3)$.
For the other vertices of $G$, we inductively choose them one by one. For each $v$, we add a new vertex $v'$ and an edge $(v,v')$ (resp., two new vertices $v',v''$ and two edges $(v,v'),(v,v'')$) if $v$ has degree $2$ (resp., indegree $0$ and outdegree $1$). The added vertices are sinks of indegree $1$. The resulting graph is clearly an instance of $\mathrm{exact3}\dstcon$.
\end{proof}

Let us demonstrate that $(\mathrm{exact3}\dstcon,m_{ver})$ is actually in  $\para\mathrm{SNL}_{\omega}$.

\begin{example}\label{3DSTCON-SNL}
Let us consider an instance $x=(G,\bar{s},\bar{t})$ with $G=(V_G,E_G)$ given to $\mathrm{exact3}\dstcon$. We assume that $G$ satisfies the aforementioned requirement of $3\dstcon$. To express this $x$, we intend to define a second-order sentence $\Psi$ as:
\begin{eqnarray*}
\lefteqn{\Psi \equiv \exists P\subseteq [n]\times V_G \forall i,j,k,h\in[n] \forall u,v,w,v_0,v_1,v_2,w_0,w_1,x_0,x_1,y \in V_G\: [ F_0(P)  } \hs{5} \\
&&  \wedge F_1(P,j) \wedge F'_2(P,v,v_0,v_1,v_2)
\wedge F'_3(P,k,u,w_0,w_1) \wedge F'_4(P,h,x_0,x_1,y) \wedge Func_2(P)],
\end{eqnarray*}
where $F_0$ and $F_1$ are the same as in Example \ref{example-SNL}, and $F'_2$, $F'_3$, and $F'_4$ are defined in the following fashion.
\begin{enumerate}\vs{-1}
  \setlength{\topsep}{-2mm}%
  \setlength{\itemsep}{1mm}%
  \setlength{\parskip}{0cm}%

\item $F'_2(P,v,v_0,v_1,v_2) \equiv (\bigwedge_{e=0}^{1} E(s,v_e)) \wedge v_0\neq v_1 \rightarrow \bigvee_{e=0}^{1} P(2,v_e)$.

\item $F'_3(P,k,u,w,w_0,w_1) \equiv  2\leq k<n \wedge u\neq s \wedge (\bigwedge_{e=0}^{1}w_e\neq t) \wedge (\bigvee_{e=0}^{1} P(k,w_e) ) \wedge (\bigwedge_{e=0}^{1} E(w_e,u) )   \wedge w_0\neq w_1 \rightarrow P(suc(k),u)$.

\item $F'_4(P,h,x,x_0,x_1,y) \equiv  1\leq h<n \wedge y\neq t \wedge (\bigvee_{e=0}^{1} P(suc(h),x_e) ) \wedge (\bigwedge_{e=0}^{1} E(y,x_e)) \wedge x_0\neq x_1  \rightarrow P(h,y)$.
\end{enumerate}
Observe that $F_0$, $F_1$, $F'_2$, $F'_3$, and $F'_4$ share no common first-order variables and that $\Psi$ satisfies the second-order variable requirements.
\end{example}

%%%%
%%%%

As a key statement concerning $\para\mathrm{SNL}_{\omega}$, we exhibit the hardness of $(\mathrm{2SAT}_3,m_{vbl})$ for $\para\mathrm{SNL}_{\omega}$ under short SLRF-T-reductions.

%%%%

\begin{proposition}\label{hardness-SNL}
Every parameterized decision problem in $\para\mathrm{SNL}_{\omega}$ is short SLRF-T-reducible to $(\mathrm{2SAT}_3,m_{vbl})$.
\end{proposition}

\begin{proof}
It suffices to prove that every problem $(A,m)$ in $\para\mathrm{SNL}_{\omega}$ can be reduced to $(\mathrm{2SAT}_k,m_{vbl})$ by an appropriate short SLRF-T-reduction, where $k$ is a certain large constant, because $(\mathrm{2SAT}_k,m_{vbl}) \equiv^{\mathrm{sSLRF}}_{T} (\mathrm{2SAT}_3,m_{vbl})$ by Lemma \ref{basic-property-SAT}(1).

Let $(A,m)\in\para\mathrm{SNL}_{\omega}$ and take a vocabulary  $\TT=\{(S_i,k_i),c_j, 0,n, suc, pred \mid i\in[d],j\in[d']\}$. We assume that $A$ is syntactically expressed by an SNL sentence $\Phi$ of the form $\exists^f P_1 \cdots \exists^f P_l \forall i_1\cdots \forall i_r \forall y_1\cdots \forall y_s \: \bigwedge_{j=1}^{t} \psi_j$ over $\TT$, as in Definition \ref{SNL-sentence}, where
each $\psi_j$ is a quantifier-free subformula composed of free variables $P_1,\ldots,P_l,i_1,\ldots,i_r,y_1,\ldots,y_s$, the constant symbols $c_1,c_2,\ldots,c_{d'}$, and the predicate symbols $S_1,\ldots,S_d$.
By the second-order variable requirements, there are at most $a+1$ variables $P_k$'s in each $\psi_j$.
For simplicity, we convert all occurrences of the logical connective ``implication'' ($\to$) into combinations of $\{\wedge,\vee\}$ and negation of literals.
We further assume that negation ($\neg$) appears only in front of those predicates by rewriting $\psi_j$ using De Morgan's laws. For convenience, let $\boldvec{\imath}=(i_1,i_2,\ldots,i_r)$ and $\boldvec{y}=(y_1,y_2,\ldots,y_s)$.

Let $x$ be any instance given to $A$. Let $\SSS_x = \{(U_i,D_i,k_i), (\bar{c}_j,V_j) \mid i\in[d],j\in [d']\}$ denote a relational structure over $\TT$, representing this particular instance $x$. By the choice of $\SSS_x$, there is a structure size of $\SSS_x$, which is  equal to the value $m(x)$.  Corresponding to $x$, we further take a domain structure $\DD_x$ for $\Phi$. By the definition of $\para\mathrm{SNL}_{\omega}$, there is a constant $\gamma\geq1$ that satisfies Definition \ref{def:SNL-omega}.

We conduct the following transformation of $\Phi$ using $\SSS_x$ and $\DD_x$. We arbitrarily fix an index $j\in[t]$. Notice that
$\psi_j$ is a Boolean combination of predicates of the following forms:    $S_k(t_{1},t_{2},\ldots,t_{a'})$, $P_k(t'_1,t'_2)$, $t''_1=t''_2$, and $s'_1\leq s'_2$, where $t_1,t_2,\ldots,t_{a},t'_1,t'_2,t''_1,t''_2,s'_1,s'_2$ are terms whose underlying variables are taken from $\{i_1,i_2,\ldots,i_r,y_1,y_2,\ldots,y_s\}$.
We want to convert $\psi_j$ into a 2CNF formulas.
Sequentially, we choose values $\boldvec{\bar{\imath}}$ and $\boldvec{\bar{y}}$ from $[e_1]\times \cdots \times [e_r]$ and $U'_1\times \cdots \times U'_s$, respectively, provided by $\DD_x$.
We assign those values to the corresponding variable symbols and write the resulting formula as  $\psi_j(\boldvec{\bar{\imath}},\boldvec{\bar{y}})$.
We first replace $P_k(\bar{t}_1,\bar{t}_2)$ in $\psi_j(\boldvec{\bar{\imath}},\boldvec{\bar{y}})$ by a new free variable symbol $z_{k,\bar{t}_1,\bar{t}_2}$, where $\bar{t}_1$ and $\bar{t}_2$ are values  obtained from $t_1$ and $t_2$ by assigning values in $(\boldvec{\bar{\imath}},\boldvec{\bar{y}})$ to all variable symbols, respectively.
Next, we determine the validity of each predicate $S_k(\boldvec{\bar{\imath}'},\boldvec{\bar{y}'})$ in $\psi_j(\boldvec{\bar{\imath}},\boldvec{\bar{y}})$ by checking whether $(\boldvec{\bar{\imath}'}, \boldvec{\bar{y}'})\in D_k$ and we then replace $S_k(\boldvec{\bar{\imath}'},\boldvec{\bar{y}'})$  by either True (1) or False (0) according to its evaluation result, where $\boldvec{\bar{\imath}'}$ and $\boldvec{\bar{y}'}$ are appropriate subsequences of $\boldvec{\bar{\imath}}$ and $\boldvec{\bar{y}}$, respectively.
Similarly, other formulas of the forms $t'_1=t'_2$ and $s'_1\leq s'_2$ are evaluated to be either True or False.
We write the obtained formula as $\psi'_j(\boldvec{\bar{\imath}},\boldvec{\bar{y}})$.
If $\psi'_j(\boldvec{\bar{\imath}},\boldvec{\bar{y}})$ is either true or false independent of the unknown values of all the introduced variables $z_{k,\bar{t}_1,\bar{t}_2}$, then we replace this formula by True or False, respectively.
Let us denote by $\psi''_j(\boldvec{\bar{\imath}},\boldvec{\bar{y}})$ the resulting formula and move to the next choice.
We then obtain the conjunction of all subformulas $\psi''_j(\boldvec{\bar{\imath}},\boldvec{\bar{y}})$ over
all values $(\boldvec{\bar{\imath}},\boldvec{\bar{y}})$ chosen from $[e_1]\times \cdots \times [e_r] \times U'_1\times \cdots \times U'_s$. We write this conjunction as $\psi_{j,conj}$.
Finally, we convert $\psi_{j,conj}$ to an equivalent 2CNF formula, say,  $\hat{\psi}_{j,conj}$.

As a concrete example of $\hat{\psi}_{j,conj}$, recall the subformula $F'_3$ given in Example \ref{3DSTCON-SNL}. This subformula can be expressed in a disjunctive form as
\begin{equation*}
k=1\vee k=n \vee u=s \vee (\bigvee_{e=0}^{1} w_e=t) \vee (\bigwedge_{e=0}^{1} \neg P_1(k,w_e) ) \vee (\bigvee_{e=0}^{1} \neg E(w_e,u) ) \vee w_0 = w_1 \vee P_1(suc(k),u).
\end{equation*}
Here, we write $P_1$ for $P$ in Example \ref{3DSTCON-SNL} to agree with the above argument.
Assume that an underlying graph $G=(V_G,E_G)$ with $\bar{n}=|V_G|$ has three distinct vertices $\{\bar{u},\bar{w}_0,\bar{v}_1\}$ satisfying that $\bar{u}\in V_G-\{\bar{s}\}$ and $\bar{w}_0,\bar{w}_1\in V_G-\{\bar{t}\}$ with $\bar{w}_0\neq \bar{w}_1$.
Assume also that $\bar{k}\in\nat$ satisfies $2\leq \bar{k}<\bar{n}$.
We fix an assignment $\sigma$ of $(k,u,w_0,w_1)$ to  $(\bar{k},\bar{u},\bar{w}_0,\bar{w}_1)$.
It then follows that $\bar{k}=1$, $\bar{k}=\bar{n}$, $\bar{u}= \bar{s}$, $(\bigvee_{e=0}^{1}\bar{w}_e=\bar{t})$, and $\bar{w}_0=\bar{w}_1$ are all false. Now, we translate $P_1(suc(k),u)$, $P_1(k,w_0)$, and $P_1(k,w_1)$ into three new variables $z_{1,\bar{k}+1,\bar{u}}$, $z_{1,\bar{k},\bar{w}_0}$, and $z_{1,\bar{k},\bar{w}_1}$, respectively.
If $\{(\bar{w}_0,\bar{u}),(\bar{w}_1,\bar{u})\}\cap E_G\neq\setempty$, then $F'_3$ is true, independent of the truth values of $z_{1,\bar{k}+1,\bar{u}}$, $z_{1,\bar{k},\bar{w}_0}$, and $z_{1,\bar{k},\bar{w}_1}$.
On the contrary, when  $\{(\bar{w}_0,\bar{u}),(\bar{w}_1,\bar{u})\} \subseteq E_G$, $F'_3$ is undefined and it is logically equivalent to
$(\neg z_{1,\bar{k},\bar{w}_0} \wedge \neg z_{1,\bar{k},\bar{w}_1}) \vee z_{1,\bar{k}+1,\bar{u}}$ because
$\bigvee_{e=0}^{1}\neg E(\bar{w}_e,\bar{u})$ is false. Note that this can be transformed into $(\neg z_{1,\bar{k},\bar{w}_0} \vee  z_{1,\bar{k}+1,\bar{u}}) \wedge (\neg z_{1,\bar{k},\bar{w}_1} \vee z_{1,\bar{k}+1,\bar{u}})$.
Therefore, we obtain
\begin{equation*}
\bigwedge_{i\in[2,|V_G|-1]_{\integer}} \bigwedge_{\bar{u}\in V_G-\{\bar{s},\bar{t}\}} \bigwedge_{(\bar{w}_0,\bar{w}_1) \in A_{\bar{u}}} \bigwedge_{e=0}^{1}
( (\neg z_{1,\bar{k},\bar{w}_0} \vee  z_{1,\bar{k}+1,\bar{u}}) \wedge (\neg z_{1,\bar{k},\bar{w}_1} \vee z_{1,\bar{k}+1,\bar{u}}) )
\end{equation*}
as $\hat{\psi}_{j,conj}$, where $A_{\bar{u}} = \{(\bar{w}_0,\bar{w}_1) \mid \bar{w}_0,\bar{w}_1\in V_G-\{\bar{t}\},  (\bar{w}_0,\bar{u}), (\bar{w}_1,\bar{u})\in E_G, \bar{w}_0\neq \bar{w}_1\}$. Since $|A_{\bar{u}}|\leq 2$, we observe that each variable of the form $z_{1,a,b}$ appears at most twice in $\hat{\psi}_{j,conj}$.

Since each disjunct in $\hat{\psi}_{j,conj}$ that contains $P_k$'s has the form $(\bigwedge_{k,i,v} P_k(i,v)) \wedge (\bigwedge_{\bar{k}',\bar{\imath}',\bar{v}'}  \neg P_{k'}(i',v'))\wedge R$, the above transformation provides $(\bigwedge_{\bar{k},\bar{\imath},\bar{v}} z_{\bar{k},\bar{\imath},\bar{v}}) \wedge (\bigwedge_{\bar{k}',\bar{\imath}',\bar{v}'} \neg z_{\bar{k}',\bar{\imath}',\bar{v}'})$.
Each variable $z_{k,\bar{t}_1,\bar{t}_2}$ appears in only subformulas that are ``undefined'' by the variable assignment, and thus it appears at most a constant number of times in $\hat{\psi}_{j,conj}$ in the form of literals. The final 2CNF formula is $\bigwedge_{j=1}^{t}\hat{\psi}_{j,conj}$.
Since there are $t$ 2CNF formulas $\hat{\psi}_{1,conj}, \ldots, \hat{\psi}_{t,conj}$, the total number of times that each new variable appears as literals in the entire conjunction $\bigwedge_{j=1}^{t}\hat{\psi}_{j,conj}$ is upper-bounded by a certain fixed constant, say, $k_*$.

Since the aforementioned transformation can be done by a DTM using only log space, this DTM serves as a Turing reduction from $(A,m)$ to $(\mathrm{2SAT}_{k_*},m_{vbl})$. By Lemma \ref{basic-property-SAT}(1), we can sL-m-reduce $(\mathrm{2SAT}_{k_*},m_{vbl})$ to $(\mathrm{2SAT}_3,m_{vbl})$. Therefore, by the transitivity of $\leq^{\mathrm{sL}}_{m}$ by Lemma \ref{reduction-property}(1), we obtain a short L-m-reduction from $(A,m)$ to $(\mathrm{2SAT}_3,m_{vbl})$, as requested.
\end{proof}

%%%%

As shown in Example \ref{3DSTCON-SNL}, $(\mathrm{exact3}\dstcon,m_{ver})$  belongs to $\para\mathrm{SNL}_{\omega}$. We remark that
$(\mathrm{exact3}\dstcon,m_{ver})$ is $\leq^{\mathrm{sL}}_{T}$-equivalent  to  $(\mathrm{2SAT}_3,m_{vbl})$
because $(\mathrm{exact3}\dstcon,m_{ver}) \equiv^{\mathrm{sL}}_{m} (3\dstcon,m_{ver})$ by Lemma \ref{exact-vs-3DSTCON} and $(3\dstcon,m_{ver}) \equiv^{\mathrm{sL}}_{T} (\mathrm{2SAT}_3,m_{vbl})$ by Proposition \ref{2SAT-DSTCON-equiv}(2). Therefore, Proposition \ref{hardness-SNL} leads to the following immediate corollary.

\begin{corollary}\label{2SAT-SNL-complete}
$(\mathrm{exact3}\dstcon,m_{ver})$ is complete for $\para\mathrm{SNL}_{\omega}$ under short SLRF-T-reductions.
\end{corollary}

%%%%
%%%%

Recall from Section \ref{sec:complete} the succinct notation $\leq_{r}\!\!(\CC)$ for any parameterized complexity class $\CC$.

\begin{theorem}\label{SNL-logical-equivalent}
The following statements are all logically equivalent.
\begin{enumerate}\vs{-1}
  \setlength{\topsep}{-2mm}%
  \setlength{\itemsep}{1mm}%
  \setlength{\parskip}{0cm}%

\item $(\mathrm{2SAT}_3,m_{vbl})\in \psublin$.

\item $\para\mathrm{SNL}_{\omega}\subseteq \psublin$.

\item $\sSLRFreduces\!\!(\para\mathrm{SNL}_{\omega}) \subseteq \psublin$.
\end{enumerate}
\end{theorem}

\begin{proof}
(2 $\Rightarrow$ 3) If $\para\mathrm{SNL}_{\omega}\subseteq \psublin$, then we obtain $\sSLRFreduces (\para\mathrm{SNL}_{\omega}) \subseteq \sSLRFreduces (\psublin)$. Note that, by Lemma \ref{reduction-property}(2), $\psublin$ is closed under short SLRF-T-reductions. Thus, it follows that $\sSLRFreduces (\para\mathrm{SNL}_{\omega}) \subseteq \psublin$.

(3 $\Rightarrow$ 1) Assume that $\sSLRFreduces (\para\mathrm{SNL}_{\omega}) \subseteq \psublin$. Since $(\mathrm{exact}3\dstcon,m_{ver})\in\para\mathrm{SNL}_{\omega}$ by Example \ref{3DSTCON-SNL}, we obtain  $(\mathrm{exact3}\dstcon,m_{ver})\in \psublin$. By Lemma \ref{exact-vs-3DSTCON}, $(\mathrm{exact3}\dstcon,m_{ver}) \equiv^{\mathrm{sL}}_{T} (\mathrm{2SAT}_3,m_{vbl})$ holds.
From these results, we can conclude that
$(\mathrm{2SAT}_3,m_{vbl})\in \psublin$ because $\psublin$ is closed under short L-T-reductions.

(1 $\Rightarrow$ 2) Assume that $(\mathrm{2SAT}_3,m_{vbl})\in \psublin$.
By Proposition \ref{hardness-SNL}, any parameterized decision problem $(A,m)$ in $\para\mathrm{SNL}_{\omega}$ can be reduced to $(\mathrm{2SAT}_3,m_{vbl})$ by an appropriate short SLRF-T-reduction. Since $\psublin$ is closed under such reductions,  our assumption $(\mathrm{2SAT}_3,m_{vbl})\in \psublin$ leads to the conclusion that $(A,m)$ also belongs to $\psublin$.
\end{proof}

Theorem \ref{SNL-logical-equivalent}(3) can be compared to the fact that $\para\nl\subseteq \LTreduces\!\!(\para\mathrm{SNL}_{\omega})$ since $(\mathrm{exact3}\dstcon,m_{ver})$ is in $\para\mathrm{SNL}_{\omega}$ by Example \ref{3DSTCON-SNL} and it is also L-T-complete for $\para\nl$. The last L-T-hardness statement comes from the fact that $\mathrm{2SAT}_3$ is L-m-complete for $\nl$ (Proposition \ref{2SAT_k-complete}) and that
$(\mathrm{2SAT}_3,m_{vbl})$ is sL-m-hard for $\para\nl_{\omega}$ (Proposition \ref{hardness-SNL}) and that, as remarked earlier,  $(\mathrm{exact3}\dstcon,m_{ver})$ is $\leq^{\mathrm{sL}}_{T}$-equivalent to  $(\mathrm{2SAT}_3,m_{vbl})$.

%%%%%%%%%%%%%%%%%
%%%%%%%%%%%%%%%%%
\section{The Working Hypothesis LSH for 2SAT$_3$}\label{sec:hypothesis}

The exponential time hypothesis (ETH) \cite{IP01} and the strong exponential time hypothesis (SETH) \cite{IPZ01} have served as a driving force to obtain better lower bounds on the time complexity of various well-known problems (see, e.g., a survey \cite{LMS11}). As their natural analogy, we have introduced in Definition \ref{LSH-definition} the linear space hypothesis (LSH) for $2\mathrm{SAT}_3$ with $m_{vbl}$. We will explore the fundamental properties of this working hypothesis.

%%%
\subsection{The Linear Space Hypothesis or LSH}\label{sec:LSH-property}

We have seen in Theorem \ref{2SAT-solvable} that $\mathrm{2SAT}$ with $n$ variables and $m$ clauses can be solved simultaneously in time polynomial in $m+n$ and using $n^{1-c/\sqrt{\log{n}}} polylog(m+n)$ space for a certain constant $c>0$;  however, $2\mathrm{SAT}$ is not yet known to be solvable in polynomial time using sub-linear space. This circumstance supports our introduction of a practical, working hypothesis---the {\em linear space hypothesis} (LSH) for $\mathrm{2SAT}_3$ with $m_{vbl}$---which asserts the insolvability of the parameterized decision problem $(\mathrm{2SAT}_3,m_{vbl})$ in polynomial time  using sub-linear space.
The choice of the problem $\mathrm{2SAT}_3$ as well as the size parameter $m_{vbl}$ seems artificial; however, this choice does not matter because, as shown in Lemma \ref{basic-property-SAT}(2) with the help of Lemma \ref{reduction-property}(2), we can replace $\mathrm{2SAT}_3$ by $\mathrm{2SAT}_k$ ($k\geq3$) and $m_{vbl}$ by $m_{cls}$ without compromising the power of LSH. More precisely, we assert:

\begin{proposition}\label{m_vbl-equivalent-m_cls}
Let $k\geq3$.
\begin{enumerate}\vs{-1}
  \setlength{\topsep}{-2mm}%
  \setlength{\itemsep}{1mm}%
  \setlength{\parskip}{0cm}%

\item LSH for $2\mathrm{SAT}_k$ with $m_{vbl}$ is logically equivalent to LSH for $2\mathrm{SAT}_3$ with $m_{vbl}$.

\item LSH for $2\mathrm{SAT}_3$ with $m_{vbl}$ is logically equivalent to LSH for $2\mathrm{SAT}_3$ with $m_{cls}$.
\end{enumerate}
\end{proposition}

\begin{proof}
(1) Assume that LSH is true for $2\mathrm{SAT}_k$ with $m_{vbl}$, where $k\geq3$. This means that $(2\mathrm{SAT}_k,m_{vbl})\notin \psublin$. By Lemma \ref{basic-property-SAT}(1), $(2\mathrm{SAT}_k,m_{vbl})\sLequiv (2\mathrm{SAT}_3,m_{vbl})$ follows. Since $\psublin$ is closed under  $\sSLRFreduces$-reductions (and thus $\sLreduces$-reductions) by Lemma \ref{reduction-property}(2), if $(2\mathrm{SAT}_3,m_{vbl})\in\psublin$, then $(2\mathrm{SAT}_k,m_{vbl})$ is also in $\psublin$, a contradiction. The converse is obvious.

(2)Assume that LSH holds for $2\mathrm{SAT}_3$ with $m_{vbl}$; that is, $(2\mathrm{SAT}_3,m_{vbl})\notin \psublin$. We want to claim that $(2\mathrm{SAT}_3,m_{cls})\notin \psublin$. Lemma \ref{basic-property-SAT}(2) shows that $(2\mathrm{SAT}_3,m_{vbl})\sLequiv (2\mathrm{SAT}_3,m_{cls})$. If $(2\mathrm{SAT}_3,m_{cls})$ belongs to $\psublin$, then $(2\mathrm{SAT}_3,m_{vbl})$ also belongs to $\psublin$ since $\psublin$ is closed under  $\sSLRFreduces$-reductions (and thus $\sLreduces$-reductions) by Lemma \ref{reduction-property}(2). This clearly contradicts LSH for $2\mathrm{SAT}_3$ with $m_{vbl}$. A similar argument proves the converse.
\end{proof}

Since we can replace $m_{vbl}$ in the definition of LSH with $m_{cls}$ by Proposition \ref{m_vbl-equivalent-m_cls}(2), we often omit any explicit  reference to $m_{vbl}$ and simply refer to the hypothesis as ``LSH for $2\mathrm{SAT}_3$.''
Assuming $\nl\subseteq \mathrm{SC}$, since $\mathrm{2SAT}_3$ falls into $\mathrm{SC}$, the parameterized problem $(2\mathrm{SAT}_3,||)$ obviously falls in $\psublin$.
Notice that there is a fixed constant $c>0$ for which $|x|\leq (3m_{vbl}(x)+c)\log|x|$ for all instances $x$ given to $\mathrm{2SAT}_3$. It therefore  follows that $(2\mathrm{SAT}_3,m_{vbl})\in \psublin$, contradicting LSH for $\mathrm{2SAT}_3$. As customary, we denote by $\logcfl$ (resp., $\logdcfl$) the complexity class $\leq^{\dl}_{m}(\cfl)$ (resp., $\leq^{\dl}_{m}(\dcfl)$).  As an immediate consequence of the above argument, we obtain:

\begin{theorem}\label{LSH-implies-L-NL}
If LSH for $\mathrm{2SAT}_3$ is true, then $\dl\neq\nl$, $\logdcfl\neq\logcfl$,\footnote{This particular consequence $\logdcfl\neq\logcfl$ was suggested earlier to the author by Markus Holzer.} and $\mathrm{SC}\neq\mathrm{NSC}$.
\end{theorem}

Since LSH for $\mathrm{2SAT}_3$ leads to, e.g., $\dl\neq\nl$ as shown above, this working hypothesis is expected to lead to finer, better consequences than what the assumption $\dl\neq\nl$ can lead to.
Although the working hypothesis LSH concerns with $\mathrm{2SAT}_3$, it may possibly carry over to $\mathrm{2SAT}$. However, it is not clear that LSH for $2\mathrm{SAT}_3$ is logically equivalent to LSH for $2\mathrm{SAT}$.

\begin{lemma}\label{extend-LSH}
Assuming that LSH for $\mathrm{2SAT}_3$ is true, the following two statements hold.
\begin{enumerate}\vs{-1}
  \setlength{\topsep}{-2mm}%
  \setlength{\itemsep}{1mm}%
  \setlength{\parskip}{0cm}%

\item $\sSLRFreduces\!\!(\mathrm{2SAT}_3,m_{vbl}) \nsubseteq \psublin$.

\item $(\mathrm{2SAT},m_{vbl})\notin \psublin$.
\end{enumerate}
\end{lemma}

\begin{proof}
(1)  Assume that  $\sSLRFreduces\!\!(\mathrm{2SAT}_3,m_{vbl}) \subseteq \psublin$. Since $(\mathrm{2SAT}_3,m_{vbl})$ obviously belongs to $\sSLRFreduces\!\!(\mathrm{2SAT}_3,m_{vbl})$,
we conclude that $(\mathrm{2SAT}_3,m_{vbl})\in\psublin$, a contradiction against LSH for $\mathrm{2SAT}_3$.

(2) Note that $(\mathrm{2SAT}_3,m_{vbl})\sLreduces (\mathrm{2SAT},m_{vbl})$ by the ``identity'' reduction. If $(2\mathrm{SAT},m_{vbl})\in \psublin$, then we obtain $(2\mathrm{SAT}_3,m_{vbl})\in \psublin$ as well. However, this is a contradiction against our assumption of LSH for $\mathrm{2SAT}_3$.
\end{proof}

As another consequence of LSH for $\mathrm{2SAT}_3$, we can show the existence of two special parameterized decision problems in the complexity class $\sSLRFreduces\!\!(\mathrm{2SAT}_3,m_{vbl})$, which  are incomparable with respect to $\sSLRFreduces$-reductions.
This indicates that the class $\sSLRFreduces\!\!(\mathrm{2SAT}_3,m_{vbl})$ has a fine, complex structure with respect to sSLRF-T-reducibility.

%%%
%%%

\begin{theorem}\label{incomparable-pair}
Assuming LSH for $\mathrm{2SAT}_3$, there are two parameterized decision problems $(A,m_A)$ and $(B,m_B)$ in the complexity class $\sSLRFreduces\!\!(\mathrm{2SAT}_3,m_{vbl})$  such that  $(A,m_A)\not\sSLRFreduces (B,m_B)$ and $(B,m_B)\not\sSLRFreduces (A,m_A)$.
\end{theorem}

\begin{proof}
Assume the working hypothesis LSH for $\mathrm{2SAT}_3$.
To make our argument simple, we set $\Sigma=\{0,1\}$ and $m_A(x)=m_B(x)=|x|$ as the standard size parameter (i.e., binary bit length) $||$.
For convenience, we further introduce the notation $\dtimespace(t(n),s(n))$ to denote the collection of all ``languages'' recognized deterministically in time $O(t(n))$ using space $O(s(n))$. Given an oracle Turing machine $M$ and a language $A$, we write $L(M,A)$ for the family of all languages recognized by $M$ using $A$ as an oracle.

The following argument is inspired by Ladner's in \cite{Lad75}.
Let us construct two sets $S_A$ and $S_B$ in $\dtimespace(n,\sqrt{n})$ and then define $A=\mathrm{2SAT}_3\cap S_A$ and $B=\mathrm{2SAT}_3\cap S_B$ so that $(A,||)\not\sSLRFreduces (B,||)$ and $(B,||)\not\sSLRFreduces (A,||)$.
Since $S_A \in\dtimespace(n,\sqrt{n})$, it follows that $(A,||)\sSLRFreduces (\mathrm{2SAT}_3,m_{vbl})$ by sequentially checking whether $x\in S_A$ by computing $S_A$ directly and then checking whether $x\in\mathrm{2SAT}_3$ in a way of making a query of $x$ to $\mathrm{2SAT}_3$. Similarly, we obtain  $(B,||)\sSLRFreduces (\mathrm{2SAT}_3,m_{vbl})$.
Thus, we conclude that $(A,||),(B,||)\in \sSLRFreduces\!\!(\mathrm{2SAT}_3,m_{vbl})$.

Consider all reduction machines $M$ satisfying that $M^{O}$ runs in time at most $|x|^{k}$ using space at most $\sqrt{|x|}$ and queries only words of length at most $k|x|+k$ for any input $x$ and any oracle $O$, where $k$ is a fixed positive constant.
We enumerate them effectively as $\{(M_1,k_1), (M_2,k_2),\ldots\}$ so that, given an index $i\in\nat^{+}$, we can generate an appropriately defined \emph{code} of $M_n$ and $k_n$ in polynomial time using $O(\log{n})$ space.

Since every $\nl$ problem can be solved deterministically
in polynomial time, we prepare a DTM $M_{2SAT}$ that recognizes  $\mathrm{2SAT}_3$ in polynomial time.
In what follows, we will construct $S_A$ and $S_B$ stage by stage to satisfy the following two requirements for each index $i\in\nat^{+}$: (i) $R_{A,i}$: $A\neq L(M_i,B)$ and (ii) $R_{B,i}$: $B\neq L(M_i,A)$.
For this purpose, we inductively construct a series $\{(S_{A,j},S_{B,j})\}_{j\in\nat}$ and then define $A_j = \mathrm{2SAT}_3\cap S_{A,j}$ and $B_j=\mathrm{2SAT}_3\cap S_{B,j}$ for any $j\in\nat$. In the end, we set $S_A = \bigcup_{j\in\nat}S_{A,j}$ and $S_B=\bigcup_{j\in\nat}S_{B,j}$.
Initially, we define $n_0=0$ and $S_{A,0}=S_{B,0}=\setempty$.

(1) Firstly, we concentrate on Stage $2i+1$ ($i\geq 0$). Assume that $n_{2i}$, $S_{A,{2i}}$, and $S_{B,{2i}}$ have been already defined.
Notice that $A_{2i}$ and $B_{2i}$ are both finite sets.
We define $C_{2i+1}$ to be $L(M_i,B_{2i})\cap \Sigma^{>n_{2i}}$.
We then assert that $(C_{2i+1}\triangle \mathrm{2SAT}_3)\cap \Sigma^{>n_{2i}}$ is an infinite set. Assuming otherwise, we now want to lead to the conclusion of $\mathrm{2SAT}_3\in\psublin$.
Since $C_{2i+1}\triangle \mathrm{2SAT}_3$ is finite in $\Sigma^{>n_{2i}}$ and $\Sigma^{\leq n_{2i}}$ is also a finite set, we conclude that $C_{2i+1}\triangle \mathrm{2SAT}_3$ is finite. This makes it possible to recognize $\mathrm{2SAT}_3$ by computing $C_{2i+1}$. By running $M_i$ with the oracle $B_{2i}$ to compute $C_{2i+1}$, we can short SLRF-T-reduce $(\mathrm{2SAT}_3,m_{vbl})$ to $(B_{2i},||)$. Since $B_{2i}$ is finite, it thus follows that $\mathrm{2SAT}_3\in\psublin$. This is a clear contradiction against our assumption of LSH for $\mathrm{2SAT}_3$.
As a consequence, there must be infinitely many strings $z$ in $(C_{2i+1} \triangle \mathrm{2SAT}_3 )\cap \Sigma^{>n_{2i}}$. Among those strings, we choose the lexicographically first string $z_0$. We choose $n_{2i+1}$ (which will be determined later) to satisfy that $n_{2i+1}>k_i|z_0|+k_i$ and $k_in^{2i+1}+k_i \leq n_{2i+1}^2$. The former requirement implies that $M_i^{B_{2i}}(z_0)$ does not make any queries of length more than $n_{2i+1}$. Let $S_{B,{2i+1}} = S_{B,{2i}}$ and $S_{A,{2i+1}} = S_{A,{2i}}\cup \{z\mid n_{2i}<|z|\leq n_{2i+1}\}$. Clearly, $z_0\in S_{A,{2i+1}}$ follows.

From $S_{A,2i+1}$ and $S_{B,2i+1}$, we obtain $A_{2i+1}$ and $B_{2i+1}$. We then claim that $z_0\in A_{2i+1}\triangle L(M_i,B_{2i+1})$. If $z_0\in A_{2i+1}$, then we obtain $z_0\in \mathrm{2SAT}_3$ and thus $z_0\notin C_{2i+1}$, yielding $z_0\notin L(M_i,B_{2i}) = L(M_i,B_{2i+1})$. This implies that $z_0\in A_{2i+1}\triangle L(M_i,B_{2i+1})$. On the contrary, when  $z_0\notin A_{2i+1}$, we obtain $z_0\notin \mathrm{2SAT}_3$ since $z_0\in S_{A,{2i+1}}$. From this, $z_0\in C_{2i+1}$ follows. Since $C_{2i+1}\subseteq L(M_i,B_{2i}) = L(M_i,B_{2i+1})$, we immediately obtain $z_0\in A_{2i+1}\triangle L(M_i,B_{2i+1})$ as well.
By the choice of $z_0$, it therefore follows  that $z_0\in A\triangle L(M_i,B)$. This makes the requirement $R_{A,i}$ satisfied.

(2) At Stage $2i$ ($i\geq1$), in contrast, we define $C_{2i}= L(M_i,B_{2i-1})\cap\Sigma^{>n_{2i-1}}$. Similarly to (1), we can show that $(C_{2i}\triangle \mathrm{2SAT}_3)\cap \Sigma^{>n_{2i-1}}$ is infinite. We then take the lexicographically first string $z_1$ in $(C_{2i} \triangle \mathrm{2SAT}_3 ) \cap \Sigma^{>n_{2i-1}}$.
Choose $n_{2i}$ (which will be defined later) so that
$n_{2i}>k_i|z_1|+k_i$ and $k_in_{2i}+k_i\leq n_{2i}^2$.
We then define $S_{A,{2i}} = S_{A,{2i-1}}$ and $S_{B,{2i}} = S_{B,{2i-1}}\cup \{z\mid n_{2i-1}<|z|\leq n_{2i}\}$. Obviously, we obtain $z_1\in S_{B,2i}$. A similar argument to (1) proves that $z_1\in B_{2i}\triangle L(M_i,A_{2i})$. This further leads to $z_1\in B\triangle L(M_i,A)$, which makes the requirement $R_{B,i}$ satisfied.

(3) To show that $S_A,S_B\in \dtimespace(n,\sqrt{n})$, since
$S_A=\{w\mid \exists i\geq0\;[n_{2i}<|w|\leq n_{2i+1}]\}$ and $S_B=\{w\mid \exists i\geq 1\;[n_{2i-1}<|w|\leq n_{2i}]\}$, it suffices to discuss how to compute the set $\{n_{j}\}_{j\in\nat}$ of numbers.

To determine the value $n_i$, let us consider the following procedure $N$.
On input $\lambda$, sequentially determine $n_0,n_1,n_2,\ldots$ by stages in the following manner. We implement an \emph{internal counter} to count the number of steps  taken to execute the following procedure. Assume that, at Stage $2i$, the number $n_{2i}$ is already constructed (in unary) on a work tape. At Stage $2i+1$, we try to find the lexicographically first string $z_0\in\Sigma^{>n_{2i}}$ for which $z_0\in L(M_i,B_{2i})\triangle \mathrm{2SAT}_3$. This can be done as follows.
Remember that $M_i$ may possibly produce query words $w$ of length more than $\sqrt{|z|}$ on a query tape. However, since we require only the information on the length of $w$, we do not need to write down $w$ on any work tape.
We lexicographically choose strings $z$ of length more than $n_{2i}$ one by one and write $z$ on a work tape.
We run $M_i^{B_{2i}}$ on $z$.
Note that we can easily determine oracle answers of $B_{2i}$ by firstly checking the length of each query word $w$ and comparing it with $\{n_j\}_{j\leq 2i-1}$ without making actual queries to any oracle.
We then run $M_{2SAT}$ on $z$.
We check if either (i) $M_i^{B_{2i}}$ accepts $z$ and $M_{2SAT}$ rejects $z$ or (ii) $M_i^{B_{2i}}$ rejects $z$ and $M_{2SAT}$ accepts $z$.
If this is the case, then we set this $z$ to be $z_0$ and additionally ``idle'' for $|z_0|^2$ steps.  Otherwise, we continue choosing the next string $z$.
As shown above, $z_0$ belongs to $A_{2i+1}\triangle L(M_i,B_{2i+1})$.
Now, using the counter, we define $n_{2i+1}$  to be the number of steps needed to obtain $z_0$ and idle for $|z_0|^2$ steps. Up to this point, we need only work space at most $|z_0|$, which is upper-bounded by $\sqrt{n_{2i+1}}$
because of the extra idling steps.
In a similar way, we assume that, at Stage $2i-1$, $n_{2i-1}$ is already constructed. At Stage $2i$, we find the lexicographically first string $z_1\in\Sigma^{>n_{2i-1}}$ satisfying that $z_1\in L(M_i,A_{2i-1})\triangle \mathrm{2SAT}_3$. This implies that $z_1\in B_{2i}\triangle L(M_i,A_{2i})$. We set $n_{2i}$ to be the number of steps needed to obtain $z_1$ and idle for $|z_1|^2$ steps.

We define a machine $N_A$ for $S_A$ by partly simulating $N$ as follows. We equip $N_A$ with an internal counter to count the number of simulation steps of $N$. On input $x$, compute $|x|$ and simulate $N$ on $\lambda$ within $|x|$ steps. If $n_j$ is constructed within $|x|$ steps and $n_j<|x|$, then we continue the simulation of $N$. If the counter reaches $|x|$ while constructing $n_j$, then we stop the simulation.  This number $j$ is an odd number, then accept $x$ or else reject $x$.
Since $N_A$ runs in time $O(|x|)$ using space $O(\sqrt{|x|})$, it thus follows that $S_A\in \dtimespace(n,\sqrt{n})$.
Similarly, we define $N_B$ by changing ``odd number'' to ``even number''.
By the definition of $N_B$, we obtain $S_B\in \dtimespace(n,\sqrt{n})$ as well.
\end{proof}

%%%%%
\subsection{Alternative Characterizations of LSH}\label{sec:alternative}

Let us rephrase the working hypothesis LSH for $2\mathrm{SAT}_3$ in a slightly different way.
We define $\delta(\mathrm{2SAT}_3)$ to express the infimum of a real number $\varepsilon\in[0,1]$ for which there exist a polynomial $p$, a polylog function $\ell$, and a DTM solving $\mathrm{2SAT}_3$ simultaneously within time $p(m_{vbl}(x))$ and at most $m_{vbl}(x)^{\varepsilon}  \ell(|x|)$ space on all instances $x$.
If we abbreviate  $\delta(\mathrm{2SAT}_3)$ as $\delta_3$, then
there are three possible cases: (i) $\delta_{3}=0$, (ii) $0<\delta_{3}<1$, and (iii) $\delta_{3}=1$. Clearly, exactly one of them must be true.
The space bound of $n/2^{O(\sqrt{\log{n}})}$ in Theorem \ref{2SAT-solvable} for $\mathrm{2SAT}$ leads to a conclusion of $0\leq \delta_{3}\leq1$, but the bound is not strong enough to imply $\delta_{3}<1$.
Proposition \ref{2SAT-vs-delta-value}, nonetheless, asserts that the hypothesis LSH for $\mathrm{2SAT}_3$ exactly matches (iii).

\begin{proposition}\label{2SAT-vs-delta-value}
The working hypothesis LSH  for $\mathrm{2SAT}_3$ is true iff $\delta_{3}=1$ holds.
\end{proposition}

\begin{proof}
As noted earlier, Theorem \ref{2SAT-solvable} guarantees that $0\leq \delta_3\leq 1$.

(If--part) Assume that $\delta_{3}=1$. By the definition of $\delta_3$, it follows that, for any value $\varepsilon\in(0,1)$, for any polylog function $\ell(\cdot)$, and for any polynomial $p$, no DTM solves $\mathrm{2SAT}_3$  in at most $p(m_{vbl}(x))$ time using at most  $m_{vbl}(x)^{\varepsilon}\ell(|x|)$ space on all instances $x$. From this consequence, we immediately obtain $(2\mathrm{SAT}_3,m_{vbl})\notin\psublin$, which is indeed LSH for $\mathrm{2SAT}_3$.

(Only If--part) On the contrary, we assume that $\delta_{3}<1$. This means that, for any constant $\varepsilon\in(\delta_{3},1)$, there exists a DTM  that solves $\mathrm{2SAT}_3$  in time polynomial in $m_{vbl}(x)$ using $m_{vbl}(x)^{\varepsilon}\ell(|x|)$ space on all instances $x$ for a certain polylog function $\ell(\cdot)$. Since $\varepsilon<1$, it is clear that this statement contradicts LSH for $\mathrm{2SAT}_3$.
\end{proof}

%%%%%%%%%%%%%

In the rest of this subsection, we wish to seek two more characterizations of the working hypothesis LSH for $\mathrm{2SAT}_3$. For this purpose, we choose two specific $\nl$ decision problems parameterized by appropriately chosen size parameters.
The first problem is a variant of a well-known NP-complete problem, called the  \emph{$\{0,1\}$-linear programming problem} ($\mathrm{LP}_{2}$).
Hereafter, a vector of dimension $n$  means an $n\times1$ matrix and a rational number is treated as a pair of appropriate integers.
For two vectors $x$ and $y$ of dimension $n$, the notation $x\geq y$ expresses  that, for any index $i\in[n]$, the $i$th entry of $x$ is at least the $i$th entry of $y$. Fix $k\in\nat^{+}$ arbitrarily.

\ms
{\sc (2,$k$)-Entry $\{0,1\}$-Linear Programming Problem} ({\sc LP$_{2,k}$}):
\renewcommand{\labelitemi}{$\circ$}
\begin{itemize}\vs{-1}
  \setlength{\topsep}{-2mm}%
  \setlength{\itemsep}{1mm}%
  \setlength{\parskip}{0cm}%

\item {\sc Instance:} a rational $m\times n$ matrix $A$ and a rational vector $b$ of dimension $n$, where $m,n\in\nat^{+}$ and each row of $A$ has at most two nonzero entries and  each column of $A$ has at most $k$ non-zero entries.

\item {\sc Question:} is there any $\{0,1\}$-vector $x$ satisfying $Ax\geq b$?
\end{itemize}

We implicitly assume that every entry value of $A$ and $b$ is expressed as an $O(\log{mn})$-bit string.
As natural size parameters, for each instance $x=(A,b)$ given to $\mathrm{LP}_{2,k}$, we take the number $n$ of columns and the number $m$ of rows of $A$ and express them as $m_{col}(x)$ and $m_{row}(x)$, respectively.

%%%%

Another NL decision problem to consider is a variant of $\dstcon$, discussed in Section \ref{sec:properties-2SAT3}, concerning a path between two particular vertices in a given directed graph.

\ms
{\sc Degree-$k$ Directed $s$-$t$ Connectivity Problem} ({\sc $k$DSTCON}):
\renewcommand{\labelitemi}{$\circ$}
\begin{itemize}\vs{-1}
  \setlength{\topsep}{-2mm}%
  \setlength{\itemsep}{1mm}%
  \setlength{\parskip}{0cm}%

\item {\sc Instance:} a directed graph $G$ whose vertices have degree (i.e., indegree plus outdegree) at most $k$, and two designated vertices $s$ and $t$ in $G$.

\item {\sc Question:} is there any path from $s$ to $t$ (called an $s$-$t$-path) of $G$?
\end{itemize}

For any instance $x=(G,s,t)$ given to $k\dstcon$, $m_{ver}(x)$ and $m_{edg}(x)$ respectively denote the number of vertices and that of edges in $G$. Obviously, $m_{ver}(x)+m_{edg}(x)\leq |x|$, where $|x|$ denotes the length of a binary encoding of $x$.

Theorem \ref{LSH-equiv} provides two alternative definitions of LSH in terms of $\mathrm{LP}_{2,3}$ and ${3}\dstcon$.

\begin{theorem}\label{LSH-equiv}
The following statements are logically equivalent.
\begin{enumerate}\vs{-1}
  \setlength{\topsep}{-2mm}%
  \setlength{\itemsep}{1mm}%
  \setlength{\parskip}{0cm}%

\item LSH for $\mathrm{2SAT}_3$.
\item LSH for $\mathrm{LP}_{2,3}$ (with either $m_{row}$ or $m_{col}$).
\item LSH for ${3}\dstcon$ (with either $m_{ver}$ or $m_{edg}$).
\end{enumerate}
\end{theorem}

This theorem allows us to use $\mathrm{LP}_{2,3}$  and ${3}\dstcon$  for LSH as ideal substitutes for $\mathrm{2SAT}_3$. Moreover, the choice of two size parameters in each case turns out to be irrelevant.

In what follows, we want to prove Theorem \ref{LSH-equiv}. To prove this theorem, we need two supporting assertions, Propositions \ref{2SAT-DSTCON-equiv} and \ref{IP-SAT-equiv}. Notice that these  propositions actually assert the results, which are slightly more general than what we truly need for the proof of Theorem \ref{LSH-equiv}.

\begin{proposition}\label{2SAT-DSTCON-equiv}
\renewcommand{\labelitemi}{$\circ$}
\begin{enumerate}%\vs{-2}
  \setlength{\topsep}{-2mm}%
  \setlength{\itemsep}{1mm}%
  \setlength{\parskip}{0cm}%

\item $(\dstcon,m_{ver}) \sLTequiv (\mathrm{2SAT},m_{vbl})$ and  $(\dstcon,m_{edg}) \sLTequiv (\mathrm{2SAT},m_{cls})$.

\item $(3\dstcon,m_{ver}) \sLTequiv (\mathrm{2SAT}_3,m_{vbl})$ and  $(3\dstcon,m_{edg}) \sLTequiv (\mathrm{2SAT}_3,m_{cls})$.
\end{enumerate}
\end{proposition}

\begin{proposition}\label{IP-SAT-equiv}
Let $k$ be any integer with $k\geq3$.
\renewcommand{\labelitemi}{$\circ$}
\begin{enumerate}\vs{-1}
  \setlength{\topsep}{-2mm}%
  \setlength{\itemsep}{1mm}%
  \setlength{\parskip}{0cm}%

\item $(\mathrm{LP}_2,m_{row}) \sLequiv (\mathrm{2SAT},m_{cls})$ and $(\mathrm{LP}_2,m_{col}) \sLequiv (\mathrm{2SAT},m_{vbl})$.

\item $(\mathrm{LP}_{2,k},m_{row}) \sLequiv (\mathrm{2SAT}_k,m_{cls})$ and $(\mathrm{LP}_{2,k},m_{col}) \sLequiv (\mathrm{2SAT}_k,m_{vbl})$ for $k\geq3$.
\end{enumerate}
\end{proposition}

Before providing the proofs of the above two propositions, we present a brief proof of Theorem \ref{LSH-equiv}.

\begin{proofof}{Theorem \ref{LSH-equiv}}
We begin with the direction (1 $\Rightarrow$ 2). Assume that LSH for $2\mathrm{SAT}_3$ is true; namely, $(2\mathrm{SAT}_3,m_{vbl})\notin \psublin$. By Proposition \ref{IP-SAT-equiv}(2), it follows that $(\mathrm{LP}_{2,3},m_{col}) \equiv^{\mathrm{sL}}_{m} (2\mathrm{SAT}_3,m_{vbl})$. Since Lemma \ref{basic-property-SAT}(2) asserts that $(\mathrm{2SAT}_3,m_{vbl})\sLequiv (\mathrm{2SAT}_3,m_{cls})$,  Lemma \ref{reduction-property}(1) further leads to the $\leq^{\mathrm{sL}}_{m}$-equivalence   $(\mathrm{LP}_{2,3},m_{row})\equiv^{\mathrm{sL}}_{m} (2\mathrm{SAT}_{3},m_{vbl})$.
If $(\mathrm{LP}_{2,3},m_{col})\in\psublin$, then we obtain $(2\mathrm{SAT}_3,m_{vbl})\in \psublin$ since $\psublin$ is closed under $\sSLRFreduces$-reductions (and thus $\sLreduces$-reductions) by Lemma \ref{reduction-property}(2). However, this contradicts our assumption. Therefore, we reach the conclusion that $(\mathrm{LP}_{2,3},m_{col})\notin\psublin$. The above argument for $(\mathrm{LP}_{2,3},m_{col})$ works also for $(\mathrm{LP}_{2,3},m_{row})$.

It is possible to show (2 $\Rightarrow$ 1) by a similar argument used for (1 $\Rightarrow$ 2).
The remaining case for $3\mathrm{DSTCON}$ is similarly treated using
Lemma \ref{2SAT-DSTCON-equiv}(2).
\end{proofof}

To complete the proof of  Theorem \ref{LSH-equiv}, it still remains to prove Propositions \ref{2SAT-DSTCON-equiv} and \ref{IP-SAT-equiv}.
For the proof of Proposition \ref{2SAT-DSTCON-equiv}, we need the following observation (Lemma \ref{unsat-condition}) that, unlike $k$SAT with $k\geq3$, any formula in $\mathrm{2SAT}$ can be characterized purely by its clause structure. A similar observation was made in the proof of \cite[Theorem 4]{JLL76}.

To describe Lemma \ref{unsat-condition}, let us recall the notion of \emph{exact 2CNF formula} from the proof of Proposition \ref{2SAT_k-complete}. It is important to remark that any clause of an exact 2CNF formula must have ``distinct'' literals.
When an exact 2CNF formula $\phi$ is given as an instance, we introduce its  corresponding directed graph $G_{\phi} =(V_{\phi},E_{\phi})$ as follows. Let $V_{\phi}$ be the set of all literals of variables in $\phi$ and let $E_{\phi} =\{(\overline{u},v),(\overline{v},u)\mid \text{ $u\vee v$ is a clause in }\phi\}$. For two vertices $u,v\in V_{\phi}$, the notation $u\leadsto v$ expresses that there is a path in $G_{\phi}$ from $u$ to $v$.

\begin{lemma}\label{unsat-condition}
An exact 2CNF formula $\phi$ is {\em unsatisfiable} iff either one of the following two cases holds: (1) there is a vertex $v\in V_{\phi}$ such that $v\leadsto \overline{v}$ and $\overline{v}\leadsto v$ in $G_{\phi}$ or (2) there are two distinct vertices $w_1,w_2\in V_{\phi}$ such that $w_1\leadsto \overline{w_1}$, $\overline{w_1}\leadsto \overline{w_2}$, and $\overline{w_2}\leadsto w_2$ in $G_{\phi}$. These paths in (1) and (2) are called {\em contradictory chains} of $\phi$.
\end{lemma}

For completeness, we include the proof of Lemma \ref{unsat-condition}. For a directed graph $G$, its \emph{underlying undirected graph} is obtained directly from $G$ by replacing all directed edges of $G$ with their corresponding undirected edges.

\begin{proofof}{Lemma \ref{unsat-condition}}
(If--part) Let $\phi$ be any exact 2CNF formula and take the corresponding directed graph $G_{\phi} = (V_{\phi},E_{\phi})$. Assume the existence of a contradictory chain $(x_1,x_2,\ldots,x_k)$ of $\phi$ that satisfies  either Condition (1) or (2) stated in the lemma. Toward a contradiction, we further assume that $\phi$ is satisfiable.
Let us take a variable assignment $\sigma$ that makes $\phi$ true.
For each index $i\in[k-1]$, since $(x_i,x_{i+1})\in E$, the corresponding clause of $\phi$ must have the form either $\overline{x_i}\vee x_{i+1}$ or $x_{i+1}\vee \overline{x_i}$. Therefore, it follows that $\sigma(x_i)=1$ implies $\sigma(x_{i+1})=1$.
From this, we conclude that (*) for any $i,j\in[k]$ with $i<j$, if $\sigma(x_i)=1$ then $\sigma(x_{j})=1$.

In the case where Condition (1) occurs, since $x_1\leadsto  \overline{x_1}\leadsto x_k$, both
$x_1$ and $x_k$ must be the same literal by the condition and there exists an index $j\in[2,k-1]_{\integer}$ for which $x_j$ coincides with $\overline{x_1}$.
If $\sigma(x_1)=1$, then (*) implies $\sigma(x_j)=1$ but we obtain $\sigma(x_j)=\sigma(\overline{x_1})=0$, a contradiction. Thus, we obtain  $\sigma(x_1)=0$, from which $\sigma(x_j)=\sigma(\overline{x_1})=1$ follows. This implies $\sigma(x_k)=1$ but $\sigma(x_j)=\sigma(\overline{x_k})=0$, another  contradiction.

In contrast, let us consider the case of Condition (2). Since $x_1\leadsto \overline{x_1}\leadsto \overline{x_k}\leadsto x_k$ for two different literals $x_1$ and $x-2$, there are two indices $i,j\in[2,k-1]_{\integer}$ with $i<j$ such that $x_i$ coincides with $\overline{x_1}$ and $x_j$ coincides with $\overline{x_k}$.  If $\sigma(x_1)=1$, then (*) implies $\sigma(x_i)=1$ but $\sigma(x_i)=\sigma(\overline{x_1})=0$, a contradiction. On the contrary, if $\sigma(x_1)=0$, then we obtain $\sigma(x_i)=\sigma(\overline{x_1})=1$.
Since $x_j\leadsto x_j$ occurs, (*) leads to $\sigma(x_j)=1$, from which we obtain $\sigma(\overline{x_k})=\sigma(x_j)=1$.
Since $x_j\leadsto x_k$, we conclude by (*) that $\sigma(x_k)=1$, a contradiction against $\sigma(\overline{x_k})=1$. Therefore, $\phi$ must be unsatisfiable.

(Only If--part)
Assume that an exact 2CNF formula $\phi$ is unsatisfiable.
Consider the directed graph $G_{\phi}$ associated with $\phi$. Group together all clauses in $\phi$ (by appropriately changing the order of the clauses) into disjoint 2CNF subformulas $\phi_1,\phi_2,\ldots,\phi_k$ so that the  underlying undirected graphs of their corresponding graphs $G_{\phi_1},G_{\phi_2},\ldots,G_{\phi_k}$
are connected components of the underlying undirected graph of $G_{\phi}$.
Assuming that Condition (1) does not hold, we want to verify that Condition (2) is true.

Let us define a new variable assignment $\sigma$ of $\phi$ by choosing an index $r$ sequentially from $[k]$ and then defining the value of $\sigma$ for all variables of the subformula $\phi_r$.
Assume that $\phi_{r}$ has the form $\bigwedge_{i=1}^{n_r}(\overline{x_i}\vee x_{i+1})$ for certain literals $x_1,x_2,\ldots,x_{n_r+1}$
with $n_r\in\nat$.
Recall that each clause $\overline{x_i}\vee x_{i+1}$ (or $x_{i+1}\vee \overline{x_i}$) of $\phi$ is translated into two edges $(x_i,x_{i+1})$ and $(\overline{x_{i+1}},\overline{x_i})$ of $G$.
Naturally, $\phi_r$ introduces a path $y$ in $G_{\phi_{r}}$. We assume that $y$ contains no cycle whose elements are all distinct variables by ignoring negations (i.e., by identifying $x$ and $\overline{x}$ as the same ``variable''), because, if there is a cycle $(x_{i_1},\ldots,x_{i_m})$ with $x_{i_1}=x_{i_m}$ for which all variables $x_{i_j}$ are mutually distinct by ignoring negations, then we can  remove this cycle from $G_{\phi_r}$ with preserving the unsatisfiability of $\phi_r$. Let $y=(x_1,x_2,\ldots,x_e)$ for a certain constant $e\geq1$.

Initially, we set $\sigma(x_1)=0$, and thus we obtain $\sigma(\overline{x_1})=1$. Sequentially, we pick $i$ from $[e]$ one by one  and concentrate on an edge $(x_i,x_{i+1})$. If $x_{i+1}$ coincides with $x_c$ (or $\overline{x_c}$) for a certain index $c<i+1$, then the value $\sigma(x_{i+1})$ must be already defined. In contrast, if $\sigma$ has not yet assigned any value to $x_{i+1}$ (as well as $\overline{x_{i+1}}$), then we newly set $\sigma(x_{i+1})=0$, and thus $\sigma(\overline{x_{i+1}})=1$ follows.
In the case where $\sigma(x_i)=0$ for all $i\in[e]$, since $\sigma(\phi_r)$ becomes $1$,  we move to the next index $r$.
Since $\phi$ is unsatisfiable, eventually we find $r$ and $i$ for which $x_i$ appears inside $\phi_r$ and $\sigma(x_i)\neq0$.
Among those $x_i$'s, we choose the smallest index $i$ forcing $\sigma(x_{i})$ to be $1$. This means that $\sigma(\overline{x_i})$ must have been defined earlier in our construction, and thus
$x_{i}$ coincides with  a certain literal $\overline{x_c}$ with $c<i$. Since Condition (1) is false, no $x_j$ ($i<j\leq e$) coincides with $x_c$. For any $j$ ($i<j\leq e$), if $\sigma(x_j)$ has not been defined, then we set $\sigma(x_j)=1$. If $\sigma(x_j)=1$ for all $j\in[i-1,e]_{\integer}$, then we obtain  $\sigma(\phi_r)=1$.
Since $\phi$ is unsatisfiable, there is an index $d$ ($i<d\leq e$) for which  $\sigma(x_d)=0$; namely, $x_d$ coincides with $\overline{x_a}$ for $a$ with $1\leq a<d$. Since $\sigma(x_a)=1$, $i\leq a<d$ follows. This implies Condition (2) by setting $w_1=x_c$ and $w_2=\overline{x_a}$.
\end{proofof}

%%%

Now, we return to Proposition \ref{2SAT-DSTCON-equiv} and describe its proof. Given a Boolean formula $\phi$, the notation $|\phi|$ expresses the size of $\phi$ when $\phi$ is expressed in binary by an appropriate encoding scheme.
For the proof that follows below, we need to convert each 2CNF formula into its logically equivalent \emph{exact 2CNF formula}.
Such a conversion has been described in the proof of Proposition \ref{2SAT_k-complete}.
The resulting formula is succinctly referred to as the \emph{exact 2CNF version} of the original 2CNF formula.

A literal is called \emph{negative} if it is the negation of a certain variable. Otherwise, it is called \emph{positive}.

%%%

\begin{proofof}{Proposition \ref{2SAT-DSTCON-equiv}}
(1) Here, we will describe the proof of the first statement of the proposition by splitting the proof
into the following two parts (i) and (ii).

(i) We first wish to prove $(\dstcon,m_{ver}) \sLTreduces (\mathrm{2SAT},m_{vbl})$ and  $(\dstcon,m_{edg}) \sLTreduces (\mathrm{2SAT},m_{cls})$. Given an instance $x=(G,s,t)$ with $G=(V,E)$ of   $\dstcon$, without loss of generality, we assume that $s$ is a {\em unique}  source (i.e., of indegree $0$) and $t$ is a sink (i.e., of outdegree $0$) because this assumption can be achieved by modifying the original graph $G$ to another graph $G'=(V',E')$ defined by $V' = (V-ID_0)\cup\{s\}$ and $E'= E-\{(u,s),(r,s),(r,v),(t,v)\mid r\in ID_0-\{s\}, u\in V',v\in V\}$, where $ID_0= \{u\in V \mid \text{ $u$ has indegree $0$ }\}$.

From the instance $x=(G,s,t)$, we define a new Boolean formula $\phi_x$ as follows.
All vertices in $G$ are treated as ``variables'' in $\phi_x$.
We prepare two fresh variables $z_s$ and $z_t$ and set $\phi_x$ as $(s\vee z_s) \wedge (s\vee \overline{z_s}) \wedge (\bigwedge_{(v,w)\in E}(\overline{v}\vee w)) \wedge  (\overline{t}\vee z_t) \wedge (\overline{t}\vee \overline{z_t})$.
Clearly, $\phi_x$ is an exact 2CNF formula.  It also follows that $m_{cls}(\phi_x) = |E|+4\leq 4m_{edg}(x)$ and $m_{vbl}(\phi_x)=|V|+2\leq 2m_{ver}(x)$. Here, we intend to claim:
\begin{center}
(*) $x$ is in $\dstcon$ iff $\phi_x$ is unsatisfiable.
\end{center}

Assuming that Claim (*) is true, we define $M$ to be a deterministic oracle Turing machine that takes $x$ as an input and queries (the encoding of) $\phi_x$ to $\mathrm{2SAT}$, which is used as an oracle. If the oracle answers ``no'', then $M$ accepts $x$; otherwise, $M$ rejects it.
By Claim (*), $M$ turns out to be a short reduction witnessing $(\dstcon,m_{ver}) \sLTreduces (\mathrm{2SAT},m_{vbl})$ as well as  $(\dstcon,m_{edg}) \sLTreduces (\mathrm{2SAT},m_{cls})$.

In what follows, we want to verify Claim (*). Let us assume that $\phi_x$ is unsatisfiable and consider the directed graph $G_{\phi_x}=(V_{\phi_x},E_{\phi_x})$ induced from $\phi_x$. Although $G_{\phi_x}$ may not be the same as $G$, it follows that (i) for any vertices $u,w\in V$, $(u,w)\in E_{\phi_x}$ (or $(\overline{w},\overline{u})\in E_{\phi_x}$) iff $(u,w)\in E$ and (ii) for any $u\in V$ and $w\in V_{\phi_x}-\{z_s,z_t\}$,  $(u,w)\in E_{\phi_x}$ (or $(\overline{u},\overline{w})\in E_{\phi_x}$) implies $w\in V$.
Lemma \ref{unsat-condition} then yields a contradictory chain $(x_1,x_2,\ldots,x_k)$ of $\phi_x$ in $G_{\phi_x}$.
Notice by the definition of $G_{\phi_x}$ that the ``anti-chain''  $(\overline{x_k},\overline{x_{k-1}},\ldots,\overline{x_1})$ is also a contradictory chain of $\phi_x$ in $G_{\phi_x}$.
For convenience, we take the shortest contradictory chain $y=(x_1,x_2,\ldots,x_k)$.
Here, we assert that this chain $y$ has the form $(\overline{s},z'_s,s,x_4,\ldots,x_{k-3}, t,z'_t,\overline{t})$ with positive literals $x_3,\ldots,x_{k-3}$ or its anti-chain, where $z'_s\in\{z_s,\overline{z_s}\}$ and $z'_t\in\{z_t,\overline{z_t}\}$.

(a) Consider the case where $y$ is of the form $\overline{v_1}\leadsto v_1\leadsto \overline{v_1}$ with $v_1$ is a positive literal. There must be a transition point in the subpath $\overline{v_1}\leadsto v_1$ from a negative literal $x_i$ to a positive literal $x_{i+1}$. By the definition of $\phi_x$, we can conclude that either $(x_i,x_{i+1}) = (\overline{z_s}, s)$ or $(x_i,x_{i+1}) = (\overline{s},z_s)$. In the former case, since $\overline{z_s}$ cannot be $v_1$, $i>1$ and $x_{i-1}=\overline{s}$ follow.
In a similar way, a transition point on the subpath $v_1\leadsto \overline{v_1}$ from a positive literal $x_j$ to a negative literal $x_{j+1}$ leads to the conclusion that either $(x_j,x_{j+1}) = (z_t,\overline{t})$ or $(x_j,x_{j+1}) = (t,\overline{z_t})$. In the former case, we obtain  $j>1$ and $x_{j-1}=t$. As a consequence, $y$ has a subpath of the form  $(\overline{s},z'_s,s,x_{i'},\ldots,x_{j'},t,z'_t,\overline{t})$ for appropriate $i'$ and $j'$.

(b) Consider the next case where $y$ is of the form $\overline{w_1}\leadsto w_1\leadsto w_2\leadsto \overline{w_2}$ with positive literal $w_1$.  An  argument similar to (a) shows the existence of an index $i$ in the subpath $\overline{w_1}\leadsto w_1$ satisfying either $(x_i,x_{i+1}) = (\overline{z_s},s)$ or   $(x_i,x_{i+1}) = (\overline{s},z_s)$. The former case leads to $x_{i-1}=\overline{s}$.
If $w_2$ is a positive literal, then there is an index $j$ on the path $w_2\leadsto \overline{w_2}$ such that either $(x_j,x_{j+1}) = (z_t,\overline{t})$ or   $(x_j,x_{j+1}) = (t,\overline{z_t})$. In the former case, we further obtain $x_{j-1}=t$. Hence, we conclude that $y$ contain a subpath of the form  $(\overline{s},z'_s,s,x_{i'},\ldots,x_{j'},t,z'_t,\overline{t})$ for appropriate $i'$ and $j'$.
When $w_2$ is negative, by contrast, we take the subpath $w_1\leadsto w_2$ instead of $w_2\leadsto \overline{w_2}$ in the above argument and obtain the same conclusion.
However, this leads to a contradiction against the shortest condition of $y$ because the subpath $w_2\leadsto \overline{w_2}$ becomes redundant.
Therefore, $y$ must be of the form $(\overline{s},z'_s,s,x_4,\ldots,x_{k-3}, t,z'_t,\overline{t})$ or its anti-chain, as requested. The subchain $(s,x_4,\ldots,x_{k-3},t)$ then forms a path from $s$ to $t$ in $G$, and thus $x$ belongs to $\dstcon$.

Conversely, if there is an $s$-$t$ path $p=(s,x_1,x_2,\ldots,x_k,t)$ in $G$, then $(\overline{s},z_s,s,x_1,\ldots,x_k,t,z_t,\overline{t})$ is a contradictory chain of $\phi_x$. Hence, by Lemma \ref{unsat-condition}, $\phi_x$ is unsatisfiable.

(ii) Next, we want to show that $(\mathrm{2SAT},m_{vbl})\sLTreduces (\dstcon,m_{ver})$ and  $(\mathrm{2SAT},m_{cls})\sLTreduces (\dstcon,m_{edg})$.
Given an instance $\phi$ to $\mathrm{2SAT}$, we first take any logically identical clauses, such as $x\vee y$ and $y\vee x$, and remove all but one such clause because those clauses are clearly redundant. Next, we transform the resulting formula to its exact 2CNF version.
For readability, we hereafter denote this newly obtained formula by the same notation $\phi$ as well. Consider the directed graph  $G_{\phi}=(V_{\phi},E_{\phi})$ induced from $\phi$.
Note that $|V_{\phi}| \leq 2m_{vbl}(\phi)$ and $|E_{\phi}|\leq 2m_{cls}(\phi)$.
It follows by Lemma \ref{unsat-condition} that $\phi\not\in \mathrm{2SAT}$ iff $G_{\phi}$ contains a contradictory chain of $\phi$. Now, our strategy is set to check the existence of such a chain in $G_{\phi}$.

Let us consider the following oracle computation. Choose recursively vertices $v$ from $V_{\phi}$ one by one and query both $(G_{\phi},v,\overline{v})$ and $(G_{\phi},\overline{v},v)$ to $\dstcon$ used as an oracle. If all oracle answers are ``yes,'' then reject the input. Otherwise, choose recursively pairs of distinct vertices $w_1,w_2\in V_{\phi}$ and query three words of $(G_{\phi},w_1,\overline{w_1})$, $(G_{\phi},\overline{w_1},\overline{w_2})$, and $(G_{\phi},\overline{w_2},w_2)$ one by one to $\dstcon$.
If all oracle answers are ``yes,'' then reject the input. Otherwise, accept the input. It is not difficult to show that the above oracle computation solves the question of whether $\phi\in\mathrm{2SAT}$.
Note that this computation requires only $O(\log|\phi|)$ work space. Since the size of each query word is at most twice as large as the size of $G_{\phi}$, the above computation
serves as the desired short L-T-reduction.

(2)
In the above construction in (i), the degree of a vertex $v$ in $G$ except for $s$ and $t$ equals the number of clauses containing literals of the variable $v$ in $\phi$. The variable $s$ (resp., $t$) and its negation appear at most the number of times equal to the degree of $s$ (resp., $t$) plus two.
Therefore, if we take any instance $x=(G,s,t)$ given to ${3}\dstcon$, then the corresponding formula $\phi_x$ is an instance of $\mathrm{2SAT}_5$. In the above construction in (ii), the number of clauses including literals of each variable $x$ in $\phi$ matches the degree of vertex $x$ as well as $\overline{x}$. This implies that any instance $\phi$ given to $\mathrm{2SAT}_3$ can be converted to a set of instances of the form  $(G_{\phi},v,w)$, which are clearly instances of ${3}\dstcon$.

Since $(\mathrm{2SAT}_5,m)\equiv^{\mathrm{sL}}_{m} (\mathrm{2SAT}_3,m)$ for any $m\in\{m_{vbl},m_{cls}\}$ by Lemma \ref{basic-property-SAT}(1), the desired result instantly follows.
\end{proofof}

In the end of this section, we present the  proof of the remaining proposition, Proposition \ref{IP-SAT-equiv}.

\begin{proofof}{Proposition \ref{IP-SAT-equiv}}
(1) We will prove this assertion in two parts (a) and (b).

(a) We first aim at verifying that $(\mathrm{LP}_2,m_{row}) \sLreduces (\mathrm{2SAT},m_{cls})$ and $(\mathrm{LP}_2,m_{col}) \sLreduces (\mathrm{2SAT},m_{vbl})$.
Let $y=(A,b)$ denote any instance given to $\mathrm{LP}_2$ with a matrix   $A=(a_{ij})_{i\in[m],j\in[n]}$ and a vector $b=(b_i)_{i\in[m]}$ for certain numbers $m,n\in\nat^{+}$.
Notice that $m_{row}(y)=m$ and $m_{col}(y)=n$.
Moreover, let $x=(x_j)_{j\in[n]}$ denote an unknown vector.
Notice that, for any row index $i\in[m]$, the $i$th row of $Ax$ has the form either (i) $a_{ij}x_j$ or (ii) $a_{ij}x_j+a_{ik}x_k$ for certain indices $j,k\in[n]$.

Let us construct a 2CNF Boolean formula $\phi$ as follows. To specify $\phi$, we need to define a set $V$ of Boolean variables and a set $C$ of 2-disjunctive clauses over $V$. In the end, if $C$ has the form $\{\xi_1,\xi_2,\ldots,\xi_{|C|}\}$, then $\phi$ is defined as $\bigwedge_{i\in|C|} \xi_i$.

Firstly, we set $V=\{x_j\mid j\in[n]\}$ by treating each $x_j$ in $x$ as a Boolean variable. Note that $|V|= m_{col}(y)$. Let us define $C$ as follows.
We start with the aforementioned case (i), where a row of $Ax$ is of the form $a_{ij}x_j$. If $a_{ij}\geq b_i>0$ (resp., $0\geq b_i>a_{ij}$) holds, then $C$ must contain $x_j$ (resp., $\overline{x_j}$); otherwise, $C$ includes both $x_j$ and $\overline{x_j}$ because there is no solution $x$ satisfying $Ax\geq b$.
Next, we consider the case (ii), where a row of $Ax$ has the form $a_{ij}x_j+a_{ik}x_k$. If $a_{ij}+a_{ik}<b_i$, $a_{ij}<b_i$, and $a_{ik}<b_i$, then $C$ must contain both $x_j$ and $\overline{x_j}$,
because $Ax\geq b$ has no solution $x$. We thus assume otherwise. There are two cases of $b_i>0$ and $b_i\leq 0$ to consider. In the case of $b_i>0$, if $a_{ij}+a_{ik}\geq b_i$, $a_{ij}<b_i$, and $a_{ik}<b_i$,  then $C$ must contain both $x_j$ and $x_k$. If $a_{ij}+a_{ik}\geq b_i$, $a_{ij}\geq b_i$, and $a_{ik}\geq b_i$,  then $C$ contains $x_j\vee x_k$. If $a_{ij}+a_{ik}<b_i$, $a_{ij}<b_i$, and $a_{ik}\geq b_i$, then $C$ contains $\overline{x_j}$ and $x_k$. If $a_{ij}+a_{ik}<b_i$, $a_{ij}\geq b_i$, and $a_{ik}\geq b_i$, then $C$ contains $x_j$ and $\overline{x_j}$. The other cases can be similarly treated.
In the case of $b_i\leq 0$, by contrast, if $a_{ij}+a_{ik}\geq b_i$, $a_{ij}<b_i$, and $a_{ik}\geq b_i$, then $C$ contains $\overline{x_j}\vee x_k$. If $a_{ij}+a_{ik}\geq b_i$, $a_{ij}\geq b_i$, and $a_{ik}\geq b_i$, then $C$ contains no corresponding clauses. If $a_{ij}+a_{ik}<b_i$, $a_{ij}\geq b_i$, and $a_{ik}\geq b_i$, then $C$ must contain both $x_j\vee x_k$ and $\overline{x_k}\vee \overline{x_j}$.
The other cases can be treated in a similar way.

From the above definition of $C$, we conclude that $|C|\leq 2m = 2 m_{row}(y)$. It is not difficult to show from the definition of $\phi$ that $Ax\geq b$ holds for a certain $x$ iff $\phi$ is satisfiable. This equivalence implies that $(\mathrm{LP}_2,m_{row}) \sLreduces (\mathrm{2SAT},m_{cls})$ and $(\mathrm{LP}_2,m_{col}) \sLreduces (\mathrm{2SAT},m_{vbl})$.

(b) We intend to show the remaining reductions of $(\mathrm{2SAT},m_{cls})\sLreduces (\mathrm{LP}_2,m_{row})$ and $(\mathrm{2SAT},m_{vbl})\sLreduces (\mathrm{LP}_2,m_{col})$. Given an instance $\phi$ of $\mathrm{2SAT}$, we consider its exact 2CNF version $\hat{\phi}$.
We take the set $V$ of all Boolean variables and the set $C$ of all 2-disjunctive clauses in  $\hat{\phi}$.
We assume that $V=\{x_1,x_2,\ldots,x_n\}$ and $C=\{c_1,c_2,\ldots,c_m\}$, where $n=|V|$ and $m=|C|$.

Let us define an $m\times n$ matrix $A=(a_{ij})_{i\in[m],j\in[n]}$ and an  $m$-dimensional vector $b=(b_i)_{i\in[m]}$ as follows. Let $i$ denote any index in $[m]$ and consider the clause $c_i$. Assume that $c_i$ is of the form $x_j\vee x_k$ for certain variables $x_j$ and $x_k$. In this case, we define $a_{ij}=a_{ik}=b_i=1$. From $Ax\geq b$, the inequality $x_j+x_k\geq 1$ follows. If $c_i$ has the form $\overline{x_j}\vee x_k$, then we set $a_{ij}=-1$, $a_{ik}=1$, and $b_i=0$ so that we obtain $-x_j+x_k\geq 0$. The case of $x_j\vee \overline{x_k}$ is similarly treated. If $c_i$ is of the form $\overline{x_j}\vee \overline{x_k}$, then we set $a_{ij}=a_{ik}=b_i=-1$, and thus we obtain $-x_j-x_k\geq -1$.

Finally, we set $y=(A,b)$. Note from the definition of $A$ and $b$ that $m_{col}(y)=|V|$ and $m_{row}(y)=|C|$. It is not difficult to show that $\phi \in \mathrm{2SAT}$ iff there exists a $\{0,1\}$-solution  $x$ of  $Ax\geq b$.  Therefore, we  conclude that $(\mathrm{2SAT},m_{vbl})\sLreduces (\mathrm{LP}_2,m_{col})$ and $(\mathrm{2SAT},m_{cls})\sLreduces (\mathrm{LP}_2,m_{row})$.

(2) In the construction of short reductions in (1), the number of occurrences of each variable in the form of literals in $\phi$ directly corresponds to the number of non-zero entries of its associated column of $A$.
This number is at most $3$.
Recall from Lemma \ref{basic-property-SAT}(2) that  $(\mathrm{2SAT}_3,m_{ver})\sLequiv (\mathrm{2SAT}_3,m_{cls})$. Using this fact, we obtain the desired result.
\end{proofof}

%%%%%%%%%%%%%%%%%%
%%%%%%%%%%%%%%%%%%
\section{Four Simple Applications of LSH for 2SAT$_3$}\label{sec:application}

To demonstrate the usefulness of the working hypothesis LSH for $\mathrm{2SAT}_3$, we will seek out four simple applications of it in the fields of search problems and optimization problems as well as automata theory.

%%
%%%
\subsection{Non-Solvability of NL Search Problems}\label{sec:NL-search-problem}

Many NL decision problems have been transformed into {\em NL search problems} by converting them   ``straightforwardly'' into a framework of search problems. However, this naive convention may not always work.  For example, 2SAT is NL-complete but the associated problem of finding a truth assignment (when variables are ordered in an arbitrarily fixed way) that satisfies a given 2CNF  formula does not look like a legitimate form of NL search problem.
Furthermore, its optimization version of $\mathrm{2SAT}$, Max2SAT, is already complete for $\mathrm{APX}$ (polynomial-time approximable NP optimization) instead of NLO (NL optimization class) under polynomial-time approximation-preserving reductions (see \cite{ACG+03}).

Following \cite{Yam13}, we will briefly describe the basic framework of NL search problems.
Generally speaking, a \emph{search problem} is expressed as $(I,SOL)$, where $I$ consists of (admissible) instances and $SOL$ is a function from $I$ to a set of solutions (called a \emph{solution space}) such that, for any pair $(x,y)\in I\circ SOL$, $|y|\leq p(|x|)$ holds for a certain fixed polynomial $p$, where $I\circ SOL$ stands for the set $\{(x,y)\mid x\in I, y\in SOL(x)\}$.
Among all such search problems, an \emph{NL search problem} $(I,SOL)$ further satisfies that $I\in \dl$ and $I\circ SOL\in\auxl$.
We denote by $\search\nl$ the collection of all NL search problems.
A DTM $M$ equipped with a write-once output tape is said to {\em solve} $(I,SOL)$  if, for any instance $x\in I$, $M$ takes $x$ as an input and produces a solution in $SOL(x)$ on the output tape if $SOL(x)\neq\setempty$, and produces a designated symbol $\bot$ (``no solution'') otherwise.

Let us present two simple applications of LSH for $\mathrm{2SAT}_3$ to the area of NL search problems.
In 1976, Jones \etalc~\cite{JLL76} discussed the NL-completeness of  a decision problem concerning {\em one-way nondeterministic finite automata} (or 1nfa's, for short).
Recall that $\lambda$ is the \emph{empty string} of length $0$. A 1nfa is a tuple $(Q,\Sigma,\delta,q_0,F)$, where $Q$ is a finite set of inner states, $\Sigma$ is an alphabet, $q_0$ is the initial state in $Q$, $F$ ($\subseteq Q$) is a set of final states, and $\delta$ is a transition function.
When the 1nfa has no $\lambda$-move (or $\lambda$-transition), $\delta$ is given as a function from $Q\times\Sigma$ to $\PP(Q)$.
An input string $x$ in $\Sigma^*$ is initially written on a read-once input tape with no endmarker. In a single step, if $M$ is in inner state $q$ scanning symbol $\sigma$ and $\delta(q,\sigma)\neq \setempty$, then $M$ nondeterministically chooses an inner state $p$ in $\delta(q,\sigma)$, changes its inner state from $q$ to $p$, and moves its tape head to the right. We say that $M$ \emph{accepts} $x$ if, starting with $q_0$, $M$ stays in a final state in $F$ just after reading the last symbol of $x$.

We modify the problem of Jones  et al. into an associated search problem, called $\search\mathrm{1NFA}$, given below. Here, we assume an efficient encoding scheme of $M$ and $1^n$ into a binary string.

\ms
{\sc 1NFA Membership Search Problem} ({\sc Search-1NFA}):
\renewcommand{\labelitemi}{$\circ$}
\begin{itemize}\vs{-2}
  \setlength{\topsep}{-2mm}%
  \setlength{\itemsep}{1mm}%
  \setlength{\parskip}{0cm}%

\item {\sc Instance:} a 1nfa $M=(Q,\Sigma, \delta,q_0,F)$ with no $\lambda$-moves, and a parameter $1^n$.

\item {\sc Solution:} an input string $x$ of length $n$ accepted by $M$.
\end{itemize}

As a meaningful size parameter $m_{nfa}$, we set $m_{nfa}(x)$ to be $|Q||\Sigma|n$ for any given instance $x=(M,1^n)$. Note that $m_{nfa}$ is almost non-zero and polynomially bounded.

\begin{theorem}\label{Search-1NFA}
Assuming that LSH for $\mathrm{2SAT}_3$, for every fixed value $\varepsilon\in(0,1/2]$, there is no $|x|^{O(1)}$-time $O(m_{nfa}(x)^{1/2-\varepsilon})$-space algorithm for $\search\mathrm{1NFA}$ on all instances $x$.
\end{theorem}

\begin{proof}
Toward a contradiction, we assume that $\search\mathrm{1NFA}$ is solvable by a certain DTM $M$ in time polynomial in $|y|$ using space at most $cm_{nfa}(y)^{1/2-\varepsilon}$ on all instances $y$,
where $c$ and $\varepsilon$ are positive constants. Our aim is to show that $({3}\dstcon,m_{ver})$ can be solved in polynomial time using sub-linear space, because this contradicts LSH for ${3}\dstcon$, which is  equivalent to LSH for $\mathrm{2SAT}_3$ by Theorem \ref{LSH-equiv}.

Let $x=(G,s,t)$ be any instance given to ${3}\dstcon$ with $G=(V,E)$ and $s,t\in V$. Assume that $s\neq t$ since the case of $s=t$ is trivial. Let $n=m_{ver}(x)$.  Associated with this $x$, we want to define a non-$\lambda$-move 1nfa $N=(Q,\Sigma,\delta,q_0,F)$.
Firstly, we set $Q=V$ and $\Sigma=[3]$, and define $q_0=s$ and $F=\{t\}$.  For each vertex  $v\in V$, let us consider its neighbor $out(v) = \{w\in V\mid (v,w)\in E\}$. We assume that  all elements in $out(v)$ are enumerated in a fixed linear order as $out(v)=\{w_1,w_2,\ldots,w_{k_v}\}$ with $0\leq k_{v}\leq 3$, where $k_v=0$ means $out(v)=\setempty$. The transition function $\delta$ is defined for such $v$ and for any $i\in\Sigma$ as $\delta(v,i)=\{w_i\}$ if $1\leq i\leq k_{v}$, and $\delta(v,i)=\setempty$ otherwise.

Suppose that $\gamma$ is a path from $s$ to $t$ in $G$ and let $\gamma=(v_1,v_2,\ldots,v_d)$ for a certain $d$ with $2\leq d\leq n$. For each index $i\in[d]$, we arbitrarily choose one index  $\ell(v_{i})$ satisfying $v_{i+1}=w_{\ell(v_{i})}\in out(v_i)$ if $out(v_i) = \{w_1,w_2,\ldots,w_{k_{v_i}}\}$, and we then define $z$ to be the string $\ell(v_1)\ell(v_2)\cdots \ell(v_{d-1})0^{n-d+1}$ of length $n$. When $N$ reads  this string $z$, it eventually enters $v_d$, which is a final state, and therefore $N$ accepts $z$. On the contrary, in the case where there is no $s$-$t$ path in $G$, $N$ never accepts any input. Therefore, it follows that (*) $G$ has an $s$-$t$ path iff $N$ accepts $z$.

Finally, we set $y=(N,1^n)$  as the desired instance to $\search\mathrm{1NFA}$ parameterized by $m_{nfa}$.  Note that $m_{nfa}(y)=|Q||\Sigma|n \leq 3|V|^2 = 3m_{ver}(z)^2$.
By Statement (*), ${3}\dstcon$ is solvable by running $M$ on $y$ in polynomial time; moreover, the space required for this computation is upper-bounded by $cm_{nfa}(y)^{1/2-\varepsilon}\leq 2c m_{ver}(x)^{1-2\varepsilon}$, which is obviously sub-linear because of $\varepsilon>0$.
\end{proof}

%%%%%

In 1995, Jenner \cite{Jen95} presented a few variants of the well-known {\em $\{0,1\}$-knapsack problem} and showed the NL-completeness of these variants. Here, we choose  one of them, which fits into the NL-search framework by a small modification.
Given a string $x$, its substring $z$ is said to be {\em unique in} $x$ if there exists exactly one pair $(u,v)$ satisfying $x=uzv$. It is easy to check, for given strings $x$ and $z$, whether or not $z$ is a unique substring of $x$.

\ms
{\sc Unique Ordered Concatenation Knapsack Search Problem} ({\sc Search-UOCK}):
\renewcommand{\labelitemi}{$\circ$}
\begin{itemize}\vs{-2}
  \setlength{\topsep}{-2mm}%
  \setlength{\itemsep}{1mm}%
  \setlength{\parskip}{0cm}%

\item {\sc Instance:} a string $w$ and a sequence $(w_1,w_2,\ldots,w_n)$ of strings over a certain fixed alphabet $\Sigma$ such that, for every index $i\in[n]$, if $w_i$ is a substring of $w$, then $w_i$ is unique.

\item {\sc Solution:} a sequence $(i_1,i_2,\ldots,i_k)$ of indices with $k\in[n]$ for which $1\leq i_1<i_2<\cdots<i_k\leq n$ and $w=w_{i_1}w_{i_2}\cdots w_{i_k}$.
\end{itemize}

It is convenient to assume that $(w_1,w_2,\ldots,w_n)$ is given in the form $w_1\# w_2\# \cdots \# w_n\#$, where $\#$ is a distinguished separator symbol  not in $\Sigma$.
Our reasonable choice of size parameter  $m_{elm}$ for $\search\mathrm{UOCK}$ is the number of all the elements $w_1,w_2,\ldots,w_n$ in the above definition  (namely, $m_{elm}(x)=n$ for each instance $x$). Obviously, $m_{elm}$ is a log-space size parameter.

\begin{theorem}\label{search-UOCK}
If LSH for $\mathrm{2SAT}_3$ holds, then, for any $\varepsilon>0$, there is no $|x|^{O(1)}$-time $O(m_{elm}(x)^{1/2-\varepsilon})$-space algorithm for  $\search\mathrm{UOCK}$ working on all instances $x$.
\end{theorem}

\begin{proof}
Notice that $m_{elm}$ is polynomially bounded and $m_{elm}(x)>0$ for any valid instance $x$ given to $\search\mathrm{UOCK}$.
Let us thus assume that there are two constants $\varepsilon,c>0$, a polynomial $p$, and a $p(|x|)$-time $cm_{elm}(x)^{1/2-\varepsilon}$-space algorithm $A$ solving $\search\mathrm{UOCK}$ on all instances $x$. We will use this algorithm $A$ to solve $({3}\dstcon,m_{ver})$ in polynomial time using sub-linear space because this obviously contradicts LSH for ${3}\dstcon$, which implies LSH for $\mathrm{2SAT}_3$  by Theorem \ref{LSH-equiv}(3).

Let $x=(G,s,t)$ denote any instance given to ${3}\dstcon$ with $G=(V,E)$. Note that $|E|\leq 3|V|$. For simplicity of our argument, without loss of generality, let $V=\{1,2,\ldots,n\}$, $s=1$, and $t=n$. Now, we define $\pair{i,j}=(i-1)n+j$ for each pair $i,j\in[n]$.
Firstly, we modify $G$ into another directed graph $G'=(V',E')$ with  $V'=\{\pair{i,j} \mid i,j\in[n]\}$ and $E'=\{(\pair{i,j},\pair{i',j'})\in V'\times V' \mid i'=i+1,(j,j')\in E\}$. It then follows that $|V'|=n^2$ and $|E'|=|V||E|\leq 3|V|^2=3n^2$ since $|E|\leq 3|V|$. Moreover, we set $s'=\pair{1,s}$ and $t'=\pair{n,t}$. This new graph $G'$ satisfies the following property, called the \emph{topological order}: for any pair $i,j\in V'$, $(i,j)\in E'$ implies $i<j$.

Consider the triplet $(G',s',t')$, which is also an instance of $3\dstcon$. From  this instance, we define $w=bin(1)\# bin(2)\# \cdots \# bin(n)\#$, where $bin(i)$ indicates the \emph{binary representation} of a natural number $i$ and $\#$ is a designated separator not in $\{0,1\}$. For each edge $(i,j)\in E'$, we further define $w_{ij}=bin(i+1)\# bin(i+2)\# \cdots \# bin(j)\#$. It follows by the definition that, for each $(i,j)$, if $w_{ij}$ is a substring of $w$, then $w_{ij}$ must be unique in $w$. Clearly, the sequence  $z_x=(w,w_{ij})_{(i,j)\in E'}$ is an instance of  $\search\mathrm{UOCK}$ satisfying that $m_{elm}(z_x) = |E'| \leq 3n^2 = 3m_{ver}(x)^2$.

Given any instance $x$, we can solve $({3}\dstcon,m_{edg})$ by running $A$ on the corresponding input $z_x$ in time polynomial in $|x|$ using space at most $cm_{elm}(z)^{1/2-\varepsilon}$,  which is upper-bounded by $3cm_{ver}(x)^{1-2\varepsilon}$. Therefore, we conclude that $(3\dstcon,m_{edg})\in \psublin$, as requested.
\end{proof}

%%%%%%
\subsection{Nonapproximability of NL Optimization Problems}

The next practical application of the working hypothesis LSH for $\mathrm{2SAT}_3$ targets the area of combinatorial NL  optimization.
We will briefly explain several key concepts of \emph{NL optimization problems}. Refer to  \cite{Tan07,Yam13} for the precise definitions of these concepts. An {\em NL optimization problem} (or an NLO problem, for short) $P$ is a quadruple $(I,SOL,mes,goal)$, where $I$ is a set of (admissible) instances in $\dl$, $SOL(x)$ is a set of solutions for each $x\in I$ with the set $I\circ SOL =\{(x,y) \mid x\in I, y\in SOL(x)\}$ from $\auxl$, $mes$ is a function in $\auxfl$ mapping  $I\circ SOL$ to $\nat^{+}$, and $goal$ is in $\{\text{\sc max},\text{\sc min}\}$.
Let $\nlo$ stand for the class of all NLO problems.
An \emph{optimal solution} $y$ for instance $x$ of $P$ must satisfy $mes(x,y)=mes^*(x)$, where $mes^*(x) = goal_{y\in SOL(x)}\{mes(x,y)\}$. The \emph{performance ratio} $R$ of a solution $y$ of $P$ for $x$ is defined as $R(x,y) = \max\{\frac{mes^*(x)}{mes(x,y)},\frac{mes(x,y)}{mes^*(x)}\}$.
We say that an NLO problem $P$ is \emph{solvable in log space} if there is a DTM that takes any instance $x\in I$ and outputs an optimal solution in $SOL(x)$ using space $O(\log{|x|})$. We write $\lo_{\nlo}$ (log-space computable NL optimization) to denote the class of all NLO problems solvable in log space.

We are focused on a problem that does not seem to be solvable using log space but can be approximated using log space.
A \emph{log-space approximation scheme} for an NLO problem $P$ is a DTM $M$ that takes any input of the form $(x,k)$ and outputs a  solution $y$ of $P$ for $x$ using space at most $f(k)\log|x|$ for a certain log-space computable function $f:\nat\to\nat$ for which the performance ratio $R$ satisfies $R(x,y)\leq 1+\frac{1}{k}$. Such a solution $y$ is called a \emph{$(1+\frac{1}{k})$-approximate solution}.
The notation $\lsas_{\nlo}$ denotes the class of all NLO problems that have log-space approximation schemes.
It then follows that $\lo_{\nlo}\subseteq \lsas_{\nlo} \subseteq \nlo$.

In 2007, Tantau \cite{Tan07} presented an NL maximization problem, called $\mathrm{Max\mbox{-}HPP}$, which falls into $\lsas_{\nlo}$. This problem was later rephrased in \cite{Yam13b} in terms of complete graphs and it was shown to be computationally hard for $\lo_{\nlo}$  under \emph{approximation-preserving exact NC$^{1}$-reduction}.

For an  $n\times n$ matrix $A$ and two indices $i,j\in[n]$, we write $A_{ij}$ to denote the $(i,j)$-entry of $A$.

\ms
{\sc Maximum Hot Potato Problem} ({\sc Max-HPP}):
\renewcommand{\labelitemi}{$\circ$}
\begin{itemize}\vs{-2}
  \setlength{\topsep}{-2mm}%
  \setlength{\itemsep}{1mm}%
  \setlength{\parskip}{0cm}%

\item {\sc Instance:} an $n\times n$ matrix $A$ whose entries are drawn from $[n]$, a number $d\in[n]$, and a start index $i_1\in[n]$, where $n\in\nat^{+}$.

\item {\sc Solution:} an index sequence $\SSS=(i_1,i_2,\ldots,i_d)$ of length $d$ with $i_j\in[n]$ for any $j\in[d]$.

\item {\sc Measure:} the total weight $w(\SSS) = \sum_{j=1}^{d-1} A_{i_j i_{j+1}}$.
\end{itemize}

We use the total number $n$  of columns of a given matrix $A$ as our desirable size parameter $m_{col}(A,d,i_1)$.
In what follows, we show that, under the assumption of LSH for $\mathrm{2SAT}_3$, $(\mathrm{Max\mbox{-}HPP},m_{col})$ cannot have polynomial-time $O(k^{1/3}\log{m_{col}(x)})$-space approximation schemes of finding $(1+\frac{1}{k})$-approximate solutions for instances $x$.

\begin{theorem}\label{Max-CPath-Weight-solver}
If LSH for $\mathrm{2SAT}_3$ is true, then there is no $|x|^{O(1)}$-time $O(k^{1/3}\log{m_{col}(x)})$-space algorithm that, on all instances $(x,k)$, finds $(1+\frac{1}{k})$-approximate solutions of $\mathrm{Max\mbox{-}HPP}$ for $x$.
\end{theorem}

To prove this theorem, we give a useful supporting lemma. Let $g:\nat\to\nat$ be any function.
An optimization problem $(I,SOL,mes,goal)$ parameterized by $m$ is said to be {\em $g(m(x))$-bounded} if $mes(x,y)\leq g(m(x))$ holds for any $(x,y)\in I\circ SOL$.

\begin{lemma}\label{LSAS-logspace-bound}
Let $c\geq 1$. Every $O(m(x)^c)$-bounded maximization problem in $\nlo$, parameterized by a log-space size parameter $m$, whose $(1+\frac{1}{k})$-approximate solutions are found using space $O(k^{\frac{1}{2c+1}}\log{m(x)})$
can be solved for a certain constant $\varepsilon>0$ in time $|x|^{O(1)}$ using space $O(m(x)^{1/2-\varepsilon})$ on all instances $x$.
\end{lemma}

\begin{proof}
Let $A=(I,SOL,mes,\text{\sc max})$ denote any maximization problem given in the premise of the lemma with a   log-space size parameter $m$.
We then choose three constants $a,b>0$ and $c\geq1$ and take  a DTM $M$ satisfying that, for any pair $(x,y)\in I\circ SOL$,  $mes(x,y)\leq a m(x)^c$ holds and $M$ on instance $(x,k)$ produces $(1+\frac{1}{k})$-approximate solutions $y$ using at most $f(k)\log|x|$ space, where $k\in\nat^{+}$ and $f(k)=bk^{\frac{1}{2c+1}}$.

In what follows, we want to solve $A$ using both polynomial time and $O(m(x)^{1/2-\varepsilon})$ space. Let $x$ be any instance in $I$ and let  $n=m(x)$ for simplicity. We set $b'=b(2a)^{\frac{1}{2c+1}}$, $k= an^c+1$, and  $g(n)=f(an^c+1)\log{n}$, which is at most $b'n^{\frac{c}{2c+1}}\log{n}$. By further  setting  $\varepsilon=\frac{1}{2c+1}$, it immediately follows that, for any number $n$,  $g(n)$ is upper-bounded by $n^{1/2-\varepsilon}$ if $b(2a)^{\frac{c}{2c+1}}\log{n}\leq n^{\frac{1}{2c+1}}$. Let $y$ denote an output of $M$ on the instance $(x,k)$. Recall from the definition of $mes$ that $mes(x,y)$ must take positive integers. The performance ratio $\frac{mes^*(x)}{mes(x,y)}$ is at most $1+\frac{1}{k}$, which equals $1+\frac{1}{an^c+1}$. Since $mes(x,y)\leq an^c$, we conclude that  $mes^*(x)\leq mes(x,y)+\frac{mes(x,y)}{an^c+1}\leq mes(x,y)+\frac{an^c}{an^c+1}$. Since $mes(x,y)\leq mes^*(x)$ and $\frac{an^c}{an^c+1}<1$, $mes(x,y)=mes^*(x)$ follows. This implies that $y$ is truly an optimal solution. Note that the space usage is at most $g(n)\leq n^{1/2-\varepsilon}$, as requested.
\end{proof}

We return to the proof of Theorem \ref{Max-CPath-Weight-solver}.

\begin{proofof}{Theorem \ref{Max-CPath-Weight-solver}}
Let us consider $(\mathrm{Max\mbox{-}HPP},m_{col})$. Toward a contradiction, we assume that there exists a polynomial-time $O(k^{1/3}\log{m_{col}(z)})$-space algorithm of finding $(1+\frac{1}{k})$-approximate solutions of each instance $z$ given to   $\mathrm{Max\mbox{-}HPP}$. Our goal is to obtain the failure of LSH for $\mathrm{2SAT}_3$, which is logically equivalent to LSH for $3\dstcon$ by Theorem \ref{LSH-equiv}(3). From our assumption, Lemma \ref{LSAS-logspace-bound} ensures that $(\mathrm{Max\mbox{-}HPP},m_{col})$ is solvable by a certain DTM, say, $M$ in polynomial time using space at most $cm_{col}(z)^{1/2-\varepsilon}$ on all instances $z$ for two fixed constants $c>0$ and $\varepsilon\in(0,1/2)$. We want to use this machine $M$ as a basis to design the desired algorithm for $({3}\dstcon,m_{ver})$ running in polynomial time using sub-linear space.

Let $x=(G,s,t)$ denote any instance given to ${3}\dstcon$ with $G=(V,E)$ and $n=|V|\geq 2$. We assume that $s\neq t$ since, otherwise, $x$ trivially belongs to ${3}\dstcon$.
We define another directed graph $G'=(V',E')$ by setting $V'=\{(i,v)\mid i\in[n],v\in V\}$ and $E'=\{((i,u),(i+1,v)), ((i,t),(i+1,t)) \mid i\in[n-1],(u,v)\in E\}$. Note that $|V'|=n^2$. Moreover, we set $s'=(1,s)$ and $t'=(n,t)$. From this new graph $G'$, we construct an instance $z=(A,n,s')$ of $\mathrm{Max\mbox{-}HPP}$ as follows,
where $A$ is an $n^2\times n^2$ matrix whose index set is $V'\times V'$.
We begin with setting $A_{s't'}=A_{t's'}=A_{vv}=1$ for any $v\in V'$, $A_{t's'}=n$, $A_{t'v}=1$ for any $v\in V'-\{s\}$, and $A_{vs'}=1$ for all $v\in V'$. For any other pair $(u,v)$ in $V'\times V'$, if  $(u,v)\in E'$, then we define $A_{vw}=n$; otherwise, we define $A_{uv}=1$. It follows that $m_{col}(z)=n^2 = m_{ver}(x)^2$. Clearly, $z=(A,n,s')$ is an instance of  $\mathrm{Max\mbox{-}HPP}$, and thus we can obtain a $(1+\frac{1}{k})$-approximate solution of $z$ by running $M$ on $z$.
As the value of $k$, we choose $k=n^2$.

Let us explain how to solve $3\dstcon$ on the instance $(G',s',t')$. If there is a path $\gamma = (v_1,v_2,\ldots,v_d)$ from $s'$ to $t'$ in $G'$ for a certain number $d\in[n^2-1]$, then $d$ must be $n$. In this case, we define $v_{ln+j}=v_j$ for all pairs $l,j\in[n-1]$. Since $v_d=t'$, we obtain  $A_{v_{ln}v_{ln+1}} = A_{t's'} = n$ for any $l\in[n-1]$. It then follows that $w(\gamma) = \sum_{i=1}^{n^2-1}A_{v_i v_{i+1}} = n^2$ and clearly this is optimal. On the contrary, let $\gamma^* =(v_1,v_2,\ldots,v_{n^2})$ be an optimal solution of $z$ with an optimal value $n^2$.  By the definition of $z$, $v_1$ must be $s'$. Moreover, for all indices $i\in[n^2-1]$, $A_{v_i v_{i+1}}=n$ must hold. We thus conclude that either $(v_i,v_{i+1})\in E'$ or $v_i=t' \wedge v_{i+1}=s'$.
This implies that $\gamma^*$ contains a subpath from $s'$ to $t'$ in $G'$.

Let $\gamma=(u_1,u_2,\ldots,u_{n^2})$ denote an outcome of $M$ on $z$. We intend to assert that $\gamma$ equals $\gamma^*$. Assume otherwise.
Since  $\sum_{i=1}^{n^2-1}A_{u_iu_{i+1}}\leq n^2-1$, the performance ratio $R(\gamma,\gamma^*)$ satisfies that $R(\gamma,\gamma^*)\geq \frac{n^2}{n^2-1}=1+\frac{1}{n^2-1}$. However, this contradicts our assumption that $R(\gamma,\gamma^*)\leq 1+\frac{1}{k}=1+\frac{1}{n^2}$.

Hence, if we run $M$ on the input $z$, then we obtain the optimal index sequence $\gamma^*$. By the above argument, if $w(\gamma)= n^2$, then an $s$-$t$ path indeed exists. Otherwise, since $A_{v_iv_{i+1}}\neq n$ for a certain $i\in[n^2-1]$,  there is no $s$-$t$ path in $G$. Since $M$ uses at most $cm_{col}(z)^{1/2-\varepsilon}$ space, the space usage of the whole procedure is at most  $cm_{col}(z)^{1/2-\varepsilon}$, which turns out to be  $cm_{ver}(x)^{1-2\varepsilon}$ because of $m_{col}(z)=m_{ver}(x)^2$. Therefore, ${3}\dstcon$ is solvable in polynomial time using sub-linear space; in other words, LSH for ${3}\dstcon$ fails, as requested.
\end{proofof}

%%%%%
\subsection{Hardness of Transforming Unary 1nfa's to 1dfa's}

The fourth example concerns with the computational complexity of transforming one type of finite automata into another type. Recall the notion of 1nfa's from Section \ref{sec:NL-search-problem}. In a similar way, we can define a \emph{one-way deterministic finite automaton} (or a 1dfa, for short), which always moves its tape head to the right.
Let us consider 1nfa's and 1dfa's that work on single-letter input alphabets. These machines are conventionally called \emph{unary 1nfa's} and \emph{unary 1dfa's}.
It is easy to show that each unary 1nfa $M$ can be converted into a unary 1dfa $M'$ that accepts the same input as the original 1nfa does. We call such a unary 1dfa an \emph{equivalent 1dfa}.
More precisely, for any given $n$-state unary 1nfa, there is always an equivalent unary 1dfa of $O(n\log{n})$ states.
A standard procedure of such transformation requires polynomial-time and $O(n)$ space (cf. \cite{HU79}). We quickly review this transformation procedure.

\begin{lemma}
For any given $n$-state unary 1nfa, there is always an equivalent unary 1dfa of $O(n\log{n})$ states.
\end{lemma}

\begin{proof}
Given an $n$-state unary 1nfa $N$, we first transform it to an equivalent 1nfa $N'$ with no $\lambda$-moves (or $\lambda$-transitions). The number of states required for this new 1nfa is $O(n\log{n})$. Write $\tilde{n}$ for this number and let $N'=(Q,\{0\},\{\triangleleft\}, \delta,q_0,F)$ with $Q=\{q_i\mid i\in[0,\tilde{n}-1]_{\integer}\}$. Starting with this $N'$, we wish to  construct the desired unary 1dfa $M=(Q',\{0\},\{\triangleleft\}, \delta',p_0,F')$. The elements of $Q'$ are inductively defined as follows. Let $p_0=\{q_0\}$.  For each index $i\in[\tilde{n}-2]$, we assume by induction hypothesis that $p_i=\{q_{j_1},q_{j_2},\ldots,  q_{j_t}\}$ with $t\leq \tilde{n}$ has been  already defined. We then define $p_{i+1} = \bigcup_{k\in[t]}\delta(q_{j_k},0)$ and set $\delta'(p_i,0)=p_{i+1}$. When $p_{i+1}$ contains an element in $F$, we place $p_{i+1}$ into $F'$. We stop this construction procedure as long as either $p_{i+1}$ becomes empty or $i+1$ becomes $\tilde{n}-1$. For such $i$, we additionally set $\delta'(p_i,0)=p_i$. Clearly, this new machine $M$ is deterministic and uses at most $\tilde{n}$ states. Moreover, it follows that $M$ accepts (resp., rejects) $x$ iff $N'$ accepts (resp., rejects) $x$.
\end{proof}

The transformation algorithm described in the above proof requires $O(n\log{n})$ space. Under LSH for $\mathrm{2SAT}_3$, nonetheless, we can demonstrate that this space bound cannot be made significantly smaller.

\begin{theorem}\label{1nfa-1dfa}
If LSH for $\mathrm{2SAT}_3$ is true, then, for any constant $\varepsilon\in[0,1)$,  there is no polynomial-time $O(n^{\varepsilon})$-space algorithm that takes an $n$-state unary 1nfa as input and produces an equivalent unary 1dfa of $O(n\log{n})$ states.
\end{theorem}

\begin{proof}
We assume that the existence of a constant $\varepsilon\in[0,1)$ and a polynomial-time $an^{\varepsilon}$-space algorithm $A$ that generates from an $n$-state  unary 1nfa an equivalent 1dfa of at most $cn\log{n}+c$ states, where $a$ and $c$ are appropriate positive constants. The size of $A$'s output may be assumed to be upper-bounded by $a'n^{e}+a'$ for  certain constants $a',e>0$. To lead to a contradiction, we want to solve $({3}\dstcon,m_{ver})$ in polynomial time using sub-linear space since this yields a contradiction against LSH for ${3}\dstcon$, and thus against LSH $\mathrm{2SAT}_3$ by Theorem \ref{LSH-equiv}.

Consider an instance $x=(G,s,t)$ given to ${3}\dstcon$ with $G=(V,E)$. Let $|V|=n$. From this graph $G$, we construct a unary 1nfa  $N=(Q,\Sigma, \delta,q_0,F)$ with $\Sigma=\{0\}$ as follows. Let $Q=V$, $F=\{t\}$, and $q_0=s$. Notice that $|Q|=m_{ver}(x)$. Moreover, we define $\delta(q,0)=\{p\in Q\mid (q,p)\in E\}$ for any $q\in V-\{t\}$ and   $\delta(t,0)=\{t\}$. It then follows that  $x\in{3}\dstcon$ iff $N$ accepts $0^k$ for any number $k\geq |Q|$.

Next, we run $A$ to construct from $N$ an equivalent unary 1dfa $M$ of, say, $m$ states, where $m\leq cn\log{n} +c$. Note that each inner state of $M$ is expressed by a certain string stored in $A$'s tape cells. Thus, to handle each inner state, we need to remember only the starting and ending cells that contain the string describing this inner state.  We generate $z=0^{\ceilings{cn\log{n}+c}+1}$ and then simulate $M$ on the input $z$ within $a'n^e+a'$ time using $O(\log{(a'n^e+a')})$ ($=O(\log{n})$) space.  If $M$ accepts $z$, then we accept $x$; otherwise, we reject $x$.
Since $|z|\geq|Q|$ and $M$ is an equivalent machine,
we  accept $x$ exactly when there exists an $s$-$t$ path in $G$. The work  space required to execute this procedure is at most $an^{\varepsilon}+c'\log{n}$, which is further upper-bounded by $c''m_{ver}(x)^{\varepsilon}$ for suitable constants $c',c''>0$.

Hence, we can solve $({3}\dstcon,m_{ver})$ in polynomial time using sub-linear space, as requested.
\end{proof}

%%%%%%%%%%%%%%%%%%
%%%%%%%%%%%%%%%%%%
\section{A Brief Discussion and Future Challenges}

The standard theory of time complexity has been concentrated on the \emph{$k$CNF Boolean formula satisfiability problem} ($k$SAT) for indices $k\geq3$; however, it has paid little attention to $2$SAT because $\mathrm{2SAT}$ is considered as a \emph{tractable} problem (i.e., a polynomial-time solvable problem). To fill the void in the study of $\mathrm{2SAT}$, this paper is particularly focused on the \emph{space complexity} of $2$SAT, while runtime is still limited to polynomials. We have demonstrated in Theorem \ref{2SAT-solvable} that a good polynomial-time algorithm can solve $\mathrm{2SAT}$ using $n^{1-c/\sqrt{\log{n}}}$ space, which is slightly larger than \emph{sub linear} in the sense of this paper (i.e., $n^{\varepsilon}$ for a certain constant $\varepsilon\in[0,1)$).

In a course of our study, we have considered a natural variant of $\mathrm{2SAT}$, $\mathrm{2SAT}_k$ for any $k\in\nat^{+}$, and we have raised a legitimate question of whether $\mathrm{2SAT}_3$ is solvable within polynomial time using only sub-linear space. In a way analogous to the \emph{exponential time hypothesis} (ETH) and the \emph{strong exponential time hypothesis} (SETH) for $k\mathrm{SAT}$ ($k\geq3$) \cite{IP01,IPZ01}, we have introduced in Section \ref{sec:new-hypothesis} a practical working hypothesis, called the \emph{linear space hypothesis} (LSH) for $2\mathrm{SAT}_3$, which asserts the impossibility of solving $\mathrm{2SAT}_3$ in polynomial time using only sub-linear space.

Throughout Section \ref{sec:application}, the hypothesis LSH for $\mathrm{2SAT}_3$ has proven to be quite useful in certain applications in the fields of $\nl$ search and $\nl$ optimization problems as well as automata theory. It is unfortunate that, even though there are five follow-up papers \cite{Yam17b,Yam18a,Yam19,Yam22a,Yam22b} published lately, we are still far away from the full understandings of the validity of this working hypothesis.

Unarguably, an ultimate goal of us is to determine whether or not LSH for $\mathrm{2SAT}_3$ is true.
Even if we cannot achieve this goal at present, we strongly hope that this hypothesis becomes a driving force to numerous discoveries of new lower bounds on computational resources needed for certain types of mathematical problems in various fields of computer science.
In what follows, for good future prospects of this research field, we intend to list seven interesting questions whose solutions may advance our understandings of the theory of space complexity.

\s
\renewcommand{\labelitemi}{$\circ$}
\begin{enumerate}\vs{-2}
  \setlength{\topsep}{-2mm}%
  \setlength{\itemsep}{1mm}%
  \setlength{\parskip}{0cm}%

\item As noted above, the biggest open question is to determine whether or not the working hypothesis LSH for $2\mathrm{SAT}_3$ is true. An affirmative solution in particular is quite difficult to obtain because, as shown in Theorem \ref{LSH-implies-L-NL}, this leads to $\dl\neq\nl$, $\logdcfl\neq\logcfl$, and $\mathrm{SC}\neq\mathrm{NSC}$. Nonetheless, a solution to this important question will surely promote our basic understanding of the power of polynomial-time sub-linear-space computation.

\item Determine whether LSH for $\mathrm{2SAT}_3$ fails iff $\para\nl\subseteq \psublin$. This is closely related to the question of whether or not $(\mathrm{2SAT}_3,m_{vbl})$ is complete for $\para\nl$ under sSLRF-T-reductions.

\item We have seen four simple applications of the hypothesis LSH for $\mathrm{2SAT}_3$ in Section \ref{sec:application}. Find other useful and practical applications of LSH for $2\mathrm{SAT}_3$ in various scientific fields. The study of various types of applications may eventually lead to the solution to the validity question of LSH for $\mathrm{2SAT}_3$.

\item The proof of Theorem \ref{2SAT-solvable}  heavily relies on the fast algorithm of Barnes \etalc~\cite{BBRS98}. Can we significantly improve the space bound of $n^{1-c/\sqrt{\log{n}}} polylog(m+n)$? To answer this question, it may be imperative to design an elaborate algorithm for $2\mathrm{SAT}$, which must be a completely different from the algorithm of Barnes et al.

\item The parameterized complexity class $\psublin$ is located in between $\para\dl$ and $\para\p$; namely, $\para\dl\subseteq \psublin \subseteq \para\p$. It is still open whether these inclusions are proper or not. There is a circumstantial evidence that answering this question may be quite difficult, because those complexity classes were shown, in certain \emph{relativized worlds}, to collapse and to be separated \cite{Yam19}.

\item Short reductions have played a special role in our study of sublinear space complexity. Concerning such reductions, prove or disprove that $(2\mathrm{SAT},m_{vbl}) \sSLRFequiv (2\mathrm{SAT},m_{cls})$ or even $(2\mathrm{SAT},m_{vbl}) \sLequiv (2\mathrm{SAT},m_{cls})$. Similarly, determine whether or not $(2\mathrm{SAT},m_{vbl}) \sSLRFequiv (2\mathrm{SAT}_3,m_{vbl})$. For more discussions on short reductions, see \cite{Yam17b,Yam22a}.

\item We have discussed the key role of $\para\mathrm{SNL}_{\omega}$ in Section \ref{sec:SNL} but its full understanding has eluded us so far. It is thus quite important to explore the structural properties of $\para\mathrm{SNL}_{\omega}$ (as well as $\para\mathrm{SNL}$ and even non-parameterized $\mathrm{SNL}$) in detail.
\end{enumerate}

%%%%%%%%%%%%%%%%%%%%%%%%%
%%%%%%%%%%%%%%%%%%%%%%%%%
\let\oldbibliography\thebibliography
\renewcommand{\thebibliography}[1]{%
  \oldbibliography{#1}%
  \setlength{\itemsep}{0pt}%
}

%%%%%%%%%%%%%%%%%
%%%%%%%%%%%%%%%%%
%%%%%%%%%%%%%%%%%%%%%%%%%%%%%%%%%%%%%%%%%%%%%%%%%%%
%%%%%%%%%%%%%%%%%%%%%%%%%%%%%%%%%%%%%%%%%%%%%%%%%%%
\end{document}